\documentclass[aps, twocolumn, superscriptaddress, pra]{revtex4-2}
\usepackage{ORI_Group_style}
\usepackage{textcomp}
\usepackage{amsmath}

\newcommand{\veps}{\varepsilon}

\newcommand{\ppop}{\hat{\pp}}
\newcommand{\sigop}{\hat{\sigma}}
\newcommand{\Sigop}{\hat{\Sigma}}
\newcommand{\Wop}{\hat{W}}
\newcommand{\kkL}{\kk_\text{L}}

\newcommand{\wT}{\w_\text{T}}
\newcommand{\ii}{\bold{i}}
\newcommand{\rrop}{\hat{\rr}}
\newcommand{\rop}{\hat{r}}
\newcommand{\rhop}{\hat{\rho}}
\newcommand{\wL}{\w_\text{L}}
\newcommand{\nth}{n_\text{th}}

\newcommand{\Matlab}{MATLAB\textregistered}

\newcommand{\sqSWAP}{\sqrt{\text{iSWAP}}}

\begin{document}

\title{Exploiting the Photonic Non-linearity of Free Space Subwavelength Arrays of Atoms}

\author{Cosimo C. Rusconi}
\affiliation{Max-Planck-Institut f\"ur Quantenoptik, Hans-Kopfermann-Strasse 1, 85748 Garching, Germany.}
\affiliation{Munich Center for Quantum Science and Technology, Schellingstrasse 4, D-80799 M\"unchen, Germany.}
\author{Tao Shi}
\affiliation{CAS Key Laboratory of Theoretical Physics, Institute of Theoretical Physics, Chinese Academy of Sciences, Beijing 100190, China.}
\affiliation{CAS Centre of Excellence in Topological Quantum Computation, University of Chinese Academy of Sciences, Beijing 100049, China.}
\author{J. Ignacio Cirac}
\affiliation{Max-Planck-Institut f\"ur Quantenoptik, Hans-Kopfermann-Strasse 1, 85748 Garching, Germany.}
\affiliation{Munich Center for Quantum Science and Technology, Schellingstrasse 4, D-80799 M\"unchen, Germany.}

\date{\today}

\begin{abstract}
Ordered ensembles of atoms, such as atomic arrays, exhibit distinctive features from their disordered counterpart. In particular, while collective modes in disordered ensembles show a linear optical response, collective subradiant excitations of subwavelength arrays are endowed with an intrinsic non-linearity. Such non-linearity has both a coherent and a dissipative component: two excitations propagating  in the array scatter off each other leading to formation of correlations and to emission into free space modes.
We show how to take advantage of such non-linearity to coherently prepare a single excitation in a subradiant (dark) collective state of a one dimensional array as well as to perform an entangling operation on dark states of parallel arrays.
We discuss the main source of errors represented by disorder introduced by atomic center-of-mass fluctuations, and we propose a practical way to mitigate its effects.
\end{abstract}

\maketitle

%================
%================
\section{Introduction}
%================
%================

Ordered atomic ensembles (arrays) have recently attracted significant attention as a new paradigm for controlling light-matter interaction which shows novel features not  shared by their disordered counterpart~\cite{Chang2018}. When the interatomic separation is subwavelength with respect to the characteristic atomic dipole transition, the optical response of atomic arrays shows a strong collective behaviour characterised by bright (superradiant) and dark (subradiant) excitations~\cite{Dicke1954,Lehmberg1970a,Lehmberg1970b}.
Superradiant states allow for efficient coupling of internal atomic states to light, while subradiant states permit long coherent storage of atomic excitations thanks to their reduced linewidth. 
Additionally, the array's collective response can be used to realise perfect reflection of light off the array~\cite{AntezzaPRL2009,*AntezzaPRA2009,BettlesPRL2016,BettlesPRA2016,Shahmoon2017,Rui2020} and to prepare topological edge modes~\cite{Bettles2017,Perczel2017}.

This combination of features contains some of the basic elements for applications in quantum information technologies, as one could store quantum information in subradiant states~\cite{AsenjoGarcia2017PRX} and read it out using superradiant excitations~\cite{Porras2008}. However, for the creation and manipulation of quantum information one also requires a non-linear optical response, \ie the dependence of light-matter interaction on the system's internal state.
Specifically, a non linear response allows to define qubits and to perform universal gate operations on them.
Optical non-linearities are usually obtained by adding a new capability to the system.
In the context of an atomic ensemble, for example, a non-linear response is obtained promoting atoms to a Rydberg state: strong dipolar interactions between two Rydberg excitations can be exploited to inhibit further absorption of photons from the ensemble (Rydberg Blockade)~\cite{Lukin2001}. Rydberg excitations have also recently been considered for atomc arrays~\cite{Bekenstein2020,MorenoCardoner2021}

In this article, we propose and analyze an alternative way to produce and coherently manipulate quantum information with ordered atomic ensembles, which does not rely on Rydberg excitations or other technologies. 
We show how one can harness the intrinsic non-linear response of subwavelength arrays of atoms in free space~\cite{Masson2020,Williamson2020,Cidrim2020,Bettles2020} to perform different tasks. Specifically, we discuss (i) a procedure to transfer an excitation from the ground to the single-excitation most-subradiant collective state of the system and (ii) a procedure to prepare an entangled state shared by two parallel arrays. Together these two tasks allow to perform a universal set of quantum gates on atomic arrays.
We discuss the nature of this intrinsic non-linear response and show that it has both a coherent and a dissipative component. 
The coherent component is rooted in the large dipole shift between closely spaced atoms; It dominates at smaller lattice spacing but diminishes for longer arrays.
The dissipative component arises from the difference in the decay rate of the subradiant mode and the (enhanced) decay rate of the double excited mode, and allows for coherent manipulation via the Zeno effect~\cite{Facchi2002}. Notably, for the case when atoms are pinned to their position, the dissipative non-linearity has more prominent effects for longer arrays.
Finally, we analyze the impact of atomic center-of-mass fluctuations on the proposed scheme. We find they represent the main source of imperfections for they are responsible for a suppression of the intrinsic dissipative non-linearity~\cite{AntezzaPRL2009,Guimond2019}. We show numerically that such detrimental effects are reduced when the time scale of atomic motion is shorter than the internal decay rate, and propose a practical way in which this regime can be obtained.

The article is organised as follows. We first consider the case where atoms are pinned to their lattice positions (\secref{sec:PinnedAtoms}). Within this simplified assumption, we present the main ideas and discuss how to perform single and two-qubit gates between the ground and the single-excitation subradiant state of atomic arrays.
In \secref{sec:Motion}, we consider the effect of the atoms' motion. We propose a way to enter the regime of fast atomic motion, and analyze the gate fidelities in this case.
In \secref{sec:Discussion}, we discuss a possible experimental implementation of our proposal in the context of neutral atoms in optical lattice and analyze limitations and additional assumptions behind our model.
We draw our conclusions in \secref{sec:Conclusions}. 
Additional non-essential details are left to the appendices.

%============================================
%============================================
\section{Arrays of Pinned Atoms}\label{sec:PinnedAtoms}
%============================================
%============================================

\begin{figure*}
	\includegraphics[width=2\columnwidth]{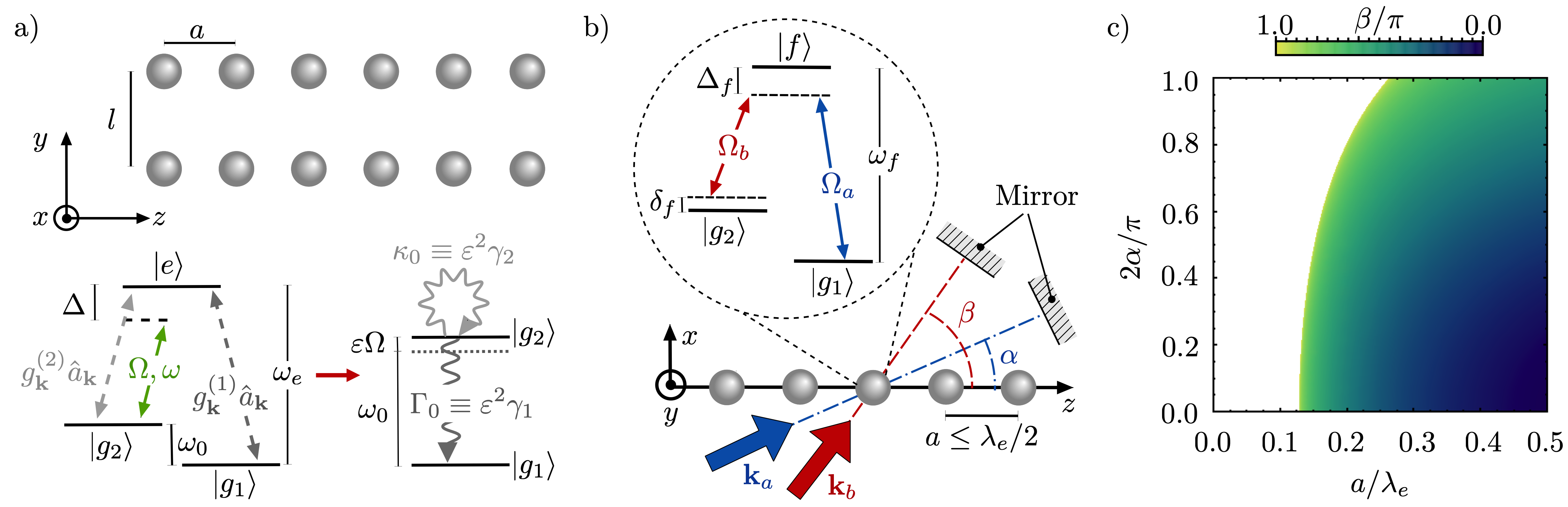}
	\caption{{\bf System description.} a) Illustration of two parallel arrays with lattice spacing $a$ and separation $l$. Internal level structure: Raman transition leads to effective two-level systems dynamics with damping $\Gamma_0=\veps^2\gamma_1$ and effective dephasing $\kappa_0\equiv \veps^2 \gamma_2$, where $\gamma_{1,2}$ is the spontaneous emission rate for the transition $\ket{e}\leftrightarrow\ket{g_{1,2}}$. b) Scheme for driving a subradiant (dark) mode of a subwavelength array. The detuning $\delta_f$ can be chosen to resonantly drive a collective mode of the array. c) Colored region: values of $\alpha$ and $a/\lambda_e$ for which a solution for $\beta$ in \eqnref{eq:q_effective} exists. Other parameters $\w_a\equiv \w_f-\Delta_f=2\w_e$, $\w_g=0.1\w_e$, and $K=\pi/a$.}\label{fig:Fig1}
\end{figure*}

We consider two parallel atomic arrays labelled ``A'' and ``B'' and placed at a distance $l$ from each other. Each array contains $N$ atoms separated by a lattice spacing $a$ (see~\figref{fig:Fig1}.a). 
In the following, we consider the case in which the arrays have unit filling and the atoms are pinned to their lattice positions.
The atoms' internal structure is described by a lambda scheme with one excited state $\ket{e}$ coupled by a dipole-allowed transition to the ground state levels $\ket{g_1}$ and $\ket{g_2}$ (\figref{fig:Fig1}.a). 
An external laser drives the atoms on the $\ket{e}\leftrightarrow\ket{g_2}$ transition with a detuning $\Delta$ and at a rate $\W$. For $\veps\equiv\W/2\Delta\ll1$, the excited state $\ket{e}$ is never populated and the atom behaves as an effective two-level system with excited state $\ket{g_2}$ and ground state $\ket{g_1}$, characterised by an effective decay rate $\Gamma_0\equiv \veps^2 \gamma_1$ and dephasing rate $\kappa_0 \equiv \veps^2 \gamma_2$.
We further assume $\ket{e}\leftrightarrow\ket{g_1}$ to be a much stronger transition than $\ket{e}\leftrightarrow\ket{g_2}$, so that $\kappa_0\ll\Gamma_0$. In the following, we thus neglect the dephsing arising when driving the $\ket{e}\rightarrow\ket{g_2}$ transition and discuss later under which conditions this approximation can be justified (see~\secref{sec:Discussion}).
The additional complication introduced by the lambda scheme, as opposed to using simple dipole-coupled levels, is instrumental for mitigating the effects of the motion as explained in \secref{sec:Motion}~\footnote{This is however not necessary for the case of pinned atoms.}.

The effective dynamics of the atoms treated as an open quantum system can be derived from the total Hamiltonian describing the dynamics of the atom-light interaction in the dipole approximation. 
Tracing out the electromagnetic field within the Born-Markov approximation,  the dynamics of the system is described by the following non-hermitian Hamiltonian~\cite{Lehmberg1970a,*Lehmberg1970b} (see also~\appref{app:ME_derivation})
\be\label{eq:Ham_2Arrays_Pinned}
	\Hop_0 = \Hop_\text{A}+\Hop_\text{B} + \Hop_\text{AB}. 
\ee
The non-hermitian Hamiltonian $\Hop_\nu$ describes the dynamics of the atoms of array $\nu=\text{A,B}$, and reads
\be\label{eq:Ham}
\begin{split}
\frac{\Hop_\nu}{\hbar} \equiv  \w_g \sum_j \spl_{\nu j}\smi_{\nu j}+\veps^2\sum_{i,j}\pare{J_{ij}-\im\frac{\Gamma_{ij}}{2}}\spl_{\nu j}\smi_{\nu i},
\end{split}
\ee 
where $\w_g$ is the splitting between $\ket{g_2}$ and $\ket{g_1}$, $\spl_{\nu j} = (\smi_{\nu j})^\dag \equiv \ket{g_{2,j}}_\nu\bra{g_{1,j}}$ and $j=1,\ldots N$ labels the atom at position $\RR_{\text{A}j}=(0,0, j a)^T$ ($\RR_{\text{B}j}=(0,l, j a)^T$) within array A (B).
Here, $J_{ij}$ ($\Gamma_{ij}/2$) is the coherent (dissipative) interaction between two atoms at sites $i$ and $j$ within the same array, and for $i\neq j$ it is given by~\citep{Lehmberg1970a,*Lehmberg1970b}
\be\label{eq:G_coupling}
	J_{ij}-\im\frac{\Gamma_{ij}}{2} = -\frac{\mu_0\w_e^2}{\hbar} \dd_{\nu i} \cdot \GG^0(\RR_{\nu i}-\RR_{\nu j},\w_e)\cdot \dd_{\nu j}
\ee
where $\GG^0(\RR,\w_e)$ is the free space electromagnetic Green's tensor of a point dipole, $\w_e$ the frequency of the $\ket{e}\leftrightarrow\ket{g_1}$ transition, and $\dd_i$ is the electric dipole moment of the atom  at site $i$. For $i=j$, we defined $\Gamma_{ii}=\gamma_1$ and we implicitly included the vacuum Lamb-shift $J_{ii}$ into the definition of $\w_g$. Hereafter, we consider all atoms to be polarized parallel to the array's direction ($z$-axis) unless otherwise specified. 
A different atomic polarization leads to results qualitatively similar to the ones presented here, as shown in~\appref{app:Perp_Polarization}.
The last term in \eqnref{eq:Ham_2Arrays_Pinned} describes the interaction between the atoms in the two arrays and reads
\be\label{eq:Ham_int}
	\frac{\Hop_\text{AB}}{\hbar}= \veps^2\sum_{i,j=1}^N\pare{g_{ij}-\im \frac{\gamma_{ij}}{2}}\pare{\spl_{\text{A}i}\smi_{\text{B}j}+\smi_{\text{A}i}\spl_{\text{B}j}}
\ee
where $g_{ij}$ ($\gamma_{ij}/2$) is the coherent (dissipative) part of the dipole coupling given by
\be\label{eq:g_coupling}
 g_{ij}-\im\frac{\gamma_{ij}}{2}\equiv -\frac{\mu_0\w_e^2}{\hbar} \dd_{\text{A}i} \cdot \GG^0(\RR_{\text{A}i}-\RR_{\text{B}j},\w_e)\cdot \dd_{\text{B}j}.
 \ee
Note that as a consequence of the Raman transition used to define the effective two level atoms, the dipole-dipole interactions in \eqnref{eq:Ham} and \eqnref{eq:Ham_int} are proportional to the Green's tensor evaluated at $\w_e$. Hence, despite the effective two level system having a characteristic wavelength $\lambda_g\equiv \w_g/2\pi c$, the array exhibit strong collective dipolar effects only for $a<\lambda_e/2$ (subwavelength condition~\cite{Porras2008,Kornovan2016,BettlesPRA2016,AsenjoGarcia2017PRX}).

To better understand the origin of the non-linear response of atomic arrays and how one can use it for preparing the arrays in a specific quantum state it is useful to consider first the case of a single array.

% -----------------------------------------------------------
\subsection{Single Array: Driving Subradiant Excitations}
% -----------------------------------------------------------

The dynamics of an isolated array of atoms in free space is described by \eqnref{eq:Ham}. Because we consider only one array, we drop the label $\nu=\text{A,B}$ and refer to the single array Hamiltonian as $\Hop_\text{1array}$ in this section. 
The eigenstates of \eqnref{eq:Ham} are simultaneous eigenstates of $\hat{N}_\text{e} \equiv \sum_j^N \spl_j\smi_j$, as both the hermitian and anti-hermitian part of \eqnref{eq:Ham} commute with $\hat{N}_\text{e}$.
Within the single excitation manifold and when the atomic array is sufficiently long ($N\gg1$), the eigenmodes of \eqnref{eq:Ham} can be understood as spin waves of a definite quasi-momentum $k$, where the value $k$ corresponds to the point in reciprocal space where the eigenmode wavefunction is peaked~\citep{Mewton2007,Porras2008,Kornovan2016,AsenjoGarcia2017PRX}.
We define the eigenmode with associated quasi-momentum $k$ as $\ket{k}\equiv\hat{\Sigma}^+_k \ket{0}$, where $\ket{0}\equiv  \ket{g_1}^{\otimes N}$ is the ground state of \eqnref{eq:Ham}, and $\Sigop_k^\pm\equiv\sum_j c_{k,j} \hat{\sigma}_j^\pm$ with $c_{k,j}\equiv\braket{e_j}{k}$.
For $a<\lambda_e/2$, there exist single excitation eigenstates of \eqnref{eq:Ham} with 
a quasi-momentum $q>\w_e/c\equiv k_e$ lying outside the light-cone of free space electromagnetic modes (hereafter $q$ labels values of the quasi-momentum which lie outside the light cone unless otherwise specified). These collective states, which are intuitively understood as an excitation propagating along the array, have been shown to decay at a rate $\sim \Gamma_0/N^3$~\citep{AsenjoGarcia2017PRX,Zhang2019}, and are referred to as collective subradiant modes or dark modes. 
Dark modes also exist in higher $n$-excitation manifolds, provided $n\ll N$, and are similarly characterized by a scaling of the decay rate $\sim\Gamma_0/N^3$. 
Notably, the $n$-excitation states $\ket{n q} \equiv (\Sigop_q^+)^n\ket{0}$ are not eigenstates of  \eqnref{eq:Ham}. This results in a non-linear structure of the dark-mode spectrum which has both a coherent and dissipative component. The coherent component is represented by a non-linear spacing of the energy levels with the number of excitation and it is quantified by the difference $\Delta_n \equiv \w_n-n\w_1$, where $\w_n$ ($\w_1$) is the energy of the most-subradiant $n$-excitation (single excitation) state. The dissipative component is represented by the enhanced decay rate of $\ket{n q}$ which scales as $\sim \Gamma_0/N$ to first order in  \eqnref{eq:Ham}~\citep{AsenjoGarcia2017PRX}. 
We will show that $\Delta_n$ approaches zero for increasing $N$, while the dissipative non-linearity grows with the array's size.

To excite a collective dark mode of a subwavelength array, it is necessary for the driving to match energy and modulation of the target state~\footnote{Note that this cannot be simply achieved with a single external laser as subwavelength excitation lie outside of the light-cone of free space electromagnetic modes.}. This condition can be achieved using a detuned Raman transition via an additional excited state $\ket{f}$. As illustrated in~\figref{fig:Fig1}.b, we consider two driving lasers with wave vectors $\kk_a,\kk_b$ forming respectively angles $\alpha,\beta$ with the array's direction ($z$-axis). For a sufficiently large detuning $\Delta_f$ such that $\ket{f}$ is never populated, the effect of the Raman lasers on the array can be modelled by an effective driving Hamiltonian
\be\label{eq:Vop}
	\Vop_\text{1array} \equiv \im\hbar \W_0 \sum_{j=1}^N \sin(K_z z_j)\pare{e^{-\im\w_d t} \spl_j -\hc},
\ee
Here, $z_j\equiv aj$, $\w_d \equiv \w_a-\w_b$, and $\W_0$ is the single atom effective Rabi-frequency which depends on the details of the two-photon transition (see~\appref{app:Driving}). 
\eqnref{eq:Vop} excites a collective spin wave of the array with an effective quasi momentum
\be\label{eq:q_effective}
	K_z \equiv k_a \cos\alpha  -k_b\cos\beta.
\ee
For $\w_f-\Delta_f = 2\w_e$, we show in \figref{fig:Fig1}.c under which conditions on the angle $\alpha$ and lattice spacing $a$ it is possible to match the driving with the most subradiant single-excitation state, namely $K_z=q_a\equiv\pi/a$.
The driving Hamiltonian \eqnref{eq:Vop} produces Rabi oscillations at frequency $\W_0\sqrt{N}$ between the atomic ground state and the collective state $\ket{\psi_{q_a}} \equiv \sqrt{2/(N+1)} \sum_{j=1}^N \sin(z_j q_a)\spl_j \ket{0}$. The state $\ket{\psi_{q_a}}$ approximates the actual dark mode $\ket{q_a}$ with an overlap-error scaling as $\sim 1/N^2$~\cite{AsenjoGarcia2017PRX}.

\begin{figure}
	\includegraphics[width=\columnwidth]{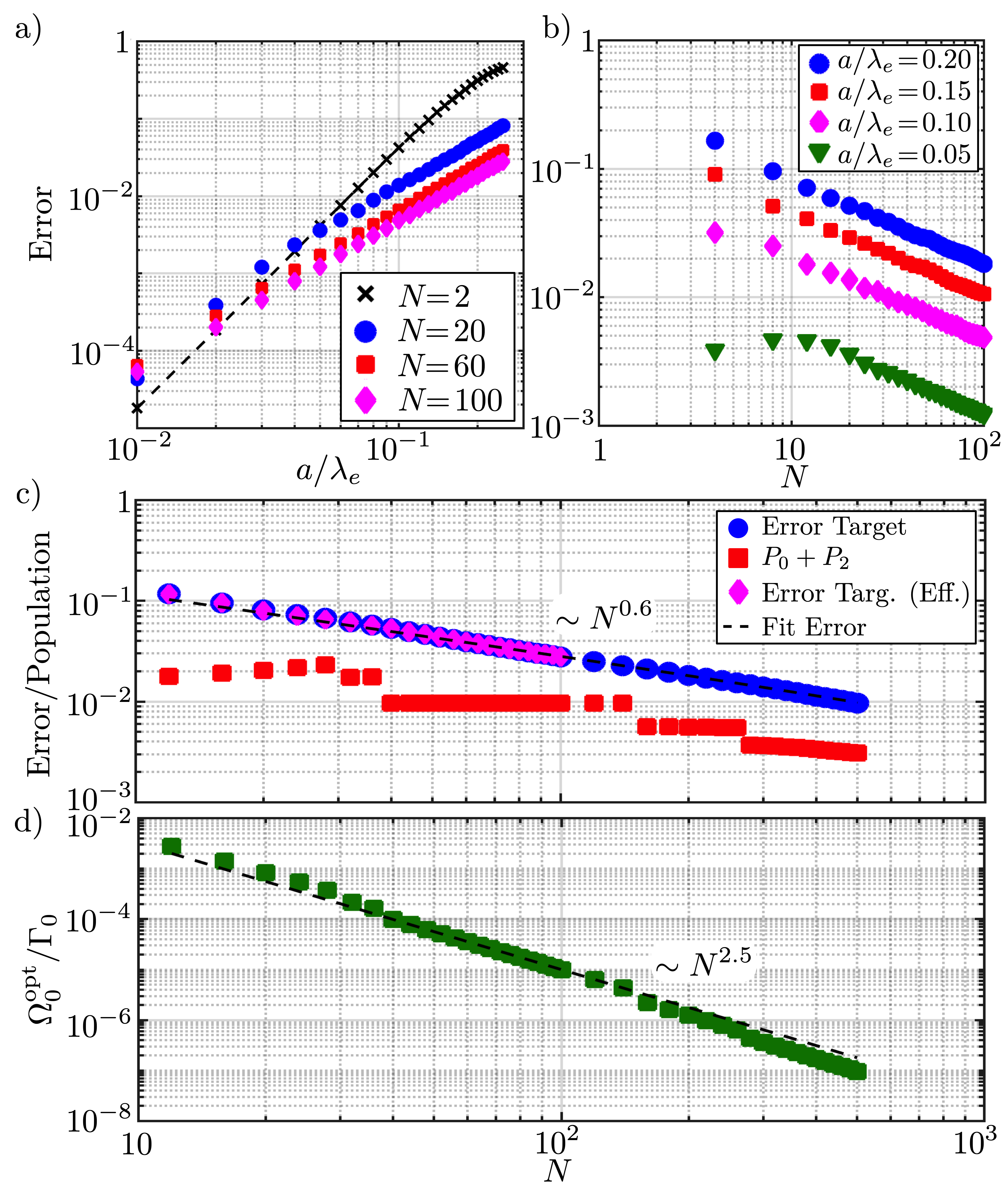}
	\caption{{\bf Single Array's Dark Mode Preparation.} a) Error for dark state preparation as function of the lattice spacing $a/\lambda_e$ (for different array's length $N$) and b) as function of the number of atoms $N$ (for different values of $a/\lambda_e$). c) Scaling with $N$ for $a=\lambda_e/4$ of different relevant quantities at the optimal driving frequency $\W_0^\text{(opt)}$: error $\epsilon$ (infidelity) (blue circles), total population in the ground state ($P_0$) and two excitation manifold ($P_2$) at the time which minimizes the error (red squares), error calculated with the effective three-level model Hamiltonian \eqnref{eq:Heff_app}  (pink diamond), and fit function $\varepsilon\approx 0.45\times N^{-0.6}$ (black dashed line). d) $N$-dependence of the optimal frequency $\W_0^\text{opt}$ at which the minimal error occurs for $a=\lambda_e/4$. Scaling $N^{-2.5}$ as guide for the eye (black-dashed line).}\label{Fig:Fig2}
\end{figure}

To study the fidelity of the subradiant state preparation, we numerically simulate the evolution of the open system under the condition of no jump occurring. Specifically, we calculate the state at time $t$ as $\ket{\psi(t)}=\exp[-\im (\Hop_\text{1array}+V_\text{1array}) t/\hbar]\ket{0}$, where we chose $\w_d$ in resonance with the energy of the target state $\ket{q_a}$ (we refer to \appref{app:Numerics} for details on the method used for the simulation).
The error in the target state preparation is calculated as $\epsilon \equiv 1-\text{max}_t[\mathcal{F}(t)]$ where $\mathcal{F}(t)\equiv |\braket{q_a}{\psi(t)}|^2$~\footnote{This gives an upper bound to the Fidelity as given by the full master equation}.
In \figref{Fig:Fig2}, we illustrate the dependence of the dark state  preparation's error on both the lattice spacing $a/\lambda_e$ and the number of atoms $N$ in the array.
The data shown in \figref{Fig:Fig2}.a,b are obtained optimizing the fidelity with respect to the Rabi-frquency in \eqnref{eq:Vop}.
\figref{Fig:Fig2}.a,b shows the two different contribution of the array's non-linear response: the error decreases for smaller lattice spacing and for larger $N$ as a consequence respectively of the dipole shift $\Delta_2$ and of enhanced decay of the double excited state as compared to the decay of the target state (Zeno effect).
The fidelity $\mathcal{F}$ improves for smaller $a/\lambda_e$ at a rate which reduces with $N$, as expected from the reduction of $\Delta_2$ for longer arrays (\figref{Fig:Fig2}.a). This effect is particularly evident when comparing the case of $N=2$ with the other lines, but it is present also for larger array sizes as shown by the crossing of the results for $N=60$ and for $N=100$ with the results for $N=20$ at $a/\lambda_e\approx0.01$. This implies that for a fixed value of $a/\lambda_e$, increasing $N$ might at first lead to an increased error. Ultimately however the error will improve for larger $N$ at fixed $a/\lambda_e$ (see the case of $a/\lambda_e=0.05$ in \figref{Fig:Fig2}.b). 
The improvement in the error with the array's size $N$ is a signature of the Zeno effect. In particular, when $N^{-7/2}\ll \W_0/\Gamma_0 \ll N^{-3/2}$, it is possible to efficiently excite $\ket{q_a}$ while the population transfer to the doubly excited state $\ket{2 q_a}$, and hence to higher excited manifolds containing $n>2$ excitations, is suppressed by the enhanced decay.
The contribution of the Zeno effect to the non-linear response is prominent for larger interatomic separations.
In \figref{Fig:Fig2}.c, we extend the results in \figref{Fig:Fig2}.b for the case of $a=\lambda_e/4$ to much larger array sizes, showing an improvement os the error for larger $N$. This improvement is only due to the Zeno effect as proven in \figref{Fig:Fig2}.d, where the optimal frequency $\W_0^\text{opt}$ as function of $N$ shows a scaling $\sim 1/N^{2.5}$ falling within the limit of the Zeno regime.
At $\W_0^\text{opt}$, the error $\epsilon$ has two major contributions coming from (i) the undesired transfer of population to levels other than the target and (ii) the finite decay rate $\Gamma_{q_a}$ of the target state $\ket{q_a}$. The latter alone would predict an error scaling $\epsilon\sim N^{-1}$ as $\Gamma_{q_a}/(\W_0^\text{opt}\sqrt{N}) \sim 1/N$. The main limitation to the state preparation fidelity is thus mainly due to population transferred to the most-subradiant two-excitation state due to imperfect Zeno-blockade (\figref{Fig:Fig2}.c). 
The dynamics of the driven array within the Zeno regime can thus be captured by a simple three level model involving only the ground state $\ket{0}$, the target state $\ket{q_a}$, and the most subradiant two-excitation state $\ket{2}$ (see~\appref{app:Heff_3levels}), as shown by the pink diamond markers in \figref{Fig:Fig2}.c.

We define logical qubit states of an array as $\ket{0}_\text{L}\equiv \otimes^N \ket{g_1}$ and $\ket{1}_L \equiv \ket{q_a}= \sum_j c_{q_a,j}\ket{g_{2,j}}$, where $c_{q_a,j}\equiv \braket{g_{2,j}}{q_a}$. 
Note also that, once the collective dark mode has been prepared, it is possible to turn off the dipole coupling between the levels $\ket{g_1}$ and $\ket{g_2}$ by turning off the laser which couples $\ket{g_2}$ to $\ket{e}$ ($\veps=0$ in \eqnref{eq:Ham} and \eqnref{eq:Ham_int}). In this way, the dark state becomes a metastable state which can be stored for long in the array. 
Let us now show how, given two parallel arrays of atoms, one can perform entangling operations between the logical states of the arrays.

% ----------------------------------------------------------------------
\subsection{Two Parallel Arrays: Entangling $\sqrt{\text{iSWAP}}$-Gate via dipole-dipole interactions}\label{ssec:PinnedAtoms_Parallel}
% ----------------------------------------------------------------------

\begin{figure*}
	\includegraphics[width=2\columnwidth]{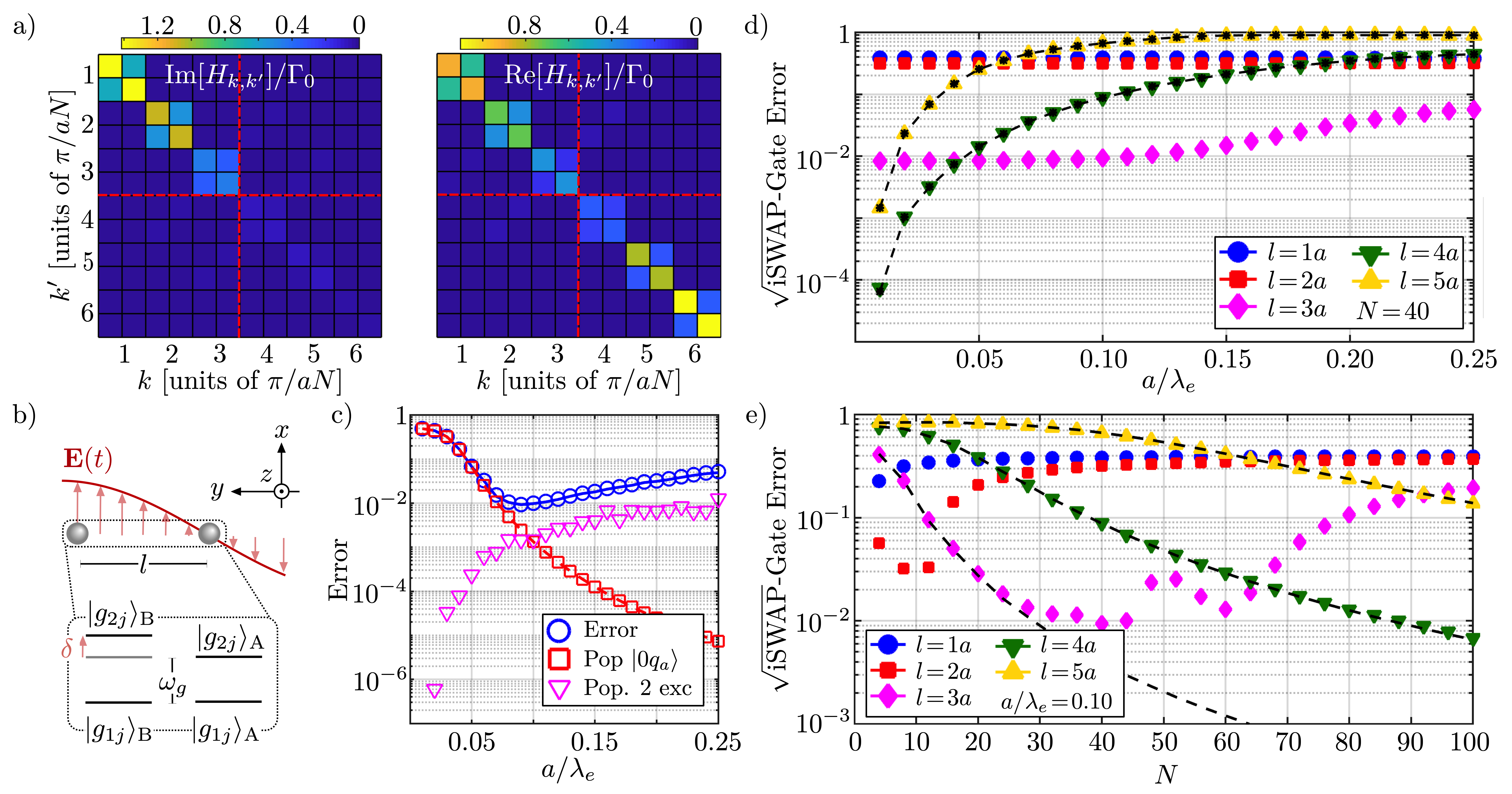}
	\caption{{\bf Parallel Arrays of Pinned Atoms.} a) Left (right) Panel: absolute value of the imaginary (real) part of the matrix elements of $\Hop_2^{(1)}$ on the basis $\{\ket{0k},\ket{k0}\}_k$, namely $H_{kk'}\equiv \bra{\phi_k}\Hop_2^{(1)}\ket{\phi_{k'}}$ where $\ket{\phi_k}
	\in\{\ket{0,k},\ket{k,0}\}$. Diagonal $2\times2$ blocks corresponds to $\Hop_k$, while of diagonal elements are cross coupling between states of different momentum. The light line is marked by red dashed lines dividing subradiant and superradiant sectors. b) Scheme for selective AC Stark Shift to detune one array with respect to the other. c) Error for preparing the state $\ket{q_a 0}$ as function of $a/\lambda_e$ for two parallel array of $N=40$ atoms. Other parameters: $l/a=1$ and $\delta/\Gamma_0=100$. d) Error $\sqSWAP$-gate as function of $a/\lambda_e$ for $N=40$ atoms. Different markers corresponds to different separation $l/a$ between the arrays as specified by the legend. e) Error $\sqSWAP$-gate as function of $N$ for $a/\lambda_e=0.10$. In both panel d) and e), black dashed lines with star markers represent the case case in which the dynamics of the states $\ket{10}_\text{L},\ket{01}_\text{L}$, and $\ket{11}_\text{L}$ is assumed to be an exponential decay at rates $\Gamma_{q_a}$ and $2\Gamma_{q_a}$.}\label{fig:Fig3}
\end{figure*}

The dynamics of two parallel arrays is described by the Hamiltonian in \eqnref{eq:Ham_2Arrays_Pinned}.
As for the case of a single array, the Hamiltonian can be diagonalized separately for each number of excitations.
For the case of a single excitation and infinite arrays, it is possible to prove that the dipole interaction between the two arrays, \eqnref{eq:Ham_int}, couples only state with the same quasi-momentum and thus we can write the single excitation Hamiltonian of two parallel arrays as $\Hop_0^{(1)}=\sum_k \Hop_k$, where $k=\pi/[a(N+1)]\ldots \pi N/[a(N+1)] $ and
\be\label{eq:H_k_block}
\Hop_k\equiv
\begin{pmatrix}
 \w_k & g_k-\im\gamma_k/2\\
 g_k-\im\gamma_k/2 & \w_k+\delta
\end{pmatrix},
\ee
on the basis $\{\ket{k 0},\ket{0k}\}$. Here, $\ket{k0}$ ($\ket{0k}$) is the state with one excitation of momentum $k$ in the array A (B), and $g_k$ ($\gamma_k/2$) is the coherent (dissipative) coupling between $\ket{k0}$ and $\ket{0k}$. 
For an infinite array, an exact formula can be derived for $g_k-\im\gamma_k/2$ (see~\appref{app:Parallel_Arrays}). In this limit, for the case of atoms polarised along the array's axis, the coupling between subradiant modes ($k>k_e$) of infinite arrays reads
\be\label{eq:g_k}
	g_k =-\frac{3\Gamma_0}{k_e a}\pare{1-\frac{k^2}{k_e^2}}K_0\pare{l\sqrt{k^2-k_e^2}},
\ee
and $\gamma_k=0$, where $K_0(x)$ is the modified Bessel function of the second kind.
\eqnref{eq:g_k} holds to a high degree of accuracy also for finite arrays provided $N\gg 1$. In particular, the coupling is coherent and falls off approximately exponentially with the array separation $l$ as described by \eqnref{eq:g_k}.
While the decoupling between states of different quasi-momentum is strictly true only for infinitely long arrays (and for the particular case of $N=2$), in \figref{fig:Fig3}.a we show that already for arrays as small as $N=6$ atoms for $l=a=\lambda_e/4$ the coupling between different $k$-blocks is negligibly small.
Such a strong suppression of the cross coupling between modes associated to different quasi-momenta $k$ can be traced back to the accuracy of approximating the exact system's eigenmodes with quasi-momentum eigenstates $\ket{k}$. Indeed, the cross coupling is larger around the light cone separating bright from dark states where this approximation is less accurate~\cite{AsenjoGarcia2017PRX}.

In \eqnref{eq:H_k_block}, we included a detuning $\delta$ between the two arrays. For arrays separated by $l<\lambda_e$, such a selective detuning could be obtained via an AC Stark-Shift. Using a standing wave and  placing the array A in a node of the field, only the atoms in the array B pick up a shift $\delta$ (\figref{fig:Fig3}.b). Alternative ways using electrostatic or magnetostatic field gradient could also be envisioned. 
When the arrays are separated by a distance comparable to $\lambda_e$, they are collectively driven according to the Hamiltonian (in a frame rotating at $\w_d$)
\be\label{eq:Vop_2array}
	\Vop_0 \equiv \im\hbar\W_0 \sum_{j=1}^N \sin(K_z z_j)\pare{ \sy_{\text{A}j} +\sy_{\text{B}j}}.
\ee
Hence, the possibility of detuning one array with respect to the other is instrumental for selectively addressing one array.
In \figref{fig:Fig3}.c, we show the scaling of the fidelity for preparing the state $\ket{10}_\text{L}\equiv\ket{q_a 0}$ of array A as function of the lattice spacing $a/\lambda_e$ for $l=a$ when array B is detuned by $\delta=100\Gamma_0$. The state preparation error (blue circles in \figref{fig:Fig3}.c) decreases for smaller lattice separation $a/\lambda_e$ owing to the larger dipole shift as for the case of a single array. Accordingly, the population transfer to the two excitation manifold also decreases with $a/\lambda_e$.  For $a/\lambda_e\lesssim 0.08$ the error increases if one further reduces the lattice separation because the detuning $\delta$ is now comparable to or smaller than the dipole coupling between $\ket{10}_\text{L}$ and $\ket{01}_\text{L}$ leading to substantial population transfer to $\ket{01}_\text{L}$.
This explanation can be confirmed with an effective four-level model obtained by projecting the full Hamiltonian \eqnref{eq:Ham} on the subspace spanned by the levels $\{\ket{00}_\text{L},\ket{10}_\text{L},\ket{01}_\text{L},\ket{\psi_2}\}$, where $\ket{\psi_2}$ is the most subradiant two-excitation state of the parallel-arrays system. The error and the population transferred to $\ket{01}_\text{L}$ as calculated by this simple model are shown in \figref{fig:Fig3}.c by the solid blue and dashed red line respectively.
For larger separation $l$ the coupling between $\ket{10}_\text{L},\ket{01}_\text{L}$ decreases leading to better state preparation fidelities at smaller values of $a/\lambda_e$ as compared to the case of $l=a$ shown in~\figref{fig:Fig3}.c.
We note that by applying this procedure sequentially to both arrays, it is possible to initialise the system into any states of the computational basis $\{\ket{00}_\text{L},\ket{01}_\text{L},\ket{10}_\text{L},\ket{11}_\text{L}\}$.

The structure of the two excitation manifold of two parallel arrays is more complicated. Two-excitation eigenstates of \eqnref{eq:Ham_2Arrays_Pinned} can still be classified between bright and dark modes depending on their decay rate, however the exact form of such states shows a strong dependence on the separation $l$ between the arrays. For large separation $l\gg a$, the two arrays are non-interacting and the most-subradiant two excitation state is simply given by $\ket{\psi_2}=\ket{11}_\text{L}\equiv\ket{q_a,q_a}$ and it is characterised by a decay rate $\Gamma_{11}$ which tends to $2\Gamma_{q_a}$ in the limit of $l/a\rightarrow \infty$. As the arrays are brought closer, the dipole-dipole interaction $\Hop_\text{AB}$ becomes stronger thus substantially modifying the form of the most-subradiant two-excitations state (see~\appref{app:Parallel_Arrays}). In particular for $l\sim a$, $\ket{11}_\text{L}$ strongly couples to a large number of states in the two-excitations manifolds, exhibiting a generally complicated dynamics.

Let us now discuss how one can realize an entangling gate between the computational states $\ket{00}_\text{L}$, $\ket{01}_\text{L}$, $\ket{10}_\text{L}$, and $\ket{11}_\text{L}$ of two parallel arrays.
The resonant dipole-dipole interaction \eqnref{eq:Ham_int} between two parallel arrays naturally implements a $\sqSWAP$ gate on the time scale $T_\text{g} \equiv \pi /4g_{q_a}$.
The system's dynamics on the computational subspace $\{\ket{00}_\text{L},\ket{10}_\text{L},\ket{01}_\text{L},\ket{11}_\text{L}\}$ realises the following mapping
\be\label{eq:TT_real_sqSWAP}
\begin{split}
\ket{00}_\text{L} \rightarrow&\, \ket{00}_\text{L},\\
\ket{10}_\text{L} \rightarrow&\, e^{-\Gamma_{q_a} T_\text{g}}\pare{\ket{10}_\text{L}-\im \ket{01}_\text{L}}/\sqrt{2},\\
\ket{01}_\text{L} \rightarrow&\, e^{-\Gamma_{q_a} T_\text{g}}\pare{\ket{01}_\text{L}-\im \ket{10}_\text{L}}/\sqrt{2},\\
\ket{11}_\text{L} \rightarrow&\, \xi \ket{11}_\text{L}.
\end{split}
\ee
Compared to the truth table of the ideal $\sqSWAP$-gate ($\Gamma=0$ and $\xi=1$), the realisation in \eqnref{eq:TT_real_sqSWAP} shows two main sources of imperfection which come from (i) the error $\epsilon_1$ due to finite decay $\Gamma_{q_a}$ of the computational states $\ket{10}_\text{L}$ and $\ket{01}_\text{L}$ and (ii) the error due to loss of population from $\ket{11}_\text{L}$.
At small interatomic separation $a/\lambda_e\ll1$, $\Gamma_{q_a} T_\text{g}\ll1$ and the error $\epsilon_1$ can be estimated as 
\be\label{eq:error_1}
	\epsilon_1 \simeq \Gamma T_\text{g} \sim \pare{\frac{a/\lambda_e}{N}}^3 \spare{K_0(\pi l/a)}^{-1}.
\ee
For fixed $l/a$, \eqnref{eq:error_1} decreases both for longer arrays and for smaller lattice spacing $a/\lambda_e$ due respectively to the scaling $\Gamma_{q_a}\sim \Gamma_0/N^3$ and to the decrease of $T_\text{g}$. 
At large separations $l/a$, $\epsilon_1$ increases according to $[K_0(l/a)]^{-1} \sim \sqrt{l/a} \exp(l/a)$ as expected for an increased gate time.
The second source of errors is due to the loss of population from the state $\ket{11}_\text{L}$ which we model by a parameter $|\xi|<1$ in \eqnref{eq:TT_real_sqSWAP}. The dynamics of $\ket{11}_\text{L}$ strongly depends on the separation $l$ between the arrays.
At large separation ($l\gg a$), $\ket{11}_\text{L}$ is an eigenstate of the parallel arrays, and it decays exponentially with a characteristic rate $\Gamma_{11}\simeq 2\Gamma_{q_a} \sim 2\Gamma_0/N^3$. In this regime, we can approximate $|\xi|\simeq \exp(-\Gamma_{11} T_\text{g})$.
For small separation ($l\sim a$), $\ket{11}_\text{L}$ couples strongly to other states in the two excitations manifold leading to large population transfer outside the computational subspace of the parallel arrays.

To esimate the performance of the proposed implementation of the $\sqSWAP$-gate, we calculate the average gate fidelity according to~\cite{Nielsen2002,Pedersen2007}
\be\label{eq:F_swap}
	\mathcal{F}_\text{G}\equiv \frac{1}{20}\pare{\text{Tr}[\mathcal{M}\mathcal{M}^\dag]+|\text{Tr}[\mathcal{M}]|^2},
\ee
where $\mathcal{M}\equiv\Pop \Udop_0 \Uop\Pop$. Here, $\Pop$ is the projector on the computational subspace of the parallel arrays, $\Uop\equiv \exp(-\im\Hop_0 T_\text{g})$ , and $\Uop_0$ is the ideal gate which acts on the computational subspace according to the truth table in \eqnref{eq:TT_real_sqSWAP} with $\xi=1$ and $\Gamma=0$, and as the identity on all the other states.
We numerically evaluate \eqnref{eq:F_swap} as a function of the system parameters $N,a/\lambda_e,$ and $l/a$ and show the results for the gate error $\epsilon_\text{tot}=1-\mathcal{F}_\text{G}$ (infidelity) in \figref{fig:Fig3}.d,e.
In \figref{fig:Fig3}.d, we plot the average gate error as a function of the lattice separation $a/\lambda_e$ for $N=40$ and different array separations $l/a$. 
For large separation $l/a$, the gate error is well described by the spontaneous decay from the computational states $\ket{10}_\text{L},\ket{01}_\text{L}$, and $\ket{11}_\text{L}$ (black dashed lines in~\figref{fig:Fig3}.d) as expected from the decoupling of $\ket{11}_\text{L}$ from other states in the two-excitations manifold. 
Accordingly, the error improves when reducing $a/\lambda_e$ as a consequence of the reduced gate operation time (see~\eqnref{eq:g_k}). In particular, when $\Gamma T_\text{g}\ll1$ the total error reads $\epsilon_\text{tot}\simeq 3 \Gamma_{q_a} T_\text{g}/5 $ and scales as in \eqnref{eq:error_1}. For the case of arrays of $N=40$ atoms as shown in~\figref{fig:Fig3}.d, the decoupling of $\ket{11}_\text{L}$ from the two excitation manifold is found to be accurate already for $l/a\geq 4$. For smaller arrays, such a decoupling occurs already for smaller values of $l/a$ as shown in \appref{app:Parallel_Arrays}.
In the opposite regime, \ie when arrays are placed close to each other, $\ket{11}_\text{L}$ couples strongly to different states in the two-excitation manifold. This leads to a large population transfer outside the computational subspace and hence to an increased error. This effect is dominant in the case of $l/a=1,2$ in~\figref{fig:Fig3}.d. In this case, the strong coupling leads to an oscillation of population between $\ket{11}_\text{L}$ and $\ket{S_2}= (\ket{2q_a,0}+\ket{0,2q_a})/\sqrt{2}$~\footnote{We found the coupling strength between $\ket{11}_\text{L}$ and $\ket{S_2}$ to be $g_2\simeq 2g_{q_a}$ for all the values $a/\lambda_e$ in the range shown. This is particularly dramatic as at the  $T_\text{g}$ the population in $\ket{S_2}$ is maximised.}. 
The case $l=3a$ in~\figref{fig:Fig3}.d, represents an intermediate case where the decoupling of $\ket{11}_\text{L}$ is not perfect and a small fraction of population is exchanged with $\ket{S_2}$. For this reason, the curve shows a slight improvement for smaller $a/\lambda_e$ due to a reduction in the gate operation time, which however saturates to a value determined by the population transferred from $\ket{11}_\text{L}$ to states outside the computational subspace.
In \figref{fig:Fig3}.e, we plot the average gate error as a function of $N$ for $a/\lambda_e=0.10$ and different array separations $l/a$. 
As for the case of \figref{fig:Fig3}.d, for large separations between the arrays, the error is solely due to the spontaneous decay from the computational states $\ket{10}_\text{L},\ket{01}_\text{L}$, and $\ket{11}_\text{L}$ as shown by the agreement between the colored markers and the black dashed lines in~\figref{fig:Fig3}.e at $l/a=4,5$.
At short separation the error is instead again dominated by the population transfer from $\ket{11}_\text{L}$ to states outside the computational subspace. 
We observe a trend, particularly evident for the case $l/a=3$, which shows a first improvement in the error for short arrays followed by a later increase in the gate error. 
As we show in \appref{app:Parallel_Arrays}, this can be interpreted as coming from the reduction of the dipole shift between $\ket{11}_\text{L}$ and $\ket{S_2}$ with growing array's length, which leads to a larger exchange of population between the two states.

The picture presented here for arrays of pinned atoms has not considered the detrimental effects which come from the atoms' center-of-mass motion around their trapping position in the lattice. As part of the system's non-linear response comes from the destructive interference of the emitted field, we expect the atomic fluctuation to drastically reduce the intrinsic dissipative non-linearity. In the following section, we discuss the motional effects and their impact on the fidelity of single and two qubit operations.

% ========================================================
% ========================================================
\section{Arrays of Fluctuating Atoms: Motional Effects}\label{sec:Motion}
% ========================================================
% ========================================================

\begin{figure}
	\includegraphics[width=\columnwidth]{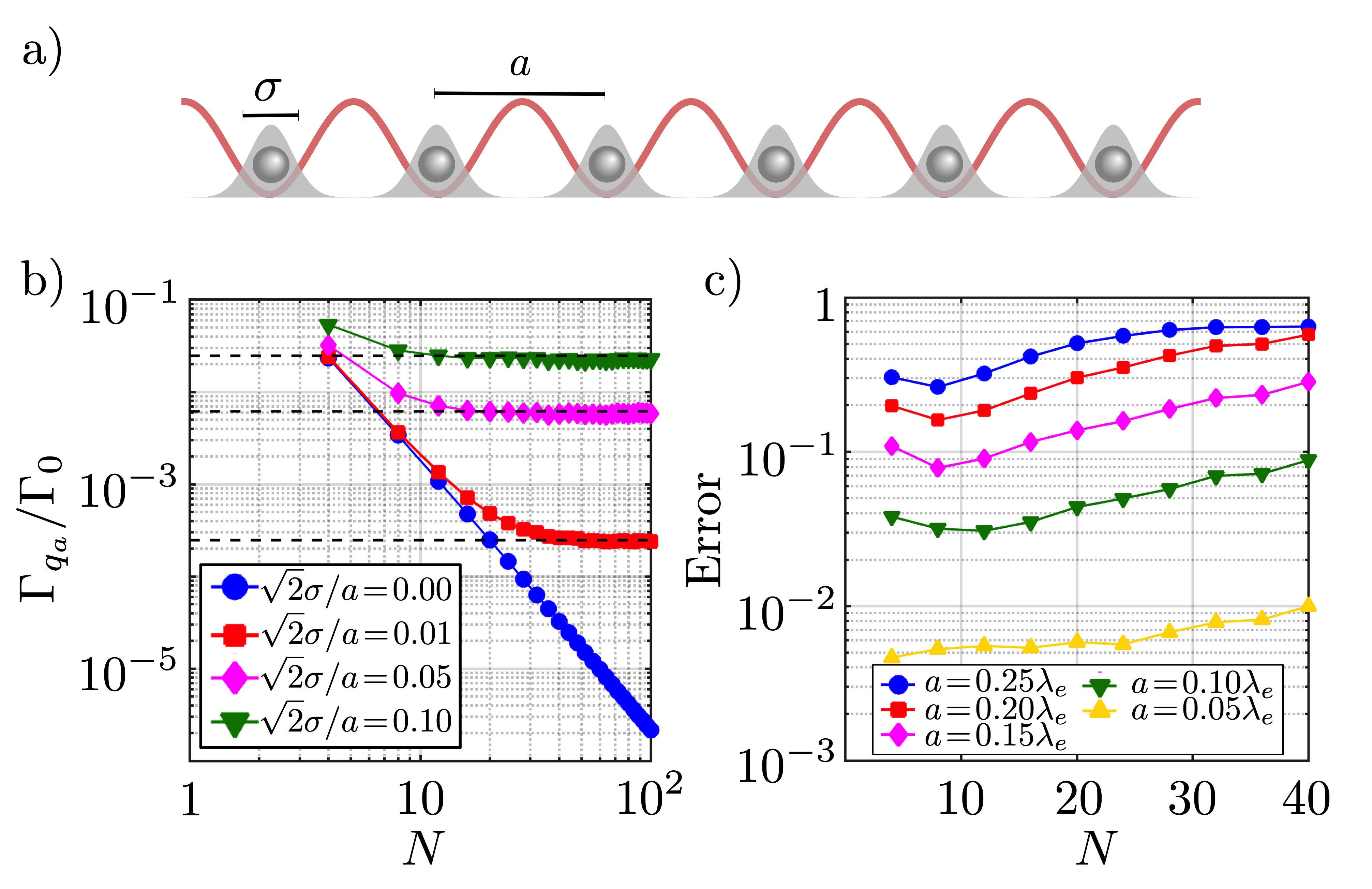}
	\caption{{\bf Arrays of moving atoms.} a) Atoms localized around their trapping positions $\RR_\ii$ with a gaussian probability distribution of width $\sigma<a$. b) Decay rate $\Gamma_{q_a}$ of the most subradiant mode for a single array as function of the number $N$ of atoms in the array as calculated by diagonalising the averaged Hamiltonian over $100$ realisation of the atomic positions (colored markers). Horizontal black dashed line corresponds to $(\sigma k_e)^2\Gamma_0$. c) Dependence on $N$ of the optimal error for preparing the most subradiant single excitation state of a single array for different values of $a$ (colored markers) and $\sqrt{2}\sigma=0.05a$.}\label{Fig:Fig4}
\end{figure}

We now consider the fluctuating motion of the array's atoms around their trapping position (see~\figref{Fig:Fig4}.a).
In this case, the open dynamics of an array of fluctuating atoms is described by the non-hermitian Hamiltonian (see~\appref{app:ME_derivation})
\be\label{eq:Ham_motion}
\begin{split}
\Hop \equiv &  \sum_\jj \bigg[\frac{\hat{\pp}_\jj^2}{2m} +\frac{1}{2}m\wT^2\rrop_\jj^2+\hbar\w_g\sigop_{21}^\jj\sigop_{12}^\jj\bigg]\\
& + \hbar\veps^2\sum_{\ii,\jj}G(\rrop_\jj,\rrop_\ii)\sigop_{21}^\jj\sigop_{12}^\ii,
\end{split}
\ee
where we modelled the mechanical degrees of freedom as harmonic oscillators of frequency $\wT$, and $m$ is the atomic mass.
The operator $\rrop_\jj$ ($\ppop_\jj$) represents the center-of-mass position (momentum) displacement relative to the trap centre $\RR_\jj\equiv(0,lj_y,aj_z)^T$, where $\jj\equiv(j_y,j_z)$, $j_y=0,1$ labels the array A,B, and $j_z=1,\ldots,N$ the position within each array. 
The second line in \eqnref{eq:Ham_motion} represents coherent and dissipative atomic dipole-dipole interactions where now the interaction strength $G(\rrop_\jj,\rrop_\ii)$ is an operator which acts on the atomic center-of-mass degrees of freedom and reads
\be\label{eq:G_ij_op}
\begin{split}
	G(\rrop_\jj,\rrop_\ii)\equiv& -\im \frac{\mu_0 \w_e^2}{\hbar}|d|^2G^0_{zz}(\RR_\jj+\rrop_\jj-\RR_\ii-\rrop_\ii,\w_e)\\
	&\times e^{\im\kk_\text{L}\cdot(\rrop_\jj -\rrop_\ii)}.
\end{split}
\ee
The exponential operator appearing as the last factor in \eqnref{eq:G_ij_op} is a result of the Raman transition used to define the effective two level system $\{\ket{g_1},\ket{g_2}\}$. 
Similarly, in the presence of atomic fluctuations, the driving Hamiltonian reads (see~\appref{app:Driving})
\be\label{eq:V_fluct}
	\Vop = \hbar\W_0 \sum_{\jj} \sin[\pare{\kk_a-\kk_b}\cdot (\RR_\jj +\rrop_\jj)] \sy_\jj.
\ee
In general, the dynamics described by \eqnref{eq:Ham_motion} and \eqnref{eq:V_fluct} leads to correlations between the internal and external degrees of freedom, which quickly complicate the simulation of the time evolution of the full system. In two limiting cases, the study of the dynamics can be greatly simplified~\cite{Porras2008}: (i) the slow atomic-motion regime $\wT\ll \Gamma_{q_a}$, and (ii) the fast atomic-motion regime $\wT\gg \Gamma_0$.

The slow-motion limit describes the situation where the time scale of the center-of-mass dynamics is much longer than the one of the slowest internal dynamics, typically proportional to the decay rate of the subradiant modes.
Under this condition, the atoms can be considered frozen at their current positions during the internal evolution, and the system's	 dynamics can be approximated solely by the internal dynamics where the coupling \eqnref{eq:G_ij_op} is evaluated with the substitution $\rrop_\jj \rightarrow \rr_\jj$ where $\rr_\jj$  is the particular value of the displacement which determines the instantaneous position of the $\jj$-th atom~\cite{Porras2008}.
The average dynamics of the system is thus determined by solving the evolution for different realisations of the atomic position and then averaging the results.
Applying this method to the one- and two-qubit gates described in \secref{sec:PinnedAtoms}, one finds poor fidelities.

The fast-motion limit describes the opposite situation in which the center-of-mass dynamics is much faster than the internal one.
While this regime is usually challenging for neutral atoms in optical trap, it might be possible to meet such condition for an appropriate choice of the parameters of the Raman transition defining the effective two level atom, such that $\wT/\gamma_1 \gg \veps^2$.
In this case, the evolution of the system can still be approximated solely by the internal dynamics described by the non-hermitian Hamiltonian in \eqnref{eq:Ham_motion} (and associated Jump operator) averaged over many different realisations of the atoms' positions~\cite{Porras2008}. This second limiting case leads to better fidelities for the one- and two-qubit operations between parallel arrays as we shall now prove.

% --------------------------------------------
\subsection{Regime of Fast Atomic Motion}\label{sec:FastMotion}
% --------------------------------------------

In the limit of fast atomic motion, $\wT\gg \Gamma_0$, we approximate \eqnref{eq:Ham_motion} by the following non-hermitian Hamiltonian
\be\label{eq:Ham_fam}
	\Hop_\text{fam} \equiv \hbar \sum_{\ii,\jj} \pare{\w_g\delta_{\ii\jj}+\veps^2\tilde{G}_{\ii\jj} } \sigop_{21}^\ii\sigop_{12}^\jj
\ee
where we defined the averaged coupling over the atomic positions $\tilde{G}_{\ii\jj}$  according to
\be\label{eq:avg_G_ij}
	\tilde{G}_{\ii\jj}\equiv \int_{\mathbb{R}^3}\!\!\text{d}\rr_\ii \,\text{d}\rr_\jj G(\rr_\ii,\rr_\jj) P(\rr_\ii)P(\rr_\jj),
\ee
where we assumed the atoms' positions to be independent variables, normally distributed according to the probability distribution 
$P(\rr) \equiv \exp(-\rr^2/2\sigma^2)/(\sqrt{2\pi}\sigma)^3$.
In the Lamb-Dicke regime $\sigma k_e\ll 1$, one can approximate \eqnref{eq:avg_G_ij} as (\appref{app:Fast_Atomic_Fluctuations})
\be\label{eq:approx_Gij_avg}
	\tilde{G}_{\ii\jj} = \left \{ 
	\begin{array}{lcr}
	-\im \gamma_1/2 & \text{for} & \ii=\jj\\
	(1-2\sigma^2k_e^2) G_{\ii\jj} & \text{for} & \ii\neq\jj
	\end{array}
	\right . ,
\ee
where $G_{\ii\jj}$ is the coupling between pinned atoms as given in \eqnref{eq:G_coupling} and \eqnref{eq:g_coupling}. 
Substituting \eqnref{eq:approx_Gij_avg} into \eqnref{eq:Ham_fam}, we obtain (in a frame rotating at $\w_g$)
\be\label{eq:Ham_fam_approx}
\begin{split}
	\Hop_\text{fam} =&(1-2\sigma^2k_e^2)\hbar\sum_{\ii,\jj} \veps^2 G_{\ii\jj} \sigop_{21}^\ii\sigop_{12}^\jj\\
	&   -\im \sigma^2k_e^2 \hbar\Gamma_0 \sum_\ii\sigop_{21}^\ii\sigop_{12}^\ii.
\end{split}
\ee
According to \eqnref{eq:Ham_fam_approx}, the fast atomic motion renormalizes the coherent and dissipative coupling between the atoms by a factor $(1-2\sigma^2k_e^2)$, and adds an independent decay rate $\sim \sigma^2k_e^2\Gamma_0$ for each atom. The main effect of the atomic motion is thus to suppress the scaling $\sim \Gamma_0/N^3$ of the dark mode decay rate, which saturates at the constant value $\sim \sigma^2k_e^2\Gamma_0$ as shown in~\figref{Fig:Fig4}.b for the case of a single array.
This effect has been pointed out before in~\cite{AntezzaPRL2009,Guimond2019}. The expression we obtain in \eqnref{eq:Ham_fam_approx} differs from the corresponding ones in~\cite{AntezzaPRL2009,Guimond2019} by a factor of two. This increased noise originates from the scattering of the Raman-laser which adds an additional contribution $\sim k_e^2 \sigma^2$ to the atomic center-of-mass diffusion (see~\appref{app:Fast_Atomic_Fluctuations}). 

Let us now consider the fidelity for preparing the dark mode of a single atomic array in the regime of fast atomic motion.
The driving Hamiltonian \eqnref{eq:V_fluct}, averaged over the atomic fluctuations, reduces to the pinned-atom expression \eqnref{eq:Vop}, with a renormalized driving frequency $\tilde{\W}_0=\W_0 \exp(-k_e^2\sigma^2)\simeq (1-\sigma^2k_e^2)\W_0$.
Proceeding as in \secref{sec:PinnedAtoms}, we simulate the Sch\"odinger evolution generated by the total averaged Hamiltonian (including the driving) and calculate the error for preparing the most subradiant single-excitation state of the system. 
We calculate $\tilde{G}_{\ii\jj}$ by averaging \eqnref{eq:G_ij_op} over one hundred realizations of the atomic positions assumed to be distributed around the lattice sites according to $P(\rr)$.
For the driving Hamiltonian there is no need to average the Rabi frequency over different realizations, because the effect of the motion leads to an overall renormalization factor in the driving, which is irrelevant after optimizing $\W_0$.
We present the results of the numerical simulations in~\figref{Fig:Fig4}.c.
As a consequence of the saturation of the dark-mode decay rate (\figref{Fig:Fig4}.b), the state-preparation fidelity decreases at large $N$ for a fixed lattice spacing $a/\lambda_e$ as evidenced in~\figref{Fig:Fig4}.c, where we consider the case $\sqrt{2}\sigma/a=0.05$. For larger values of $\sigma$ the dependence on $N$ is qualitatively the same, albeit with larger noise, and the small improvement at small $N$ is lost (see cases $N=4,8,12$ in \figref{Fig:Fig4}.c). 
The results in \figref{Fig:Fig4}.b,c suggest that, due to the suppression of the Zeno effect in the presence of atomic fluctuations, it is not advantageous to use larger arrays.

\begin{figure}
	\includegraphics[width=\columnwidth]{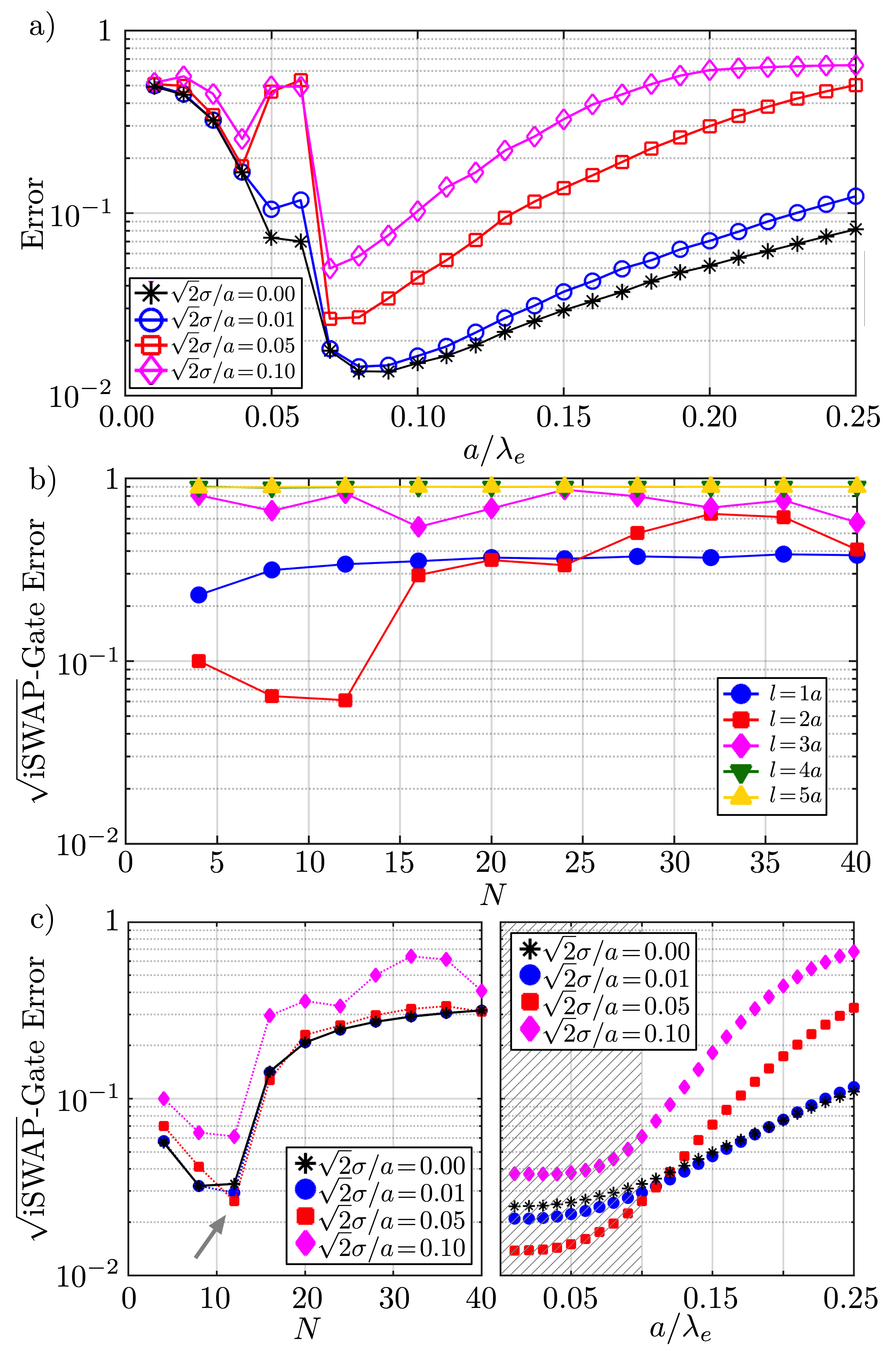}
	\caption{{\bf Gate Error for parallel arrays of fluctuating atoms.} a) Error for the preparation of the state $\ket{10}_\text{L}$ as function of $a/\lambda_e$, for different values of $\sigma$. Other parameters: $N=20$, $l=a$, $\delta=100\Gamma_0$. b) $\sqrt{\text{iSWAP}}$-gate error as function of $N$ for $a/\lambda_e=0.10$ and $2\sigma=0.1a$. Different colored markers corresponds to different values of $l$ (see legend). c) Left panel: $\sqrt{\text{iSWAP}}$-gate error as function of $N$ for $l=2a$ and $a/\lambda_e=0.10$. Right panel: $\sqrt{\text{iSWAP}}$-gate error as function of $a/\lambda$ for $l=2a$ and $N=12$. In both panels different values of $\sigma$ corresponds to different coloured markers (see legend). In all panel we calculated the coupling in \eqnref{eq:Ham_fam} by averaging over 100 realisation according to $P(\rr)$.}\label{Fig:Fig5}
\end{figure}

Let us now analyze the effects of fast atomic motion on the fidelity of the single- and two-qubit gates in a system of two parallel arrays. Because the decay rate of the single excitation dark mode is clamped by the fluctuations after $N\gtrsim40$ for the case $\sqrt{2}\sigma=0.01a$ (\figref{Fig:Fig4}.b), we investigate the gate fidelity only for small arrays up to a maximum of $N=40$ atoms.
The results of the numerical simulation for the gate fidelity of two parallel arrays are shown in \figref{Fig:Fig5}.
In \figref{Fig:Fig5}.a, we compare,  for different values of $\sigma$, the minimal error for preparing the state $\ket{10}_\text{L}$ of two parallel arrays of $N=20$ atoms separated by a distance $l=a$ as function of the lattice spacing $a/\lambda_e$.
As expected, the error grows with $\sigma$ as a consequence of the contribution $(\sigma k_e)^2 \Gamma_0$ to the dark-mode decay rate. This effect is particularly prominent at large values of $a/\lambda_e$ where the main contribution to the non-linearity comes from the Zeno effect. For small $a/\lambda_e\lesssim 0.05$, values corresponding to different $\sigma$ yield similar results because the main contribution to the error comes from populating the state $\ket{01}_\text{L}$, as for the case of pinned atoms (see discussion in \secref{ssec:PinnedAtoms_Parallel}).
In \figref{Fig:Fig5}.b, we compare the $\sqrt{\text{iSWAP}}$-gate error, calculated according to \eqnref{eq:F_swap}, as a function of $N$ for different value of the separation $l$ between the arrays, assuming a position noise characterized by $\sqrt{2}\sigma=0.10a$. 
The gate fidelity deteriorates for increasing $l$, much faster than for the case of pinned atoms [Cf. \figref{fig:Fig3}.e]. This behaviour is consistent with the increased decay rate of the array's dark mode, because we expect the free space decay to have a larger contribution to the total error at larger $l$ due to an increased gate time $T_g$.  
In \figref{Fig:Fig5}.c, we study in more detail the particular case of $l=2a$, which yields better fidelities than other values of $l$ in~\figref{Fig:Fig5}.b~\footnote{We consider only integer values of $l/a$. Better results can be obtained if this assumption is relaxed.}.
In particular, we compare the scaling of the error with both $N$ (left panel) and $a/\lambda_e$ (right panel) for different values of position noise $\sigma$. While larger noise typically leads to an increase in the gate error, we find that for $N=12$ moderate noise might improve the gate fidelity as compared to the pinned-atom case (see gray arrow in the right panel of \figref{Fig:Fig5}.c). 
The left panel of \figref{Fig:Fig5}.c shows that at larger lattice separation positional noise always deteriorate the gate fidelity, as expected for a reduced Zeno blockade. However, for $a\lesssim 0.10\lambda_e$, a value of $\sqrt{2}\sigma=0.05a$ leads an improvement in the gate error over the pinned-atom case.
This improvement is attributed to a change in the effective coupling rate between the states $\ket{11}_\text{L}$ and $\ket{S_2}$, which leads to a larger population in $\ket{11}_\text{L}$ at $t=T_g$. We do not further investigate this effect, as we believe the values of the error calculated from the effective model \eqnref{eq:Ham_fam} do not yield correct estimates for $a/\lambda_e\lesssim 0.1$ (hatched region in \figref{Fig:Fig5}.c) as we argue in the following. 

% ---------------------------------------------------------------------------
\subsection{Limitation of the current description of the atomic fluctuations and alternative approaches}\label{sec:Limitation_AtomicMotion}
% ---------------------------------------------------------------------------

The treatment of the atomic fluctuations presented in the previous section does not take into account corrections due to a finite velocity of the atoms~\cite{Cooper1974,Power1974,Berman1997,Zhu2016,Bettles2020}. Finite atomic velocity might lead to excitation of the atomic motional state, which would result in a reduction in the fidelity of the one and two qubit gates between arrays. In the following, we take into account effects of finite atomic velocity using a perturbative treatment of the atomic fluctuations~\footnote{The perturbative approached followed here was presented in~\cite{Guimond2019}. In the present case the form of the dipole coupling rate is difference due to the Raman transition.}.  
As in~\secref{sec:FastMotion}, we consider a subwavelength atomic array ($a<\lambda_e/2$) where atoms are well localized around their trapping position, namely $r_0\ll a$, with $r_0\equiv \sqrt{\hbar/2m \wT}$ the center-of-mass zero point motion. Under this assumption and for low thermal phonon number $\nth < 1$, the system is in the Lamb-Dicke regime, $\eta \equiv \sigma k_e\ll1$, for $\sigma\equiv r_0\sqrt{2\nth+1}$ and we approximate the coupling between internal and external degrees of freedom by a power expansion of $G(\rr)$ to second order in $\eta$~\cite{Guimond2019},
\be\label{eq:Ham_Motion_Expansion}
	\Hop\simeq \Hop_0 + \Hop_\text{I1}+\Hop_\text{I2}.
\ee
Here, $\Hop_0$ describes the dynamics in the absence of mechanical effects of light and is simply given by \eqnref{eq:Ham_2Arrays_Pinned} with the addition of the mechanical energy of the atoms. 
The term $\Hop_\text{I1}$ ($\Hop_\text{I2}$) represents the first (second) order correction in the atomic center-of-mass displacement. They have the general form
\bea
	\Hop_\text{I1} &\equiv& \veps^2 \sum_{\ii,\jj} \pare{\rrop_\ii-\rrop_\jj}\cdot \grad G(\RR_\jj,\RR_\ii) \sigop_{21}^\jj\sigop_{12}^\ii ,\label{eq:Ham_I1}\\
	\Hop_\text{I2} &\equiv& \frac{\veps^2}{2}\!\! \sum_{\jj\ii}\sum_{\alpha,\beta=x}^z\! \pare{\rop_{\alpha\ii}-\rop_{\alpha\jj}}\pare{\rop_{\beta\ii}-\rop_{\beta\jj}}\label{eq:Ham_I2}\\
	& & \times \pa{\alpha}\pa{\beta}G(\RR_\ii,\RR_\jj)\sigop_{21}^\jj\sigop_{12}^\ii.\nonumber
\eea
Let us here remark that \eqnref{eq:Ham_Motion_Expansion} reduces to \eqnref{eq:Ham_fam_approx}, as expected, if one traces over the center-of-mass degrees of freedom assuming the atoms in a thermal state of their mechanical motion.

The driving Hamiltonian \eqnref{eq:V_fluct} up to second order in the center-of-mass fluctuations reads
\be\label{eq:V_Motion_Expansion}
	\Vop \simeq \Vop_0 + \Vop_1 + \Vop_2.
\ee
Here, $\Vop_0$ is the driving Hamiltonian for arrays of pinned atoms as given in \eqnref{eq:Vop_2array}. The term $\Vop_1$ ($\Vop_2$) represents the first (second) order correction in the Lamb-Dicke parameter. They read (see~\appref{app:LambDicke})
\be\label{eq:V1_main}
\begin{split}
\Vop_1 \equiv& -\im \hbar\W_0 \sum_{j=1}^N \cos(K_z aj_z)\Big[\eta_z\pare{\bdop_{zj}+\bop_{zj}}\\
& +\eta_x \pare{\bdop_{xj}+\bop_{xj}}\Big]\sy_j,
\end{split}
\ee
and
\be\label{eq:V2_main}
\begin{split}
\Vop_2 \equiv& \im \hbar\W_0 \sum_{j=1}^N \sin(K_z aj_z)\Big[\eta_z^2\pare{\bdop_{zj}+\bop_{zj}} ^2\\
&+2\eta_z\eta_x\pare{\bdop_{zj}+\bop_{zj}}\pare{\bdop_{xj}+\bop_{xj}}\\
&+\eta_x^2 \pare{\bdop_{xj}+\bop_{xj}}^2 \Big]\sy_j.
\end{split}
\ee
Here, we assumed the driving laser to be directed orthogonal to the $y$-axis. Moreover, we defined $\eta_{x,z}\equiv K_{x,z}r_0$, where $\xop_j \equiv r_0(\bdop_{xj}+\bop_{xj})$, $\zop_j \equiv r_0(\bdop_{zj}+\bop_{zj})$, and $K_x \equiv k_a \sin\alpha -k_b \sin\beta$, while $K_z$ is given in \eqnref{eq:q_effective}.

\begin{figure}
	\includegraphics[width=\columnwidth]{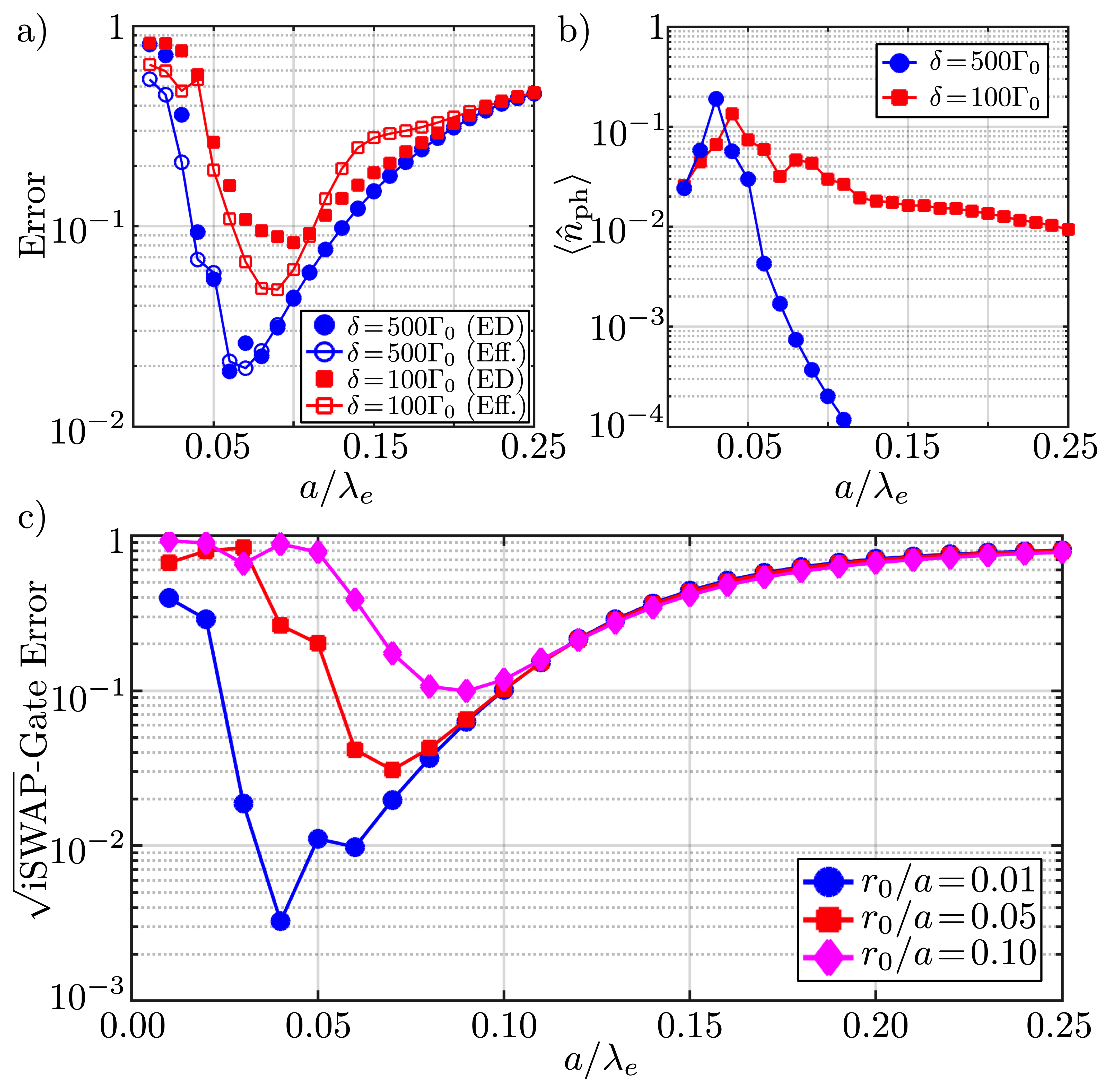}
	\caption{{\bf Gate error for arrays of moving atoms: perturbative treatment.} a) Error for preparing the dark state $\ket{10}_\text{L}$ for two parallel arrays separated by $l=a$. Solid markers refer to exact digonalisation (ED) results while empty markers refer to the effective model (Eff.) where the motion is adiabatically eliminated in its ground state to second order in $\eta$. b) Total average center-of-mass phonon population for $\delta/\Gamma_0=500$ (circles) and $\delta/\Gamma_0=100$ (squares). For both panels a) and b) we used $\eta=2\pi\times 0.05 a/\lambda_e$.
	c)  $\sqSWAP$-gate error for parallel arrays of fluctuating atoms separated by $l=2a$ as function of $a/\lambda_e$. Different values of atomic center of mass fluctuations $r_0/a$ are compared (see legend).
For all the plot the following values of additional parameters have been used : $n_\text{th}=0$, $N=2$, phonon Hilbert space dimension $d=2$, and $\wT/\Gamma_0=100$.}\label{fig:Fig6}
\end{figure}

We now proceed to study the dynamics generated by \eqnref{eq:Ham_Motion_Expansion} and \eqnref{eq:V_Motion_Expansion} in the fast-motion regime $\w_T\gg\Gamma_0$ for the case of single- and two-qubit gates between parallel arrays. We consider here the case where the atoms are initially prepared  in their motional ground state ($\nth=0$).
In particular, we compare the results obtained from exact diagonalisation of the total Hamiltonian \eqnref{eq:Ham_Motion_Expansion} and \eqnref{eq:V_Motion_Expansion}, and the results obtained from an effective model in which the atomic motion has been adiabatically eliminated keeping contribution up to second order in $\eta$ (\appref{app:LambDicke}). 
This effective model contains both contributions from processes which do not change the motional state of the atoms (first order in $\Hop_\text{I2}$ and $\Vop_2$) and from second order processes via intermediate excited states of the atomic motion (second order in $\Hop_\text{I1}$ and $\Vop_1$).
In~\figref{fig:Fig6}.a, we show the error for preparing the state $\ket{10}_\text{L}$ for a system of two parallel arrays of $N=2$ atoms separated by a distance $l=a$ and trapped with a center-of-mass frequency $\wT/\Gamma_0=100$. We assume one array to be detuned from the target one by $\delta$.
The results of exact diagonalization and of the effective model show excellent agreement at large lattice spacings $a/\lambda_e$. For smaller interatomic separations, the difference between the exact and effective model is due the excitation of the center-of-mass motion from atomic recoil (\figref{fig:Fig6}.b). 
In~\figref{fig:Fig6}.c, we show the dependence on $a/\lambda_e$ of the error in performing a $\sqSWAP$ operation between two arrays of $N=2$ atoms separated by a distance $l=2a$. We ascribe the decrease of fidelity at small interatomic separations to excitation of the center-of-mass motion. We observe that the values of $a/\lambda_e$ at which the center-of-mass motion is excited depends on how tight the trap is: for smaller $r_0/a$ (tighter trap) fluctuations are excited for smaller $a/\lambda_e$.

\figref{fig:Fig6}.a-c show that, at small interatomic separation, the main source of error is due to the excitation of center-of-mass fluctuations caused by the terms \eqnref{eq:Ham_I1}, \eqnref{eq:Ham_I2}, \eqnref{eq:V1_main} and \eqnref{eq:V2_main}.
Precisely, the excitation of the atomic fluctuations is predominantly due to the atomic recoil involved in the dipole-dipole interaction [\eqnref{eq:Ham_I1} and \eqnref{eq:Ham_I2}] because these terms represents the largest coupling between internal and external degrees of freedom. The mathematical origin of this large coupling is in the faster divergence of the derivatives of $G_{\ii\jj}$ for small separation. 
Because of the enhancement in the the coupling rate in \eqnref{eq:Ham_I1} and \eqnref{eq:Ham_I2} for small atomic separation, the perturbative approach to the atomic fluctuations developed here is valid only for sufficiently small $\eta$ such that $\eta k_e^{-1}|\pa{\alpha}G(\rr)/G(\rr)|\ll1$ and $\eta^2 k_e^{-2}\pa{\alpha}\pa{\beta} G(\rr)/G(\rr)\ll1$ for $\alpha,\beta=x,y,z$.
Furthermore the critical values of $\eta$ for which the perturbative approach is justified depends on $N$: for fixed $\eta$ one observes the appearence of unphysical eigenvalues of \eqnref{eq:Ham_Motion_Expansion} with positive imaginary part as $N$ is increased.

The perturbative model, while not sufficiently accurate to correctly describe the effects of motion for long arrays, validates the results of the fast atomic motion regime for $a\gtrsim 0.1\lambda_e$. Furthermore, it indicates that for smaller interatomic separations the center-of-mass motion is expected to play a major role in the sytem dynamics, a feature not captured by the model in~\secref{sec:FastMotion}.

% ================================
% ================================
\section{Discussion}\label{sec:Discussion}
% ================================
% ================================

\begin{figure}
	\includegraphics[width=0.9\columnwidth]{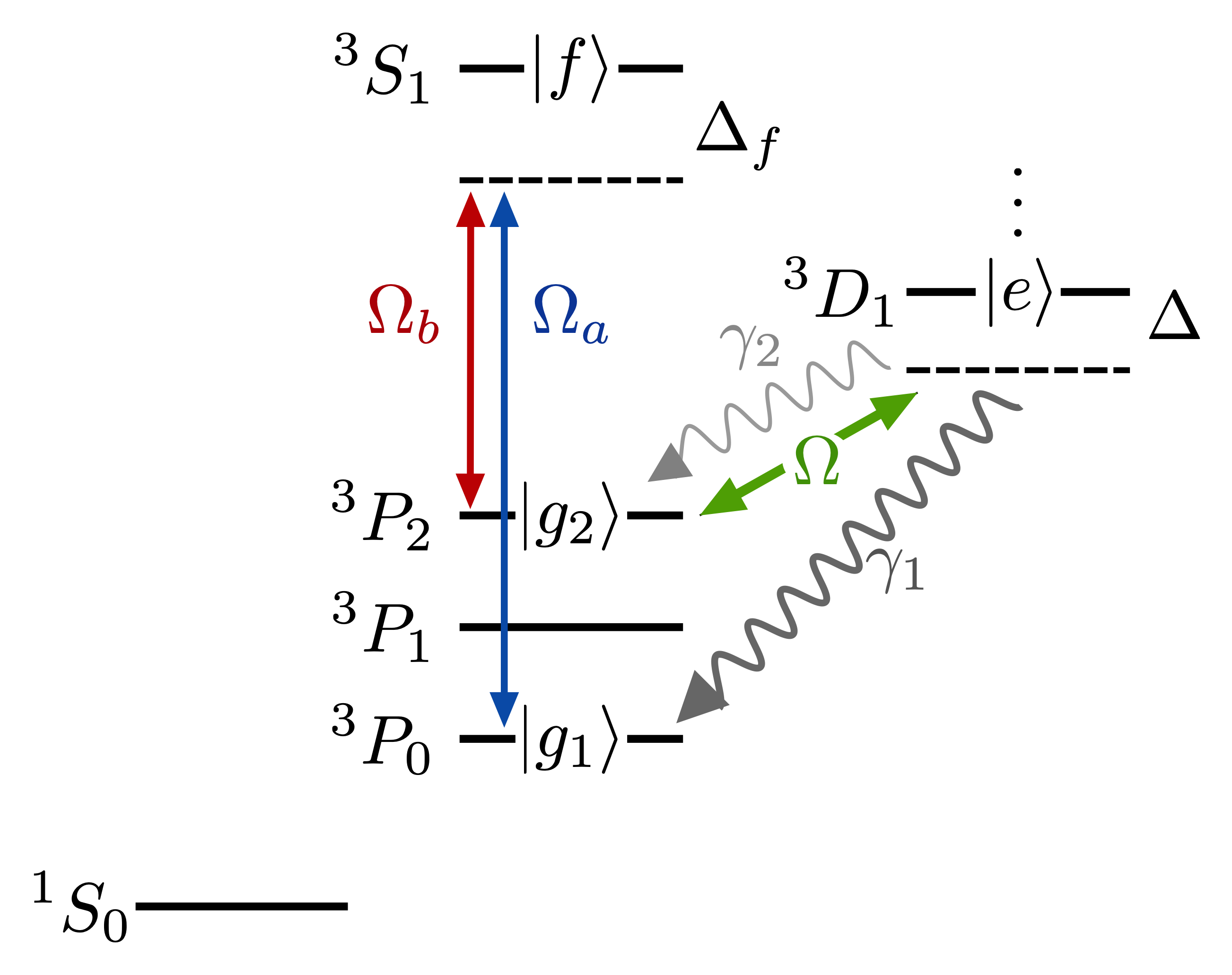}
	\caption{{\bf Relevant level structure (not to scale) of alkaline-earth-metal atoms.} We highlight the transitions and levels used to implement our scheme with the same notation of \figref{fig:Fig1}.}\label{fig:Fig7}
\end{figure}

The results presented in the previous sections are obtained under several assumptions:  (i) the possibility to trap atoms in an optical lattice which is subwavelength with respect to the transition $\ket{e}\leftrightarrow\ket{g_1}$, (ii) a particular atomic structure which comprises two distinct Raman transitions connecting $\ket{g_2}$ to $\ket{g_1}$ via the intermediate levels $\ket{e}$ and $\ket{f}$ where $\w_f \geq \w_e$, (iii) a negligible dephasing rate of $\ket{g_2}$ ($\kappa_0\ll\Gamma_0$), (iv) fast atomic center-of-mass fluctuations $\wT \gg \Gamma_0$, and (v) the Lamb-Dicke regime for the atomic motion.
Let us now show how these requirements can be met using ultracold alkaline-earth atoms in an optical lattice. The relevant level structure for such atoms is shown in \figref{fig:Fig7} where we marked transitions and levels used to implement the scheme illustrated in~\figref{fig:Fig1}.
The ground-state manifold levels $\ket{g_{1,2}}$ are encoded in the levels $^3P_0$ and $^3P_2$ respectively, while the excited state $\ket{e}$ is encoded in $^3D_1$, and the auxiliary state $\ket{f}$ used for the excitation of the array's collective dark modes is encoded in $^3S_1$.
Decay from $^3P_0$ to $^1S_0$ is a forbidden transition which happens at a rate $\approx 10 \text{mHz}$ for Sr, while decay from $^3P_2$ to $^1S_0$ has an even longer predicted lifetime~\cite{Ludlow2015}. The two levels $\ket{g_1}$ and $\ket{g_2}$ can thus be considered stable.
Alkaline-earth atoms exhibit long range dipole interactions on the transition $^3P_0- ^3\!D_1$, which combined with the possibility of creating optical lattices using transitions from $^3P_J$ to higher excited states allows for the creation of deep-subwavelength arrays. In the case of bosonic Strontium, for instance, the transition wavelength between $\ket{e}$ and $\ket{g_1}$ is $2.6~\mu\text{m}$ which allows to attain a subwavelength array with lattice spacing $a/\lambda_e \approx 0.08$ for an optical lattice with wavelength $\lambda_\text{opt}\simeq 400~\text{nm}$~\cite{Olmos2013}.
Additionally, the transition between $^3P_0- ^3\!\!D_1$ has a linewidth $\gamma_1/2\pi\approx 290 \text{kHz}$ which is broader than the linewidth $\gamma_2/2\pi\approx 10 \text{kHz}$ for $^3P_2- ^3\!D_1$~\cite{Zhou2010} resulting in a weak effective dephasing $\kappa_0/\Gamma_0 \approx 3\times 10^{-3}$ of the level $\ket{g_2}$ after the elimination of the level $\ket{e}$.
The Raman-driving of deep-subwavelenght arrays can be realised with $\w_f/\w_e \simeq 3.8$ using the rapid transitions between $^3S_1-  ^3\!P_2$ ($\gamma/2\pi\approx 45~\text{MHz}$) and between $^3S_1- ^3\!P_0$ ($\gamma/2\pi\approx 10~\text{MHz}$)~\cite{Zhou2010}.
Finally, let us consider the requirement on the atomic motion.
We assumed the atoms to be sufficiently cold initially to be well localized at their optical lattice sites (in~\secref{sec:Limitation_AtomicMotion} we even assumed them in their motional ground state). This condition can be achieved via known cooling schemes for alkali--earth atoms which requires trapping depths of the order $\sim 10-10^3 ~\text{kHz}$ for standard cold-atoms experiments~\cite{Ludlow2015}. Note that the recoil energy of the transition $\ket{e}\leftrightarrow\ket{g_{1,2}}$ is weaker than for the scattering of photons from the optical lattice, thus ensuring the system is in the Lamb-Dicke regime $\eta=r_0 k_e \ll1$ with respect to the relevant dipole transition.
For these values of the trap depth, the trap frequency is of order $\wT\sim 100\text{kHz}$ which is comparable to the decay rate $\gamma_1$ of the transition $\ket{e}\leftrightarrow \ket{g_1}$. Hence, it should be feasible to enter the fast atomic motion regime ($\wT/\Gamma_0 < 1$) as soon as $\W/\Delta\lesssim 0.1$.
Lattices of cold alkaline-earth atoms thus represent a promising platform for the realization of the scheme proposed in \secref{sec:PinnedAtoms} and \secref{sec:Motion}.

It is important to point out that the dephasing arising from the Raman driving (\figref{fig:Fig1}.a) imposes, in principle, an upper limit to the length $N$ of an array, which corresponds to the case in which the decay rate of the most subradiant single excitation state of the array equals the dephasing rate $\kappa_0$. For arrays longer than this upper limit, dephasing plays an important role and cannot be neglected. However, we note that incoherent processes caused by dephasing in the Raman scheme could be avoided by using a cycling transition for the $\ket{e}\leftrightarrow\ket{g_1}$, and a two-photon driving for the transition $\ket{e}\leftrightarrow\ket{g_2}$ as proposed in~\cite{Porras2008}. Additionally, the limit on the array length imposed by the dephasing is generally negligible while considering the effect of the atomic motion which has a much stronger effects on the optical response of long arrays (see~\secref{sec:Motion}).

Finally, let us remark that the scheme proposed here can be extended to more than two parallel arrays by generalizing the idea of selective detuning to several parallel arrays. This can be done by borrowing techniques used in super-resolved fluorescence microscopy~\cite{Betzig2006,Hell2007}. For instance, the doughnut-shaped Laguerre-Gaussian mode $(p,l)=(0,1)$ has a dark central region which is not diffraction limited. Illuminating the system with such a beam and placing the target array in the dark spot allows to selectively tune the other arrays out of resonance. 

% ==================================
% ==================================
\section{Conclusions}\label{sec:Conclusions}
% ==================================
% ==================================

Our findings can be summarised in four main points. 
First, we have discussed how to coherently excite dark modes of subwavelength arrays using a Raman laser. This techniques represents a novel alternative to optical phase imprinting techniques~\cite{Plankensteiner2015,He2020a,He2020b}.  
Second, we described how to realise a universal set of gates based on dipole-blockade between qubit states. We found that the intrinsic non-linear response of collective dark modes of arrays of pinned atoms leads to fidelities of $~99\%$ for sufficiently large arrays or small interatomic separation.
Third, we considered the effects of atomic motion, and showed that its detrimental effects on the non-linear optical response of atomic arrays can be partially mitigated in the fast atomic motion regime. 
Additionally, the center-of-mass fluctuations place a bound on the size $N$ of the array as well as on the lattice spacing $a/\lambda_e$. Surpassing this bound by either considering longer arrays or shorter atomic separations does not lead to an improvement in the fidelity of gate operations or worse might even increase their errors. 
It is worth mentioning that while working in the fast atomic motion limit allows to reduce detrimental motional effects on the array's collective response, it comes at the cost of an increased gate time and of additional diffusion of the atomic center-of-mass motion due to scattering of the driving photons. At our present understanding, this cost is however a necessary one to pay to partially recover the collective non-linear response of subwavelength atomic arrays.
It is an interesting question for future investigations whether a clever pulse scheme can be devised such as to produce the desired internal dynamics while disentangling it from the external one after a pulse cycle, as for the case of trapped ions~\cite{Sorensen1999,Sorensen2000}. 
To this aim one could think to use the collective mechanical modes of an array of atoms which arise due to the mechanical forces mediated by the dipole-dipole interactions~\cite{Shahmoon2019,Shahmoon2020}.
Another interesting direction, which we are currently pursuing, is the development of better theoretical models which would allow to extend the results of \secref{sec:Motion} to larger arrays while accurately taking into account possible excitation of the center-of-mass motion.
Finally, we showed that realisation of the proposed scheme for manipulating quantum information with arrays of cold alkaline-earth atoms seems possible in the near future. 

\begin{acknowledgments}
  C.C.R.\ would like to thank Ana Asenjo-Garcia, Giuseppe Calaj\'o, Henrik Dreyer, Giacomo Giudice, Johannes Kn\"orzer, Daniel Malz, Nicola Pancotti, and Daniel Robaina for enlightening discussions. The moral support of Claudio Benzoni during these particularly difficult times is also gratefully acknowledged.
  C.C.R is particularly thankful to Daniel H\"ummer for providing insightful comments on the manuscript.
  C.C.R.\ and J.I.C.\ acknowledge funding from ERC Advanced Grant QUENOCOBA under the EU Horizon 2020 program (Grant Agreement No. 742102). T.S. was supported by the NSFC (Grants No. 11974363).
\end{acknowledgments}

\appendix

% ============================================================
% ============================================================
\section{Derivation of the Effective Atomic Dynamics}\label{app:ME_derivation}
% ============================================================
% ============================================================

In this Appendix, we derive the effective internal dynamics of the atoms starting from the full Hamiltonian describing the interaction between the atoms and the free-space electromagnetic field within the dipole approximation.
We include the atomic center-of-mass motion for the case of harmonically trapped atoms. These general expressions reduce to the one of \secref{sec:PinnedAtoms}, if the position of the atom is treated as a simple c-number.

The total Hamiltonian of the system when the atoms are driven by an external laser on the transition  $\ket{g_2}\leftrightarrow\ket{e}$ reads
\be\label{eq:H_full}
	\Hop_\text{tot} \equiv \Hop_\text{rad} + \Hop_\text{at} + \Hop_\text{int} + \Hop_\text{L}(t)
\ee
In \eqnref{eq:H_full}, we defined $\Hop_\text{rad}$ ($\Hop_\text{at}$) the energy of the field (atoms), $\Hop_\text{int}$ the atom-field interaction in the dipole approximation, and $\Hop_\text{L}(t)$ the external laser driving. Specifically, these different terms are defined as
\begin{widetext}
\bea
	\Hop_\text{rad}&\equiv& \sum_{\kk,\boldsymbol{\veps}_\kk} \hbar\w_\kk \adop_{\kk\boldsymbol{\veps}_\kk}\aop_{\kk\boldsymbol{\veps}_\kk},\\
	\Hop_\text{at} &\equiv& \sum_{\jj} \spare{\frac{\pp_\jj^2}{2m}+\frac{1}{2}m\wT^2 \rrop_\jj^2 +\hbar \w_e \ketbra{e_\jj}{e_\jj}+\hbar\w_g\ketbra{g_{2,\jj}}{g_{2,\jj}}},\\
	\Hop_\text{int} &\equiv & \sum_{\jj}\sum_{\kk,\boldsymbol{\veps}_\kk} \spare{\aop_{\kk\boldsymbol{\veps}_\kk} e^{\im \kk\cdot (\RR_\jj+\rrop_\jj)}\pare{g_{\kk\boldsymbol{\veps}_\kk}^{(1)}\ketbra{e_\jj}{g_{1,\jj}}+g_{\kk\boldsymbol{\veps}_\kk}^{(2)}\ketbra{e_\jj}{g_{2,\jj}}}+\hc},\label{eq:H_int}\\
	\Hop_\text{L}(t)	&\equiv & \frac{\hbar\W}{2}\sum_{\jj} \cpare{e^{\im[ \kkL \cdot (\RR_\jj+\rrop_\jj)- \wL t]}\ketbra{e_\jj}{g_{2,\jj}}+e^{-\im[\kkL\cdot (\RR_\jj+\rrop_\jj) -\wL t]}\ketbra{g_{2,\jj}}{e_\jj}}.\label{eq:H_L}
\eea
Here, $\kkL$ ($\wL=c|\kkL|$) is the laser wave vector (frequency), $\w_\kk\equiv c |\kk|$, and upon introducing the dipole $\dd_\nu$ for the transition $\ket{e}\!\!\leftrightarrow\!\!\ket{g_\nu}$ ($\nu=1,2$), we defined the coupling constants
\bea
	g_{\kk\boldsymbol{\veps}_\kk}^{(\nu)}&\equiv& \dd_\nu \cdot \boldsymbol{\veps}_\kk \sqrt{\frac{\hbar \w_\kk}{2\veps_0 V}},\\
	\hbar\W &\equiv& 2\dd_2 \cdot \boldsymbol{\veps}_\text{L} E_\text{L},
\eea
where $\veps_\text{L}$ ($E_\text{L}$) is the polarisation (modulus) of the driving electric field and $V$ the quantisation volume. 
The derivation of \eqnref{eq:Ham_motion} starting from \eqnref{eq:H_full} is based on the following two steps. 
First, we adiabatically eliminate the excited state $\ket{e}$ and obtain an effective Hamiltonian describing the dynamics of two level systems of levels $\ket{g_1},\ket{g_2}$ interacting with the electromagnetic field~\cite{Marzoli1994,GonzalezTudela2015,Hung2016}. 
The adiabatic elimination can be carried out independently for each atom. The Schr\"odinger equation for the single-atom state $\ket{\Psi(\rr)}\equiv \psi_1(\rr)\ket{g_1}+\psi_{2}(\rr)\ket{g_2}+\psi_e(\rr)\ket{e}$ reads (in a frame rotating at the driving frequency)
\bea
	\im \hbar \Dot{\psi}_e(\rr) &=& \pare{\Hop_\text{cm}+\hbar\Delta}\psi_e(\rr)+\frac{\hbar\W}{2}e^{-\im \kkL\cdot \rr}\psi_2(\rr)+\sum_{\kk,\boldsymbol{\veps}_\kk} \adop_{\kk\boldsymbol{\veps}_\kk} e^{-\im \kk\cdot \rr}\spare{g_{\kk\boldsymbol{\veps}_\kk}^{(1)*}e^{-\im(\w_e-\Delta)t}\psi_1(\rr)+g_{\kk\boldsymbol{\veps}_\kk}^{(2)*}e^{-\im\wL t}\psi_2(\rr)},\label{eq:psi_e}\\
	\im\hbar\Dot{\psi}_2(\rr) &=& \Hop_\text{cm} \psi_2(\rr) + \frac{\hbar\W}{2}e^{\im \kk\cdot\rr}\psi_e(\rr) +\sum_{\kk,\boldsymbol{\veps}_\kk} g_{\kk\boldsymbol{\veps}_\kk}^{(2)}e^{\im \wL t}\aop_{\kk,\boldsymbol{\veps}_\kk} e^{\im \kk\cdot \rr}\psi_e(\rr),\label{eq:psi_2}\\ 
	\im \hbar \Dot{\psi}_1(\rr) &=& \Hop_\text{cm} \psi_1(\rr)+\sum_{\kk,\boldsymbol{\veps}_\kk}  g_{\kk\boldsymbol{\veps}_\kk}^{(1)}e^{\im(\w_e-\Delta)t}\aop_{\kk\boldsymbol{\veps}_\kk} e^{\im \kk\cdot \rr} \psi_e(\rr),\label{eq:psi_1}
\eea
where $\Delta\equiv \w_e -(\w_g+\wL)$ and $\Hop_\text{cm} \equiv \ppop^2/2m +m\wT^2 \rrop^2/2$. We set $\Dot{\psi}_e(\rr)=0$ in \eqnref{eq:psi_e} and solve for $\psi_e(\rr)$ neglecting the contribution of of $\Hop_\text{cm}$ under the assumption $\Delta \gg \wT$. We obtain an effective equation for the dynamics of $\psi_1(\rr)$ and $\psi_2(\rr)$. Generalising this procedure to the case of $N$ atoms is straightforward, and one obtains that a Hamiltonian which yields the effective equations of motion for $\psi_{1j}(\rr)$ and $\psi_{2j}(\rr)$ reads
\be\label{eq:Heff_1at}
\begin{split}
	\Hop_\text{T}^\text{eff} \equiv& \Hop_\text{rad}+ \sum_\jj\bigg\{ \bigg[\frac{\pp_\jj^2}{2m}+\frac{1}{2}m\wT^2 \rrop_\jj^2 +\hbar\w_g\sigop_{22}^\jj\bigg] -\sum_{\nu,\mu=1}^2\sum_{\kk,\kk',\boldsymbol{\veps}_\kk,\boldsymbol{\veps}_\kk'} \frac{g_{\kk\boldsymbol{\veps}_\kk}^{(\mu)}g_{\kk'\boldsymbol{\veps}_{\kk'}}^{(\nu)*}}{\Delta}e^{-\im[(\kk-\kk')\cdot (\RR_\jj+\rrop_\jj)-\wL(\nu-\mu)t]}\adop_{\kk'\boldsymbol{\veps}_{\kk'}}\aop_{\kk\boldsymbol{\veps}_\kk}\sigop^\jj_{\nu\mu}\\
	&-\frac{\hbar\W}{2\Delta}\sum_{\kk,\boldsymbol{\veps}_\kk}\spare{\aop_{\kk\boldsymbol{\veps}_\kk} e^{\im(\kk-\kkL)\cdot (\RR_\jj+\rrop_\jj)}\pare{g_{\kk\boldsymbol{\veps}_\kk}^{(1)}e^{\im(\w_e-\Delta)t}\sigop^\jj_{21}+g_{\kk\boldsymbol{\veps}_\kk}^{(2)}e^{\im \wL t}\sigop^\jj_{22}}+\hc}\bigg\},
\end{split}	
\ee
where we redefined $\w_g$ to include the AC Stark shift $\W^2/2\Delta$.
We stress that such procedure yields a light matter coupling which can be tuned via $\W/\Delta$.
Second, we eliminate the photonic degrees of freedom assumed to be in the vaccum state and obtain a Born-Markov master equation for the atomic variables. Master equations for the internal and center-of-mass dynamics of atoms interacting with electromagnetic field have been derived for single or independent atoms in the context of laser cooling~\cite{Stenholm1986,Javanainen1986,*Javanainen1988}. In the case of strong dipole interactions, such master equation can be extended to include coherent coupling and interference~\cite{Palmer2010}. Following a similar derivation, the master equation for the effective two-level atom reads
\be\label{eq:ME_tot}
	\pa{t} \rhop = -\im \hbar^{-1} \pare{\Hop \rhop - \rhop \Hop^\dag} + \mathcal{J}_{12}(\rhop)+\mathcal{J}_{22}(\rhop).
\ee
Here, $\Hop$ is the non-hermitian Hamiltonian of the system and reads
\be\label{eq:H_nh_tot}
	\Hop = \sum_\jj \bigg[\frac{\pp_\jj^2}{2m}+\frac{1}{2}m\wT^2 \rrop_\jj^2 +\hbar\w_g\sigop_{22}^\jj\bigg] + \hbar\pare{\frac{\W}{2\Delta}}^2\sum_{\ii,\jj}\spare{G(\rrop_\jj,\rrop_\ii)\sigop_{21}^\jj\sigop_{12}^\ii+F(\rrop_\jj,\rrop_\ii)\sigop_{22}^\jj\sigop_{22}^\ii},
\ee
where we defined the operators
\bea
	G(\rrop_\jj,\rrop_\ii) &\equiv& -\frac{\im}{\hbar^2}\sum_{\kk,\boldsymbol{\epsilon}_\kk} g_{\kk\boldsymbol{\veps}_\kk}^{(1)}g_{\kk\boldsymbol{\veps}_\kk}^{*(1)}e^{\im (\kk-\kkL)\cdot[(\RR_\jj+\rrop_\jj)-(\RR_\ii+\rrop_\ii)]}\!\int_0^\infty\!\!\text{d}\tau\, e^{\im(\w_e-\Delta-\w_\kk)\tau},\label{eq:G_def}\\
	F(\rrop_\jj,\rrop_\ii) &\equiv& -\frac{\im}{\hbar^2}\sum_{\kk,\boldsymbol{\epsilon}_\kk}g_{\kk\boldsymbol{\veps}_\kk}^{(2)}g_{\kk\boldsymbol{\veps}_\kk}^{*(2)}e^{\im (\kk-\kkL)\cdot[(\RR_\jj+\rrop_\jj)-(\RR_\ii+\rrop_\ii)]}\!\int_0^\infty\!\!\text{d}\tau\, e^{\im(\w-\w_\kk)\tau}.\label{eq:F_def}
\eea
Taking the continuum limit for $\kk$ and making use of the identity $\int_0^\infty\!\! \text{d}\w\, \exp(\pm \im \w x) =  \pi \delta(x)\mp\im \text{P.V.}(1/x)$, where $\text{P.V.}(1/x)$ stands for the Cauchy principal value of $1/x$, we can write \eqnref{eq:G_def} and  as
\be\label{eq:Gij_intermediate}
	G(\rrop_\jj,\rrop_\ii)=\frac{-\im}{2\epsilon_0\hbar(2\pi c)^3} \int\!\! \text{d}\nn\, \W(\nn)\! \int_0^\infty\!\!\! \text{d}\w_\kk\, \w_\kk^3 e^{\im (\kk-\kkL)\cdot[(\RR_\jj+\rrop_\jj)-(\RR_\ii+\rrop_\ii)]} \spare{\pi\delta(\w_e-\Delta-\w_\kk)+\im \text{P.V.} \pare{\frac{1}{\w_e-\Delta-\w_\kk}}},
\ee
where $\W(\nn) = \sum_{\boldsymbol{\veps}_\kk}\pare{\dd_1\cdot \boldsymbol{\veps}_\kk}^2$ describes the dipole emission pattern, $\nn$ is a unit vector which points in the direction of $\kk$, and $\boldsymbol{\veps}_\kk \perp  \kk$. 
Before proceeding we need to make an important remark.
As it is, \eqnref{eq:Gij_intermediate} does not lead to the correct (collective) shift induced by the electromagnetic field on the atoms as originally derived in~\cite{Bethe1947} using other techniques. Comparing \eqnref{eq:Gij_intermediate} with the equivalent expressions in~\cite{Lehmberg1970a,*Lehmberg1970b}, which yields results consistent with~\cite{Bethe1947}, one notes that the integral on $\w_\kk$ should be extended to the whole real line. The difference between \eqnref{eq:Gij_intermediate} and Eq.(15) in \cite{Lehmberg1970a} originates from the rotating wave approximation we assumed in \eqnref{eq:H_int}~\cite{Agarwal1973,Fleming2010}. The largest contribution to the frequency integral in \eqnref{eq:Gij_intermediate}, comes from values of $\w_\kk$ around $\w_e-\Delta \simeq \w_e$. In the optical domain such frequency is very large as compared to any other frequencies in the system. It is thus justified to extend the limit of integration $\int_0^\infty\text{d}\w_\kk \simeq \int_{-\infty}^\infty\text{d}\w_\kk$ in \eqnref{eq:Gij_intermediate} which then leads to the correct value of the atomic interaction~\footnote{The same kind of approximation was originally adopted in~\cite{Fermi1932} to preserve causality in the field-mediated interaction between two atoms. Such \emph{ad hoc} correction is also discussed more recently in~\cite{Ujihara2002,Haizhen2004,Dolce2006}}
\be
	G(\rrop_\ii,\rrop_\jj)=-\im \frac{\mu_0 \w_e^2}{\hbar}e^{\im\kk_\text{L}\cdot(\RR_\jj+\rrop_\jj-\RR_\ii -\rrop_\ii)}\Big[\dd_1 \cdot \GG^0(\RR_\jj+\rrop_\jj-\RR_\ii-\rrop_\ii,\w_e)\cdot \dd_1\Big].\label{eq:G_ij}
\ee
In a similar way, starting from \eqnref{eq:F_def} one can show that
\be
	F(\rrop_\ii,\rrop_\jj)=-\im \frac{\mu_0 \w_{e2}^2}{\hbar}e^{\im\kk_\text{L}\cdot(\RR_\jj+\rrop_\jj-\RR_\ii -\rrop_\ii)}\Big[ \dd_2 \cdot \GG^0(\RR_\jj+\rrop_\jj-\RR_\ii-\rrop_\ii,\wL)\cdot \dd_2\Big].
\ee
where we approximated $\w_e+\Delta\simeq \w_e$ and $\w_e-\w_g-\Delta \simeq \w_e-\w_g\equiv \w_{e2}$, and where the components of the free-space electromagnetic Green's function,  $G_{\alpha\beta}^0 \equiv \bold{e}_\alpha\cdot\GG_0(\rr,\w)\cdot\bold{e}_\beta$ (for $\alpha,\beta=x,y,z$), read
\be\label{eq:G0}
G^0_{\alpha \beta}(\rr,\w) = \frac{e^{\im k r}}{4\pi k^2 r^3}\Big[\pare{k^2 r^2 +\im k r -1}\delta_{\alpha\beta}-\pare{k^2 r^2 +3\im k r-3}\frac{r_\alpha r_\beta}{r^2}\Big].
\ee
The last two terms in \eqnref{eq:ME_tot} are the superoperators representing quantum jumps associated respectively to a decay from $\ket{g_2}$ to $\ket{g_1}$ or a dephasing of $\ket{g_2}$. They read
\bea
	\mathcal{J}_{12}(\rhop)&=& \sum_{\jj,\ii} \pare{\frac{\W}{2\Delta}}^2\gamma_1\, e^{-\im k_{e2}\hat{\kk}_\text{L} \cdot(\RR_\ii+\rrop_\ii)}\spare{\int_S\!\! \text{d} \nn\, \W(\nn)\, e^{\im k_e \nn\cdot(\RR_\ii+\rrop_\ii)}\sigop_{12}^\ii \rhop \sigop_{21}^\jj e^{-\im k_e \nn \cdot(\RR_\jj+\rrop_\jj)}}e^{\im k_{e2}\hat{\kk}_\text{L} \cdot (\RR_\jj+\rrop_\jj)},\label{eq:Jump_1}\\
	\mathcal{J}_{22}(\rhop)&=& \sum_{\jj,\ii}\pare{\frac{\W}{2\Delta}}^2 \gamma_2\, e^{-\im k_{e2}\hat{\kk}_\text{L} \cdot(\RR_\ii+\rrop_\ii)}\spare{\int_S\!\! \text{d} \nn\, \W(\nn)\, e^{\im k_{e2} \nn\cdot(\RR_\ii+\rrop_\ii)}\sigop_{22}^\ii \rhop \sigop_{22}^\jj e^{-\im k_{e2} \nn \cdot(\RR_\jj+\rrop_\jj)}}e^{\im k_{e2}\hat{\kk}_\text{L} \cdot (\RR_\jj+\rrop_\jj)},\label{eq:Jump_2}
\eea
\end{widetext}
where the integral is carried out on the unit sphere, $\gamma_1 \equiv |\dd_1|^2\w_e^3 /(3\pi \veps_0 \hbar c^3)$, and $\gamma_2 \equiv |\dd_2|^2\w_{e2}^3 /(3\pi \veps_0 \hbar c^3)$. Let us now interpret the processes in \eqnref{eq:Jump_1} and \eqnref{eq:Jump_2}. The process in \eqnref{eq:Jump_1} describes the correlated emission of a photon between each pair of atoms $\ii,\jj$ associated to the corresponding recoil of the atomic center-of-mass wavefunction. We observe that the atoms undergo two different recoils: one associated with the absorption of a photon from the laser on the transition $\ket{e}\leftrightarrow\ket{g_2}$ and a second recoil associated with the spontaneous emission of a photon on the transition $\ket{e}\leftrightarrow\ket{g_1}$. 
The first recoil happens always along the same direction fixed by $\hat{\kk}_\text{L}\equiv \kkL/|\kkL|$ where $|\kkL|\simeq k_{e2}$, while the second occurs around a random direction as prescribed by the dipole emission pattern $\W(\nn)$. Similar interpretation holds for \eqnref{eq:Jump_2}.
In the following, we assume $\gamma_2\ll\gamma_1$. Within this approximation the term proportional to $F(\rrop_\ii,\rrop_\jj)$ in \eqnref{eq:H_nh_tot} and the contribution $\mathcal{J}_{22}(\rhop)$ are negligible and the atomic evolution can be approximated by the non hermitian Hamiltonian
\be\label{eq:H_nh}
\begin{split}
	\Hop =& \sum_\jj \bigg[\frac{\ppop_\jj^2}{2m}+\frac{1}{2}m\wT^2 \rrop_\jj^2 +\hbar\w_g\sigop_{22}^\jj\bigg]\\
	& + \veps^2\sum_{\ii,\jj} \hbar G(\rrop_\jj,\rrop_\ii)\sigop_{21}^\jj\sigop_{12}^\ii,
\end{split}
\ee
and the action of stochastic quantum jump according to \eqnref{eq:Jump_1}.

% -----------------------------------
\subsection{Limit of Pinned Atoms}
% -----------------------------------

The case of atoms pinned to their lattice site is readily obtained from \eqnref{eq:H_nh} by setting $\rrop_\jj=0$. The center-of-mass motion is thus decoupled from the internal motion and the mechanical energy contribution in \eqnref{eq:H_nh}  can be neglected. Following this procedure one obtains the non-hermitian Hamiltonian
\be
	\frac{\Hop_0}{\hbar} = \w_g\sum_\jj\sigop_{22}^\jj +\veps^2\sum_{\ii,\jj} G_{\jj\ii}\,\sigop_{21}^\jj\sigop_{12}^\ii
\ee
where $G_{\jj\ii}$ is simply given by \eqnref{eq:G_ij} where the center-of-mass fluctuations $\rr_\jj$ have been neglected.
The complex coupling rate $G_{\jj\ii}$ is not the same as the free-space dipole-coupling rate between two two-level systems at positions $\RR_\ii$ and $\RR_\jj$, because of the phase factor $e^{\kk_\text{L}\cdot(\RR_\jj-\RR_\ii)}$ due to the Raman driving [Cfr. \eqnref{eq:G_ij} and Eq.(22) and Eq.(26) in~\cite{Lehmberg1970a} (beware that we used a different notation as compared to~\cite{Lehmberg1970a})].
However, $G_{\jj\ii}$ reduces to the usual form of the free-space dipole coupling for the case of pinned atom when we consider the Raman laser in \eqnref{eq:H_L} to be directed orthogonal to the plane containing the  arrays, \ie $\kk_\text{L}\parallel \uex$, for arrays lying in the $yz$-plane. We assumed this to be the case in all results presented in~\secref{sec:PinnedAtoms}. 

% ===================================
% ===================================
\section{Derivation of the Driving Hamiltonian}\label{app:Driving}
% ===================================
% ===================================

We now discuss in detail the conditions to drive single excitations dark modes of a single atomic array. The results generalise easily to the case of parallel arrays discussed in the text.
We consider an additional highly excited level $\ket{f}$ for each atom, and drive a two-photon transition between $\ket{g_1}$ and $\ket{g_2}$ through $\ket{f}$ (see~\figref{fig:Fig1}.b). When the two driving lasers are detuned from the intermediate level $\ket{f}$, this latter is negligibly populated during the process and can be adiabatically eliminated, yielding the effective driving Hamiltonian (in a frame rotating at the driving frequency $\w_d$)
\be\label{eq:V}
	\Vop = \hbar\W_0 \sum_{j=1}^N \sin[\pare{\kk_a-\kk_b}\cdot (\RR_j +\rrop_j)] \sy_j,
\ee
where $\kk_a$ ($\kk_b$) is the wave-vector associated to the laser driving the transition $\ket{f}\leftrightarrow\ket{g_1}$ ($\ket{f}\leftrightarrow\ket{g_2}$), $\sy_j\equiv\im(\sigop_{12}^j-\sigop_{21}^j)$ and $\W_0\equiv\W_a\W_b/2\Delta_f$ is the effective Rabi frequency of the two-photon transition. Note that to derive \eqnref{eq:V}, we assumed the system to be drive both from the left and from the right, a situation which can be achieved using a mirror to reflect the driving laser as illustrated in~\figref{fig:Fig1}.b. 

% -----------------------------------
\subsection{Limit of Pinned Atoms}
% -----------------------------------

For the case of atoms pinned to their lattice position, \eqnref{eq:V} reduces to 
\be\label{eq:V0}
	\Vop_0 =\hbar\W_0 \sum_{j=1}^N \sin(K_z a j_z) \sy_j,
\ee
where we further assumed the array to be aligned along the $z$-axis, $\RR_\jj = Z_j\uez$, $Z_j\equiv aj_z$, and we defined $K_z\equiv k_a \cos\alpha -k_b \cos\beta$  as in \eqnref{eq:q_effective}. Here, $k_a=k_f-\Delta_f/c$, $k_b=k_f-\Delta_f/c-\w_g/c$, where $k_f\equiv \w_f/c$ (see~\figref{fig:Fig1}.b).
For a given interatomic separation $a$, the values of $\alpha$ and $\beta$ which allows to tune the driving in \eqnref{eq:V0} to match the most subradiant single excitation state,  can be obtained by setting $K_z=\pi/a$. We write
\be\label{eq:beta}
	\beta = \frac{k_a}{k_b} \cos\alpha - \frac{k_e}{2k_b}\frac{\lambda_e}{a},
\ee
and assuming $\w_f=p\w_a+\Delta_f$ ($p\in \mathbb{Q}$), we solve for $\beta$ varying $a/\lambda_0$ and $\alpha$. The results are shown in \figref{Fig:Fig2}.b for $p=2$ demonstrating that a Raman transition through a higher level allows for exciting dark modes for arrays with small $a/\lambda_e$.
Let us finally remark that larger values of $p$ allow to drive subradiant modes of arrays with smaller interatomic separation.

% =================================================
\section{Motional Averaging in the limit of Fast Atomic Fluctuations}\label{app:Fast_Atomic_Fluctuations}
% =================================================

In the limit $\wT\gg \gamma_1$, the dynamics of an array of fluctuating atoms on the time scale of the internal dynamics can be approximated by an effective master equation for the sole internal degrees of freedom, where the coupling coefficients have been averaged over the center-of-mass motional state.
Specifically, the non hermitian Hamiltonian \eqnref{eq:H_nh} in the fast-atomic-motion limit is approximated as
\be
	\frac{\Hop_\text{fam}}{\hbar} =\w_g\sum_\jj\sigop_{22}^\jj +  \veps^2 \sum_{\ii\jj} \tilde{G}_{\ii\jj}\,\sigop_{21}^\jj\sigop_{12}^\ii
\ee
where $\tilde{G}_{\ii\jj}\equiv\avg{G(\rrop_\ii,\rrop_\jj)}_\text{cm}$, and the average is taken with respect to the probability distribution of the center of mass position of atoms at sites $\ii$ and $\jj$.
As in the main text, we assume the position fluctuation of the atoms to be independently and equally distributed according to 
\be\label{eq:Prob_atom_fluct}
	P(\rr_\ii) = \frac{e^{-\rr_\ii^2/2\sigma^2}}{(\sqrt{2\pi}\sigma)^3}.
\ee

In the following,  we evaluate the expression for the averaged coupling rate  $\tilde{G}_{\ii\jj}$ for the case of a single array of atoms polarised along the array direction in the limit of tightly trapped atoms, $\sigma\ll a$.
We start from \eqnref{eq:Gij_intermediate} and take the average over the atoms fluctuation with respect to the distribution \eqnref{eq:Prob_atom_fluct}. Using the result
\be
	\avg{e^{\im \kk\cdot(\rr_i - \rr_i)}}_\text{cm} = \delta_{ij}+(1-\delta_{ij})e^{-k^2 \sigma^2},
\ee
we obtain, separating real ($\tilde{J}_{ij}$) and imaginary ($\tilde{\Gamma}_{ij}$) parts,
\begin{widetext}
\bea
	\tilde{\Gamma}_{ij} &=& -\frac{\im \pi}{2\epsilon_0\hbar(2\pi c)^3} \int\!\! \text{d}\nn\, \W(\nn)\! \int_0^\infty\!\!\! \text{d}\w_\kk\, \w_\kk^3 e^{\im (\kk-\kkL)\cdot(\RR_j-\RR_i)}\spare{\delta_{ij}+(1-\delta_{ij})e^{-|\kk-\kkL|^2 \sigma^2}}\delta(\w_e-\Delta-\w_\kk)\label{eq:gammaij_avg}\\
	\tilde{J}_{ij} &=&\frac{\pi}{2\epsilon_0\hbar(2\pi c)^3} \int\!\! \text{d}\nn\, \W(\nn) \text{P.V.}\int_0^\infty\!\!\! \text{d}\w_\kk\, \w_\kk^3 e^{\im (\kk-\kkL)\cdot(\RR_j-\RR_i)}\spare{\delta_{ij}+(1-\delta_{ij})e^{-|\kk-\kkL|^2 \sigma^2}}\  \pare{\frac{1}{\w_e-\Delta-\w_\kk}.}\label{eq:Jij_avg},
\eea
For the case of $i=j$, one has $\tilde{\Gamma}_{ii}=\gamma_1$ $\forall i$, and the divrgent Lamb-shift $\tilde{J}_{ii}$.
For the case $i\neq j$ is it convenient to analyze \eqnref{eq:gammaij_avg} and \eqnref{eq:Jij_avg} separately.
Starting from the dissipative coupling rate $\tilde{\Gamma}_{ij}$, one proceed by changing to spherical coordinates, from which the radial integral can be immediately performed to yield
\be
	\tilde{\Gamma}_{ij} = - \im \frac{3\gamma_1}{8\pi}e^{-2k_e^2\sigma^2}\int_0^{2\pi}\!\!\!\text{d}\phi\!\!\int_0^\pi\!\!\!\text{d}\theta\,\sin^3\theta\, e^{\im k_e (Z_i-Z_j)\cos\theta}e^{2k_e^2\sigma^2\cos\phi\sin\theta},
\ee
where we defined $\kk_e \equiv k_e(\cos\phi\sin\theta,\sin\phi\sin\theta,\cos\theta)^T$, $\kkL\equiv k_\text{L}\uex\simeq k_e \uex$, assumed the array to be aligned along the $z$-axis. 
In the limit of $\sigma k_e \ll 1$, keeping only terms up to order $(k_e \sigma)^2$ we obtain
\be\label{eq:Gamma_ij_avg}
	\tilde{\Gamma}_{ij} = - \frac{3}{2}\gamma_1 (1-2k_e^2\sigma^2)\cpare{\frac{\sin[k_e(Z_i-Z_j)]}{[k_e (Z_i-Z_j)]^3}-\frac{\cos[k_e(Z_i-Z_j)]}{[k_e(Z_i-Z_j)]^2}} \simeq (1-2k_e^2\sigma^2) \Gamma_{ij},
\ee
where $\Gamma_{ij}$ is the dissipative coupling rate between two pinned atoms at site $i$ and $j$ as given by the imaginary part of \eqnref{eq:G_coupling}.
To evaluate the averaged collective dipole shift \eqnref{eq:Jij_avg}, it is convenient to rewrite it in Cartesian coordinates
\be\label{eq:J_integral}
	\tilde{J}_{ij} = \frac{\pi}{2\epsilon_0\hbar(2\pi)^3}\, \text{P.V.}\int_{\mathbb{R}^3}\!\!\!\text{d}^3\kk\, \frac{k}{k_e-k} e^{\im (\kk-\kkL)\cdot(\RR_j-\RR_i)}e^{-|\kk-\kkL|^2 \sigma^2}.
\ee
The integral in \eqnref{eq:J_integral} has been approximated in~\cite{AntezzaPRL2009,Perczel2017} for the case of an infinitely long array. By noticing that $e^{2k_e^2\sigma^2}\tilde{J}_{ij}$ is independent on $\sigma$ one can show that a good approximation for $\tilde{J}_{ij}$ reads,
\be\label{eq:Jij_approx}
	\tilde{J}_{ij} \simeq e^{-2k_e^2 \sigma^2} J_{ij} \simeq (1-2k_e^2 \sigma^2)J_{ij},
\ee
where in the last passage we assumed $k_e^2 \sigma^2\ll 1$, and $J_{ij}$ is the coherent coupling rate between two pinned atoms at site $i$ and $j$ given in \eqnref{eq:G_coupling}.
\end{widetext}

Let us point out that the method of averaging over the atomic fluctuations  albeit yielding for the coupling $\tilde{G}_{\ii\jj}$ similar results as the ones obtained in~\cite{AntezzaPRL2009,AntezzaPRA2009,Perczel2017} following a renormalisation procedure, it follows a fundamentally different approach. In particular, in \cite{AntezzaPRL2009,AntezzaPRA2009,Perczel2017} the free space coupling is properly renormalized yielding finite results for $\tilde{J}_{ii}$. The procedure followed here is not a renormalisation:  the coupling is still between point-like atoms, which however are randomly distributed according to \eqnref{eq:Prob_atom_fluct}.

% --------------------------------------------
\subsection{Averaged Driving Hamiltonian}
% --------------------------------------------

Under the assumption of fast atomic motion, we can approximate the driving Hamiltonian \eqnref{eq:V}, as
\be
	\Vop_\text{fam} \equiv \hbar\W_0 \sum_{j=1}^N \avg{\sin[\pare{\kk_a-\kk_b}\cdot (\RR_j +\rrop_j)]}_\text{cm} \sy_j,
\ee
which yields
\be
	\Vop_\text{fam} = \hbar\W_0 e^{-|\kk_a-\kk_b|^2\sigma^2/2} \sum_{j=1}^N \sin[\pare{\kk_a-\kk_b}\cdot \RR_j] \sy_j.
\ee
The effect of atomic motion, in the limit $\wT\gg \gamma_1$, is thus a renormalisation of the Rabi frequency $\W_0$.

%=============================================================
%=============================================================
\section{Perturbative Treatment of Atomic Fluctuations: Lamb-Dicke Expansion}\label{app:LambDicke}
%=============================================================
%=============================================================

When the atoms are prepared in a deep optical lattice ($r_0<a$), where $r_0$ is the zero point motion of the center of mass, and if the optical lattice is subwavelength ($a<\lambda_e/2$), the system is in the Lamb-Dicke regime ($\eta\equiv r_0 k_e \ll 1$). Within this regime, we expand the dipole coupling in power series up to second order in $\hat{\rr}_j$. After the expansion, we write \eqnref{eq:H_nh} as
\be\label{eq:Ham_exp_app}
	\Hop \simeq \Hop_0 +\Hop_\text{I1} + \Hop_\text{I2}.
\ee
Here, $\Hop_0$ represents the term to zero-order in $\rrop_\jj$, where center-of-mass dynamics and internal dynamics are decoupled. It reads
\be\label{eq:H_0}
\begin{split}
	\Hop_0 \equiv& \sum_{j=1}^N \bigg[\frac{\hat{\pp}_j^2}{2m} +\frac{1}{2}m\wT^2\rrop_j^2+\hbar\w_g\sigop_{21}^j\sigop_{12}^j\bigg]\\
	& + \veps^2\sum_{i,j}\hbar G_{ji}\sigop_{21}^j\sigop_{12}^\ii,
\end{split}
\ee
where as before we defined $\veps^2G(\RR_i,\RR_i)\equiv-\im \veps^2\gamma_1/2 \equiv -\im \Gamma_0/2$.
The term $\Hop_\text{I1}$ ($\Hop_\text{I2}$) represents the first (second) order correction in the atomic center-of-mass displacement. They read
\bea
	\Hop_\text{I1} &\equiv&\veps^2 \sum_{i,j} \pare{\rrop_i-\rrop_j}\cdot \grad G_{ji} \sigop_{21}^j\sigop_{12}^i ,\label{eq:H_I1_app}\\
	\Hop_\text{I2} &\equiv& \frac{\veps^2}{2} \sum_{j,i}\sum_{\alpha,\beta} \pare{\rop_{\alpha i}-\rop_{\alpha j}}\pare{\rop_{\beta i}-\rop_{\beta j}}\label{eq:H_I2_app}\\
	& & \times\pa{\alpha}\pa{\beta}G_{ij}  \sigop_{21}^j\sigop_{12}^i.\nonumber
\eea
Analogously, the expansion of \eqnref{eq:Jump_1} in powers of $\rrop_\jj$ yields the approximated quantum jump superoperator to second order in $\eta$.
Substituting \eqnref{eq:G_ij} into \eqnref{eq:H_I1_app} and \eqnref{eq:H_I2_app}, one obtains the following expressions for the higher order corrections
\begin{widetext}
\bea
	\Hop_\text{I1}&=& \sum_{i\neq j} \Big[\im \kkL\cdot(\rrop_i-\rrop_j)G^0_{ij}+\pare{\rrop_i-\rrop_j}\cdot \grad G^0_{ij}\Big]\sigop_{21}^j\sigop_{12}^i,\label{eq:H_I1_explicit}\\
	\Hop_\text{I2} &=& \inv{2}\sum_{i\neq j} \bigg\{-\Big[\kk_\text{L}\cdot\pare{\rrop_i-\rrop_j}\Big]^2G^0_{ij} +2\im\Big[ \kk_\text{L}\cdot\pare{\rrop_i-\rrop_j}\Big] \pare{\rrop_i-\rrop_j}\cdot \grad G^0_{ij}\label{eq:H_I2_explicit}\\
	& & +\sum_{\alpha,\beta=x}^z \pare{\rop_{\alpha i}-\rop_{\alpha j}}\pare{\rop_{\beta i}-\rop_{\beta j}} \pa{\alpha}\pa{\beta} G^0_{ij} \bigg\} \sigop_{21}^j\sigop_{12}^i.\nonumber
\eea
Here, we defined $G^0_{i,j}$ as $G(\rr_i,\rr_j)\equiv e^{\im\kkL\cdot(\RR_i+\rrop_i-\RR_j-\rrop_j)}G^0_{ij}$ where $G(\rr_i,\rr_j)$ is given in \eqnref{eq:G_ij}.
We note that in the limit $\wT\gg \Gamma_0$, the perturbative description,  \eqnref{eq:Ham_exp_app}, is consistent with the fast atomic motion approximation given in \appref{app:Fast_Atomic_Fluctuations}.
In fact, to first approximation, we take the trace over the center-of-mass degrees of freedom assuming the the atoms to be in a thermal state, and we obtain the following effective non-hermitian Hamiltonian
\be\label{eq:Heff_0_avg}
	\Hop_\text{eff,0} \equiv \Tr\spare{\rhop_\text{th} \pare{\Hop_0 +\Hop_\text{I1} + \Hop_\text{I2}}} \simeq  \hbar \w_g \sum_j \sigop_{21}^j\sigop_{12}^j + \veps^2\sum_{i,j}G_{ji}\sigop_{21}^\jj\sigop_{12}^i - 2 \veps^2 k_e^2 \sigma^2 \sum_{i\neq j} G_{ij}\sigop_{21}^j\sigop_{12}^i,
\ee
where, we approximated $k_\text{L}\simeq k_e$, and used the results in the supplemental material of~\cite{Guimond2019} to approximate $\sum_{\alpha\beta}\pare{\rop_{\alpha i}-\rop_{\alpha j}}\pare{\rop_{\beta i}-\rop_{\beta j}}\pa{\alpha}\pa{\beta} G^0_{ij} \simeq -(k_e\sigma)^2 G_{ij}$. Here, $\sigma \equiv r_0\sqrt{2n_\text{th}+1}$ and $n_\text{th}$ the thermal mean phonon number.
\eqnref{eq:Heff_0_avg} is evidently equivalent to the Hamiltonian averaged over the atomic positions given by \eqnref{eq:Ham_fam}, with \eqnref{eq:Gamma_ij_avg} and \eqnref{eq:Jij_approx}.

% -----------------------------------------------------------------
\subsection{Perturbative Expansion of the Driving Hamiltonian}
% -----------------------------------------------------------------

Within the Lamb-dicke regime, we can expand the driving modulation in \eqnref{eq:V} up to second order in power of $\rr_j$ as $\Vop\simeq \Vop_0 + \Vop_1 + \Vop_2$. The three contributions to the driving Hamiltonian read 
\bea
	\Vop_0 &\equiv&\hbar\W_0\sum_{\jj}\sin(K_z aj_z)\sy_\jj,\label{eq:V0_app}\\
	\Vop_1 &\equiv& \hbar\W_0 \sum_{j=1}^N \cos(K_z aj_z)\Big[\eta_z\pare{\bdop_{zj}+\bop_{zj}}+\eta_x \pare{\bdop_{xj}+\bop_{xj}}\Big]\sy_j,\label{eq:V1}\\
\Vop_2 &\equiv& - \frac{\hbar\W_0}{2} \sum_{j=1}^N \sin(K_z aj_z)\Big[\eta_z^2\pare{\bdop_{zj}+\bop_{zj}} ^2+2\eta_z\eta_x\pare{\bdop_{zj}+\bop_{zj}}\pare{\bdop_{xj}+\bop_{xj}}+\eta_x^2 \pare{\bdop_{xj}+\bop_{xj}}^2 \Big]\sy_j.\label{eq:V2}
\eea
Here, we assumed the two lasers to be directed orthogonal to the $y$-axis, defined $\eta_{x,z}\equiv K_{x,z}r_0$, where $\xop_j \equiv r_0(\bdop_{xj}+\bop_{xj})$, $\zop_j \equiv r_0(\bdop_{zj}+\bop_{zj})$, and
\bea
	K_x &\equiv& k_a \sin\alpha -k_b \sin\beta,\label{eq:K_x}\\
	K_z &\equiv& k_a \cos\alpha -k_b \cos\beta.\label{eq:K_z}
\eea
As illustrated in \figref{fig:Fig1}.b, $k_a=k_f-\Delta_f/c$, $k_b=k_f-\Delta_f/c-\w_g/c$, where $k_f\equiv \w_f/c$.

% ------------------------------------------------------------------
\subsection{Adiabatic elimination of the center-of-mass motion: Effective Hamiltonian for the internal dynamics}\label{app:Heff_motion}
% ------------------------------------------------------------------

In the regime $\wT\gg \Gamma_0$, we can approximate the dynamics of \eqnref{eq:Ham_exp_app} and Eq.~(\ref{eq:V0_app}-\ref{eq:V2}) by adiabatically eliminating the center-of-mass motion.
%, assumed in its ground state. Such effective dynamics correctly
Combining the expression for the Hamiltonian and the driving up to second order in the Lamb-Dicke parameter we obtain $\Hop_\text{t}=\Hop_\text{t}^{(0)}+\Hop_\text{t}^{(1)}+\Hop_\text{t}^{(2)}$, where $\Hop_\text{t}^{(i)}\equiv \Hop_i +\Vop_i$ contains the array and driving Hamiltonian up to order $\eta^i$ ($i=0,1,2$).
Assuming the array to be in its motional ground state, we calculate the effective Hamiltonian for the atoms' internal dynamics as
\be\label{eq:def_Heff}
	\Hop_\text{eff} = \Pop \pare{\Hop_\text{t}^{(0)}+\Hop_\text{t}^{(2)}}\Pop - \Pop \Hop_\text{t}^{(1)}\Qop \frac{1}{\Hop_0}\Qop\Hop_\text{t}^{(1)}\Pop \equiv \Hop_\text{eff,0} + \Vop_\text{eff} + \Wop_\text{eff}
\ee
Here, we introduced the projectors $\Pop=\id_\text{int}\otimes \ketbra{\bold{0}}{\bold{0}}$ and $\Qop\equiv \id_\text{int}\otimes \sum_{\nn\neq\bold{0}} \ketbra{\nn}{\nn}$, where $\id_\text{int}$ is the identity operator acting on the internal atomic degrees of freedom and $\ket{\nn}$ is the Fock state containing $\nn$ phonons in the array. 
Let us now derive an expression for the different terms on the right-hand side of \eqnref{eq:def_Heff}.

We define, $\Hop_\text{eff,0}\equiv \Pop (\Hop_0+\Hop_\text{I2})\Pop$, which is easily obtained from \eqnref{eq:H_0} and \eqnref{eq:H_I2_explicit}, and reads
\be\label{eq:Heff_0}
	\frac{\Hop_\text{eff,0}}{\hbar} = \w_g\sum_{j} \sigop^j_{21}\sigop^j_{12}+(1-2\eta^2)\veps^2\sum_{i, j} G_{ji} \sigop^i_{12}\sigop_{21}^j.
\ee
\eqnref{eq:Heff_0} corresponds to \eqnref{eq:Heff_0_avg} for $\sigma=r_0$, and is thus equivalent to the limit of fast moving atom \eqnref{eq:Ham_fam}.
The second term on the right-hand side of \eqnref{eq:def_Heff} is defined as $\Vop_\text{eff}\equiv\Pop(\Vop_0+\Vop_2)\Pop$, and it can be written as
\be\label{eq:Veff}
	\Vop_\text{eff} =\hbar\W_0\spare{1-\eta^2\pare{\frac{K_z}{k_e}}^2-\eta^2\pare{\frac{K_x}{k_e}}^2} \sum_j \sin(K_z a j_z)\sy_{j},
\ee 
where $\sigop_y^j \equiv \im (\sigop_{12}^j - \sigop^j_{21})$.
The contribution of \eqnref{eq:Veff} corresponds to an averaging of the driving modulation over the atomic center-of-mass state.
The last term in \eqnref{eq:def_Heff}, $\Wop_\text{eff}\equiv- \Pop \Hop_\text{tot}^{(1)}\Qop \Hop_0^{-1} \Qop \Hop_\text{tot}^{(1)}\Pop$, represents the coupling between different collective states mediated by the atomic motion. It reads
\bea
	\Wop_\text{eff} &=&- \eta^2 \sum_{m}\sum_{i\neq j}\sum_{ j\neq m} \bigg[\sum_{\alpha=y,z}G_\alpha'(R_{ji})G_\alpha'(R_{jm})\bigg]\pare{\sigop_{21}^j\sigop_{12}^i+\sigop_{21}^i\sigop_{12}^j}\frac{1}{\wT+\Pop\Hop_0\Pop}\pare{\sigop_{21}^j\sigop_{12}^m+\sigop_{21}^m\sigop_{12}^j}\label{eq:Weff_1}\\
	& & -\eta^2 \W_0 \sum_{i\neq j} \spare{\sum_{\alpha=yz}\frac{K_\alpha}{k_e}G_\alpha'(\RR_{ji})}\!\spare{\sigop^j_y\frac{\cos\pare{K_z  a j_z}}{\wT+\Pop\Hop_0\Pop}\pare{\sigop_{21}^j\sigop_{12}^i +\hc}+\pare{\sigop_{21}^j\sigop_{12}^i +\hc}\frac{\cos\pare{K_z  a j_z}}{\wT+\Pop\Hop_0\Pop}\sigop^j_y}\label{eq:Weff_2}\\
	& & -\eta^2\W_0^2 \sum_j \cos^2(K_z aj_z)\sigop_y^j \frac{1}{\wT+\Pop\Hop_0\Pop}\sigop_y^j,\label{eq:Weff_3}
\eea
where we defined $G'_\alpha(\RR_{ij})\equiv k_e^{-1} \pa{\alpha}G_{ij}$.
The interpretation of the different contributions in $\Wop_\text{eff}$ is the following. The term \eqnref{eq:Weff_1} is an off-diagonal contribution to the dipole coupling in \eqnref{eq:H_0} for it couples different eigenstates of $\Pop\Hop_0\Pop$ within the same $n$-excitation manifold. The term in \eqnref{eq:Weff_2} can be interpreted as a distorsion of the driving modulation, which couples the initial state to states other than the targeted dark mode. The last term \eqnref{eq:Weff_3} is correction to second order in the driving $\Vop_1$ and thus excites two photons transitions.

\end{widetext}

% ==================================
% ==================================
\section{Effective Three-Level Dynamics of a Driven Array of Pinned Atoms}\label{app:Heff_3levels}
% ==================================
% ==================================

In the following, we show that the complex  many body dynamics of driven single arrays of pinned atoms is effectively confined to a handful of levels. We do so  by deriving an effecting model for the driven dynamics. 

\begin{figure}
	\includegraphics[width=\columnwidth]{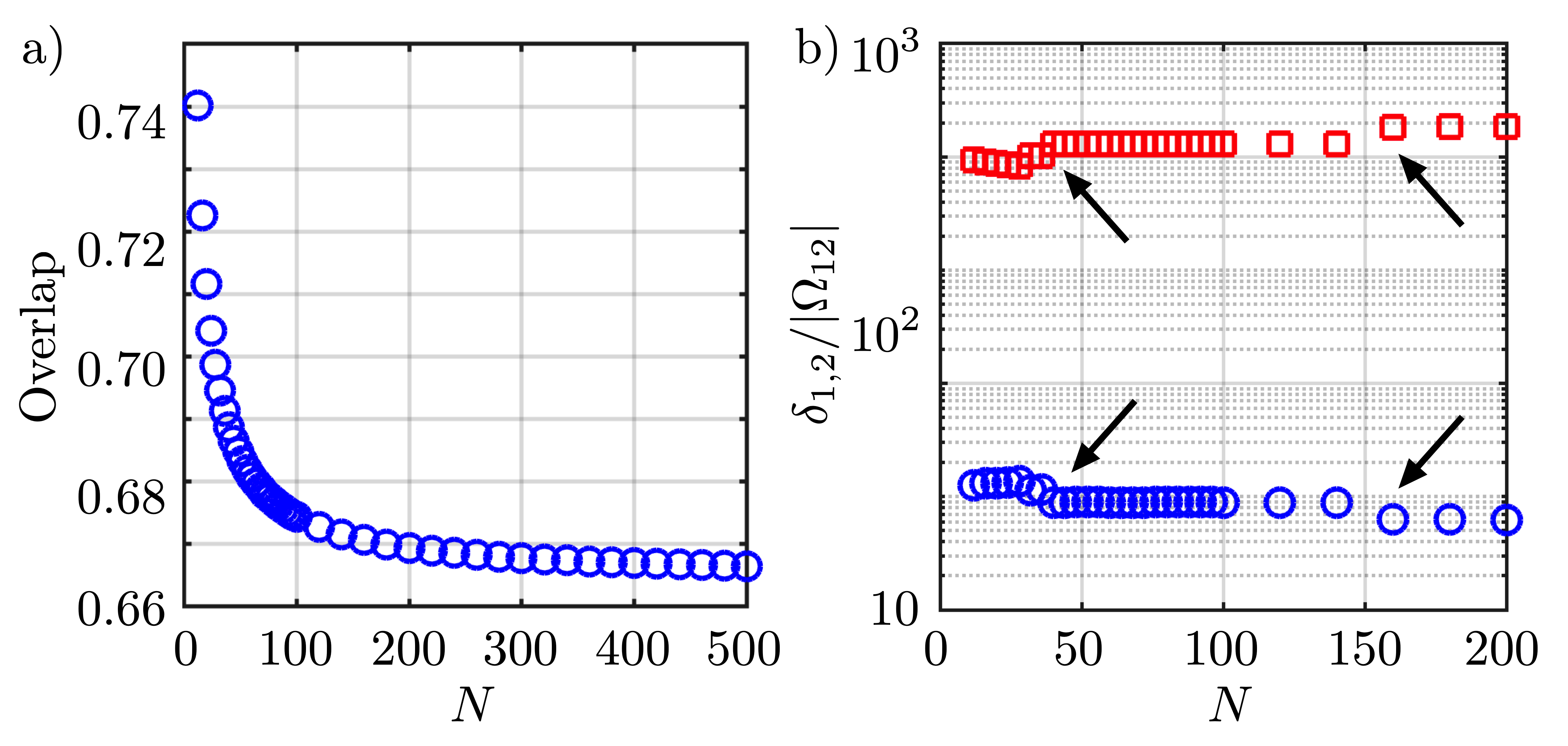}
	\caption{{\bf Effective three-level model.} a) Overlap $|\bra{\bar{2}}(\Sop^+)^2\ket{0}|^2$ as function of $N$.  b) Dependence of $\delta_1$ (blue circles) and $\delta_2$ (red squares) on the array length $N$. The arrows point at the sudden jumps in the detuning which corresponds to the jumps in the population shown in \figref{Fig:Fig2}.c. In both panels we set $a=\lambda_e/4$.}\label{Fig:EffectiveModel}
\end{figure}

Within the blockade regime the dynamics of a single driven array is effectively confined to the three-state manifold containing the ground state $\ket{0}$ and the most subradiant one- and two-excitation state. The remaining one- and two-excitation states are not populated due to the mode matching of the driving, while the population transferred to the three excitation manifold is negligible due to the non-linearity of the array (Dipole blockade and Zeno effect).
We note that the array's non-hermitian Hamiltonian \eqnref{eq:Ham} can be diagonalized in terms of right $\ket{n_\mu}$ and left $\ket{\bar{n}_\mu}$ eigenvectors ($n=1,\ldots,N$).  The set of left and right eigenvectors, which do not form separately orthogonal basis, satisfy the bi-orthogonality relation  
\be\label{eq:bi-orth}
	\braket{\bar{n}_\mu}{m_\nu} = \delta_{nm}\delta_{\mu\nu},
\ee
where we univocally fix the normalization of the left and right eigenvector by the additional requirement $\braket{n_\mu}{n_\mu}=1 \,\forall n,\mu$.
In the following, it is convenient to define $\ket{n_1}\equiv \ket{n}$ ($\ket{\bar{n}_1}\equiv \ket{\bar{n}}$) to represent the most-subradiant right (left) eigenstate with $n$ excitations.
The effective three-states non-hermitian Hamiltonian to second order in the coupling $\Vop$ is obtained as~\cite{Sternheim1972}
\be\label{eq:Heff_nonherm}
\begin{split}
\Hop_\text{eff}' =& \Pop'(\Hop_\text{1array}+\Vop_\text{1array})\Pop'\\
&-\Pop' \Vop_\text{1array} \Qop'(E_0 + \Hop_\text{1array})^{-1}\Qop'\Vop_\text{1array}\Pop',
\end{split}
\ee
where $\Hop_\text{1array}$ is the Hamiltonian of a single array as given in \eqnref{eq:Ham}, $\Vop_\text{1array}$ is given in \eqnref{eq:Vop}, we defined the projectors $\Pop'\equiv \ketbra{0}{0}+\ketbra{1}{\bar{1}}+\ketbra{2}{\bar{2}}$, $\Qop' = \id -\Pop'$, and the average complex energy of the three-level manifold $E_0$, assuming the driving laser to be resonant with $\ket{1}$.
On the bi-orthogonal basis $\{\ket{0},\ket{1},\ket{2},\ket{\bar{1}},\ket{\bar{2}}\}$, the effective three-state non-hermitian Hamiltonian in the frame rotating at the driving frequency thus reads
\be\label{eq:Heff_app}
	\frac{\Hop_\text{eff}'}{\hbar} = 
	\begin{pmatrix}
	0 &  \W_{01} & 0\\
	\W_{10} & \tilde{\delta}_1-\im \tilde{\Gamma}_1/2 & \W_{12}\\
	0 & \W_{21} & \tilde{\delta}_2-\im \tilde{\Gamma}_2/2
	\end{pmatrix}.
\ee
Here, we defined the coupling rates $\W_{nm}\equiv \bra{\bar{n}}\Vop_\text{1array}\ket{m}$ and we neglected the complex energy shift of the ground state as well as a direct coupling between the gound state and the two excitation state. Both these latter quantities are in fact much smaller than the remaining matrix elements already for arrays of few atoms, and decrease for increasing array size.
The diagonal elements in \eqnref{eq:Heff_app} already contain the correction due to coupling to higher excitation  manifolds. They read
\bea
	\tilde{\delta}_1-\im\frac{\tilde{\Gamma}_1}{2} &\equiv&\! -\im\frac{\Gamma_1}{2}-\! \sum_{\mu=2}^{\binom{N}{2}} \frac{\bra{\bar{1}}\Vop\ketbra{2_\mu}{\bar{2}_\mu}\Vop \ket{1}}{E_0+(\Delta_{2,\mu}-\im\Gamma_{2,\mu}/2)},\label{eq:correction_1}\\
	\tilde{\delta}_2-\im\frac{\tilde{\Gamma}_2}{2} &\equiv&\! \delta_2 -\im\frac{\Gamma_2}{2}-\! \sum_{\mu=2}^{N}\!\frac{\bra{\bar{2}}\Vop\ketbra{1_\mu}{\bar{1}_\mu}\Vop \ket{2}}{E_0+\!(\Delta_{1,\mu}-\im\Gamma_{1,\mu}/2)}\nonumber\\
	& &  -\sum_{\mu=1}^{\binom{N}{3}}\!\frac{\bra{\bar{2}}\Vop\ketbra{3_\mu}{\bar{3}_\mu}\Vop \ket{2}}{E_0+\!(\Delta_{3,\mu}\!-\!\im\Gamma_{3,\mu}/2)},\label{eq:correction_2}
\eea
where $\Delta_{n,\mu}$ is the detuning of the driving from the energy of the $\mu$-th state in the $n$-excitation manifold of $\Hop_\text{1array}$, and $\Gamma_{n,\mu}$ its decay rate. In \eqnref{eq:correction_1} and \eqnref{eq:correction_2} the indexes $\mu$ label states within each $n$-excitation manifold ordered for increasing values of the decay rate  $\Gamma_{n,\mu}$. We also defined $\delta_n\equiv \Delta_{n,1}$ and $\Gamma_n\equiv \Gamma_{n,1}$.
The effective three-level model in \eqnref{eq:Heff_app} captures the correct dynamics of the system in the Zeno regime as shown by the pink diamond in \figref{Fig:Fig2}.c.
Furthermore, via the effective description in \eqnref{eq:Heff_app} we can understand the discontinuous behaviour as function of $N$ of the total population in the ground state and the subradiant two-excitations state. 
We point out that the state obtained  by creating two subradiant collective excitations on  top of each other in the array, $\ket{2k_N}\sim(\Sop_{k_N}^+)^2\ket{0}$, where $\Sop_{k_N}^+\equiv\sum_{j=1}^N \sqrt{2/(N+1)}\sin(z_j k_N )\spl_j$, has a large overlap with the two-excitation most-subradiant state of the array even in the limit $N\gg1$ (\figref{Fig:EffectiveModel}.a). Population transfer from the single-excitation subradiant state to the two-excitation subradiant state under the action of the driving is prevented by the detuning $\delta_2$ of the latter with respect to the laser driving. The dependence of $\delta_{1,2}$ on the array's length $N$ is discontinuous as shown in \figref{Fig:EffectiveModel}.b. This discontinuity is a consequence of the discreteness of the array and it is reflected by the discontinuous jumps in the total population transferred to $\ket{0}$ and $\ket{2}$ observed in \figref{Fig:Fig2}.c.

Before concluding this section, let us remark that the effective Hamiltonian derived according to~\eqnref{eq:Heff_nonherm} is generally different from the one that would have been obtained, had we performed an adiabatic elimination on the full master equation describing the array's dynamics. The two approaches give consistent results only if the additional correction obtained following the latter  (more rigorous) method can be neglected. The good agreement between the results obtained from  \eqnref{eq:Heff_app} and the simulation of the full Hamiltonian in \eqnref{eq:Ham}, justifies a posteriori the simple approach followed here. 

% =============================================
% =============================================
\section{Description of Two Parallel Arrays of Pinned Atoms}\label{app:Parallel_Arrays}
% =============================================
% =============================================

The  non-hermitian Hamiltonian describing the open dynamics (conditioned on no jump occurring) for two parallel arrays of atoms is given in~\eqnref{eq:Ham_2Arrays_Pinned}.
As for the case of a single array, the hermitian and anti-hermitian parts of \eqnref{eq:Ham_2Arrays_Pinned} commute with the total number of excitations $\hat{N}_\text{e}$. The Hamiltonian of two parallel arrays can thus be block diagonalized in blocks which act only on a manifold of given number of excitations $n$,
\be
	\Hop_0 = \sum_{n=1}^{2N} \Hop_0^{(n)}.
\ee
Here, $\Hop_0^{(n)}$ is the Hamiltonian acting on the $n$-excitations manifold.
In the following, we look separately at the eigenvalues and eigenvectors of one- and two-excitation block of $\Hop_0$.

% ----------------------------------------
\subsection{Single Excitation Structure}
% ----------------------------------------

We consider first the case of periodic boundary conditions, where the single-excitation block $\Hop_0^{(1)}$ can be diagonalized exactly. We later discuss the relevant case of open boundary conditions.
Within the single excitation subspace we can replace the spin operators in $\Hop_0^{(1)}$ with bosonic operators $\bop_{\nu,j},\bdop_{\nu,j}$ where $\nu=\text{A,B}$ labels the array. In terms of Fourier modes $\bop_k\equiv \sum_{j=0}^{N} \exp(\im Z_j k)\bop_j$ and $\bdop_k\equiv \sum_{j=0}^{N} \exp(-\im Z_j k)\bdop_j$, $\Hop_0^{(1)}$ has the block diagonal form 
\be
	\Hop_0^{(1)} = \sum_k \Hop_k 
\ee
where each block $\Hop_k$ is a 2-level operators containing only operators with a well defined quasi momentum $k=\pi/[a(N+1)],\ldots,\pi N/[a(N+1)]$. Each $k$-block reads
\be\label{eq:H_k}
\begin{split}
	\frac{\Hop_k}{\hbar}=&\sum_{\nu=\text{A,B}}\pare{\w_{\nu,k}-\im \frac{\Gamma_k}{2}} \bdop_{\nu,k}\bop_{\nu,k}\\
	& + \pare{g_k-\im\frac{\gamma_k}{2}} \pare{\bdop_{\text{B}k}\bop_{\text{A}k}+\bdop_{\text{A}k}\bop_{\text{B}k}},
\end{split}
\ee
where the two frequencies $ \w_{\text{B},k}$ and $ \w_{\text{A},k}$ may differ if the atoms of one array are detuned from the ones of the other array. The expression for $\pare{g_k-\im\gamma_k/2}$ appearing in~\eqnref{eq:H_k} is given in~\tabref{Tab:Coupling_q-space} for different direction of the atomic polarization (assuming all atoms to be polarized parallel to each other). The $k$-space expression of the couplings can be derived as follows.
\begin{table*}
\caption{Momentum space expression of the conservative $g_k$ and dissipative $\gamma_k$ coupling between two arrays within the single excitation subspace. Here $J_n(\rho)$ ($Y_n(\rho)$) and $K_n(\rho)$ are respectively the $n$-th order Bessel function of the first (second) kind and the modified Bessel function of the second kind. We defined $\rho\equiv l\sqrt{k^2-k_e^2}$ when $k>k_e$ and $\rho\equiv l\sqrt{k_e^2-k^2}$ when $k<k_e$.}\label{Tab:Coupling_q-space}
\begin{ruledtabular}
\begin{tabular}{lll}
Polarization & $|k| > k_e$ & $|k| < k_e$\\
\hline
\\
$\dd_\text{A}=\dd_\text{B}=d \uex$ & 	$g_k = -3\Gamma_0 \spare{K_0\pare{\rho}
	 -\frac{\rho K_1\pare{\rho}}{(k_el)^2}
	+\,\frac{\rho^2 K_2\pare{\rho}}{(k_el)^2}}/k_e a$ &  $g_k =3\pi \Gamma_0\spare{Y_0\pare{\rho}-\frac{\rho Y_1\pare{\rho}}{(k_e l)^2}+\frac{\rho^2 Y_2\pare{\rho}}{(k_e l)^2}}/4k_e a$\\
	& $\gamma_k=0$ & $\gamma_k =3\pi\Gamma_0\spare{J_0\pare{\rho}-\frac{\rho J_1\pare{\rho}}{(k_el)^2} +\frac{\rho^2 J_2\pare{\rho}}{(k_el)^2}}/2k_e a$\\
	\\
	\hline
	\\
$\dd_\text{A}=\dd_\text{B}=d \uey$	& $g_k = -3\Gamma_0\spare{K_0(\rho)-\rho K_1(\rho)/(k_el)^2}/k_e a$  & $g_k =-3\pi\Gamma_0\spare{Y_0\pare{\rho}-\rho Y_1(\rho)/(k_el)^2}/4k_ea$\\
	& $\gamma_k=0$ & $\gamma_k=3\pi\Gamma_0\spare{J_0(\rho)-\rho J_1(\rho)/(k_el)^2}/2 k_ea$\\
	\\
	\hline
	\\
	$\dd_\text{A}=\dd_\text{B}=d \uez$ & $g_k = -3\Gamma_0\pare{1-k^2/k_e^2}K_0(\rho)/k_e a$ & $g_k = 3\pi\Gamma_0\pare{1-k^2/k_e^2}Y_0(\rho)/4 k_e a$\\
	& $\gamma_k=0$ & $\gamma_k=3\pi\Gamma_0 \pare{1-k^2/k_e^2} J_0(\rho)/2k_ea$\\
\\
\end{tabular}
\end{ruledtabular}
\end{table*}
Let us first define the mixed representation of the free-space electromagnetic Green's function as
\be\label{eq:G0_MixedRep}
\begin{split}
	G^0_{\alpha\beta}(\rr,\w_0) =&\frac{1}{(2\pi)^3} \int_{\mathbb{R}^3}\!\!\!\text{d}\pp\, \frac{\delta_{\alpha\beta}-p_\alpha p_\beta/k_e^2}{\pp^2-k_e^2} e^{\im \pp\cdot \rr}\\
	 \equiv & \frac{1}{2\pi}\int_{\mathbb{R}}\!\!\!\text{d}p_z\, \tilde{G}^0_{\alpha\beta}(x,y,p_z,\w_e) e^{\im p_z z}.
\end{split}
\ee
Substituting \eqnref{eq:G0_MixedRep} into
\be
	g_{ij}-\im\frac{\gamma_{ij}}{2} \equiv -\frac{\mu_0\w_0^2}{\hbar} \dd_{\text{A}i} \cdot \GG^0(\rr_{\text{A}i}-\rr_{\text{B}j},\w_e)\cdot \dd_{\text{B}j},
\ee
and using the identity
\be
	\sum_R e^{-\im k R} = \frac{2\pi}{a} \sum_Q \delta(k+Q),
\ee
where $Q\in \{2n\pi/a\}_{n\in\mathbb{Z}}$ is the reciprocal lattice vector of the atomic array, one obtains
\be\label{eq:g_general1}
	g_k-\im \frac{\gamma_k}{2} = -\frac{\mu_0 \w_e^2}{a \hbar} \sum_Q \tilde{G}^0_{\alpha\beta}(x,y,k+Q,\w_e).
\ee
Before proceeding to evaluate $\tilde{G}^0_{\alpha\beta}(x,y,k+Q)$ we remark that it is sufficient to keep only the term $Q=0$ in \eqnref{eq:g_general1}. Higher values of $Q$ represent contribution coming from modes beyond the first Brillouin zone (umklapp processes) which can be neglected since $a<\lambda_e/2$ and $\w_e\gg \Gamma_0$.
The coupling of an excitation of quasi-momentum $k$ between two parallel arrays within the single excitation manifold thus reads
\be\label{eq:g_general2}
	g_k-\im \frac{\gamma_k}{2} = -\frac{\mu_0 \w_e^2}{a \hbar} \tilde{G}^0_{\alpha\beta}(x,y,k),
\ee
and depends on the polarization direction $\alpha,\beta$ of the atoms in the arrays.
The 2D-fourier transform
\be \tilde{G}^0_{\alpha\beta}(x,y,k,\w_0) = \int_{\mathbb{R}^2} \frac{\text{d}\pp_\perp}{(2\pi)^2} \frac{\delta_{\alpha\beta} -p_\alpha p_\beta}{\pp_\perp^2+k^2-k_e^2}e^{\im \pp_\perp\cdot \rr_\perp},
\ee
where $\pp_\perp \equiv p_x\uex + p_y \uey$ and $\rr_\perp = x \uex + y\uey$, can be evaluated by expressing the integral in polar coordinate and evaluating the polar integral first. To evaluate the radial integral it is necessary to distinguish between the two cases (i) $|k| > k_e$ and (ii) $|k|<k_e$. The first case represent the coupling between collective mode lying outside the light cone and thus its dissipative contribution vanish. The second case corresponds to mode within the light cone and it thus have a strong dissipative contribution (see~\tabref{Tab:Coupling_q-space}). 
In the subradiant part of the spectrum $\gamma_k=0$, thus the eigenstates of \eqnref{eq:H_k} are the usual dressed state with frequency $\w_\pm =\delta/2 \pm \sqrt{\delta^2+4g_k^2}/2$ and decay rate $\Gamma_k$.
Within the light cone $\gamma_k$ and $g_k$ are comparable and one needs to diagonalized a $2\times2$ complex symmetric Hamiltonian. The results of such diagonalization is similar to the previous case upon substituting $g_k$ with $g_k-\im\gamma_k/2$.

\begin{figure}
	\includegraphics[width=\columnwidth]{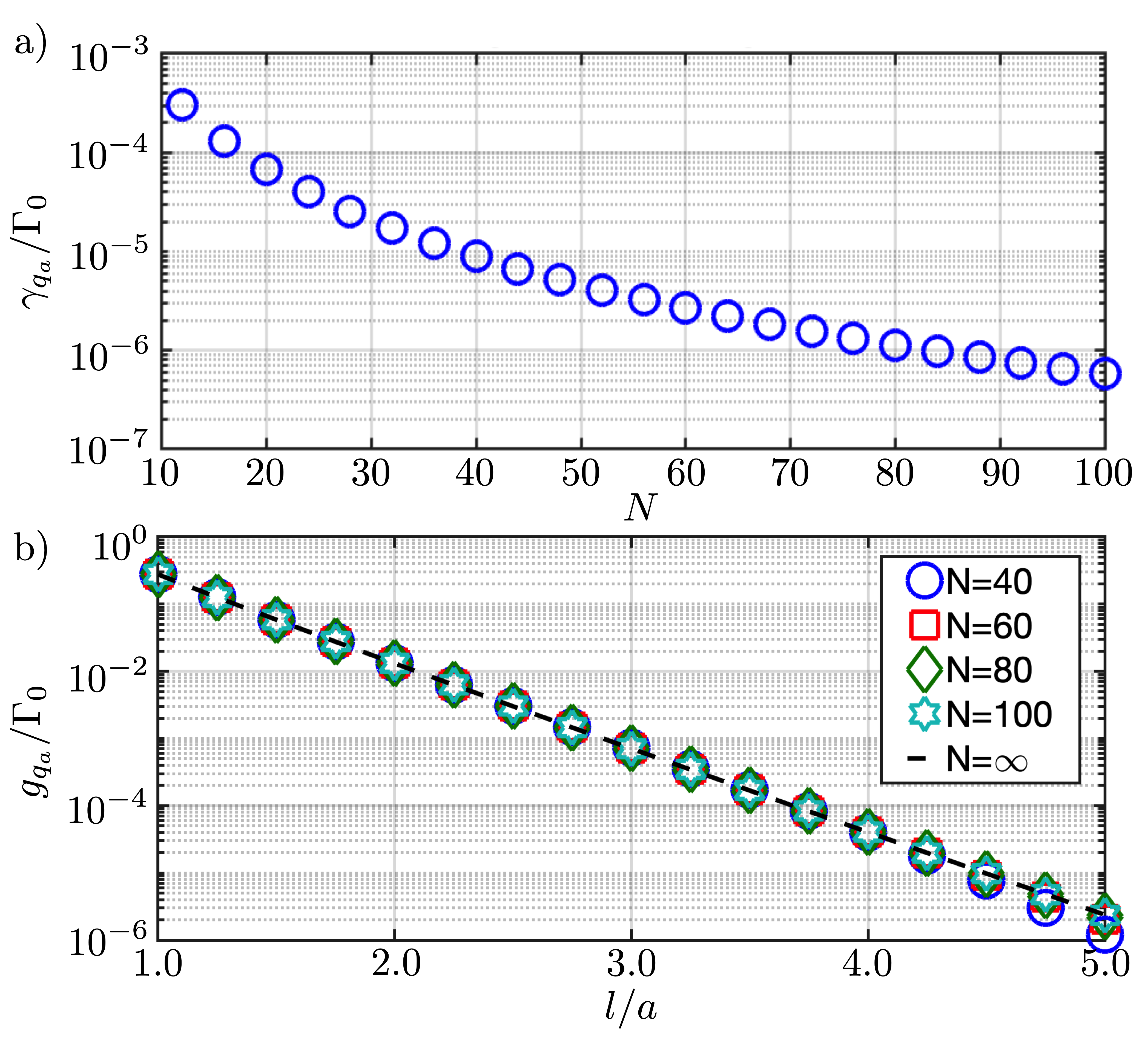}
	\caption{{\bf Single excitation coupling rate between dark modes of parallel arrays.} Scaling of $g_k-\im\gamma_k/2$ for atoms polarized along $\uez$ for $a=\lambda_e/4$. a) $\gamma_{q_a}$ in function of $N$ for $l=a$. b) $g_{q_a}$ in function of $l$ for different array length $N$ (markers) and for an infinite array (dashed black line).}\label{Fig:Scaling_gq}
\end{figure}

For open boundary conditions, the decoupling between collective excitations with different wave number is not exact and in the subradiant sector the coupling acquires a small dissipative part (\figref{Fig:Scaling_gq}.a).
Remarkably for $l=a\leq \lambda_e/4$ the block diagonal picture still holds  accurately for array as short as $N=6$ (see \figref{fig:Fig3}.a in the main text). In \figref{Fig:Scaling_gq}.b, we compare the dependence of $g_k$ for $k=q_a$ on the separation $l$ for different array sizes $N$ with the case $N=\infty$ in \tabref{Tab:Coupling_q-space}. For small separations the agreement is good, while at larger separation shorter arrays deviates from the ideal case.
For finite array size, the dressed state of \eqnref{eq:H_k} in the subradiant sector have different decay rate due to a non-zero $\gamma_k$. In the limit $\gamma_k\ll g_k,\delta$, the dressed complex energies to first order in $\gamma_k$ read $E_\pm \equiv \delta\pm g_k/\W_k +\im (\Gamma_0\pm \gamma_k g_k/\W_k)$ where $\W_k\equiv \sqrt{\delta^2+4g_k^2}$.

% ------------------------------------------------------
\subsection{Two Excitation Structure and Dynamics}
% ------------------------------------------------------

\begin{figure}
	\includegraphics[width=\columnwidth]{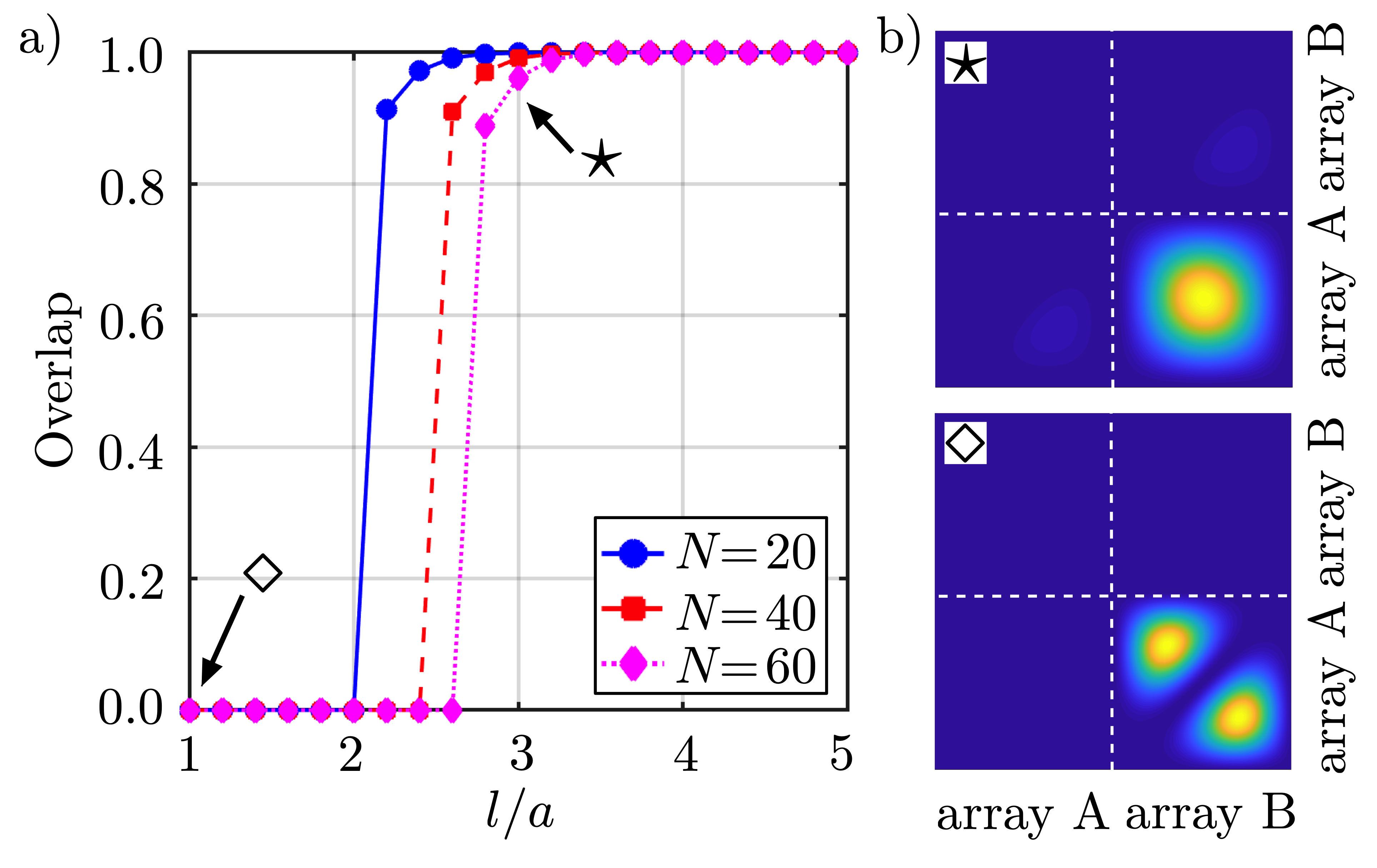}
	\caption{{\bf Two-excitations subradiant-state structure.} a) Overlap between $\ket{11}_\text{L}$ and the two-excitation most-subradiant state of two parallel arrays $\ket{\psi_2}$. We plot the overlap as funcition of $l/a$ for $a/\lambda_e=0.1$ and for $N=20$ (blue circles), $N=40$ (red squares), and $N=60$ (pink diamonds). b) Porbability distribution of the two-excitation most-subradiant state for $N=60$ and $a/\lambda_e=0.1$ at $l/a=1$ (bottom panel) and $l/a=3$ (top panel).}\label{fig:Structure_2exc}
\end{figure}

Let us now study the two excitation structure of two parallel atomic arrays. Here, we focus in particular on the subradiant sector of the two-excitations manifold.
Due to the dipole interaction \eqnref{eq:Ham_int} between the arrays, we cannot make general statements on the form of the first few dark modes as it was done for the case of an isolated array~\cite{AsenjoGarcia2017PRX}. In particular, the structure of the subradiant states strongly depends on the separation $l$ between the arrays.
In the limit of separation $l/a\gg1$, the arrays do not interact. The two-excitations most-subradiant state of the system is thus given by the state $\ket{11}_\text{L}=\ket{q_a,q_a}$, which contains one subradiant excitation in each array and is a characterised by a decay rate $\Gamma_{11}=2\Gamma_{q_a}$.
When the arrays are brought closer they start to interact thus modifying the structure of the ground state. In particular, at small separation $l\sim a$, we expect the two excitations to exhibit anti-bunching behaviour between the two arrays due to strong dipole-dipole interaction. In~\figref{fig:Structure_2exc}.a, we plot the overlap between $\ket{11}_\text{L}$ and the two-excitations most-subradiant state of the system $\ket{\psi_2}$.
In~\figref{fig:Structure_2exc}.b we plot the probability distribution of $\ket{\psi_2}$ on the sites of each array and show that while for $l=a$ it shows anti-bunching behaviour (bottom panel) at large separation converges towards $\ket{11}_\text{L}$ (top panel).

\begin{figure}
	\includegraphics[width=\columnwidth]{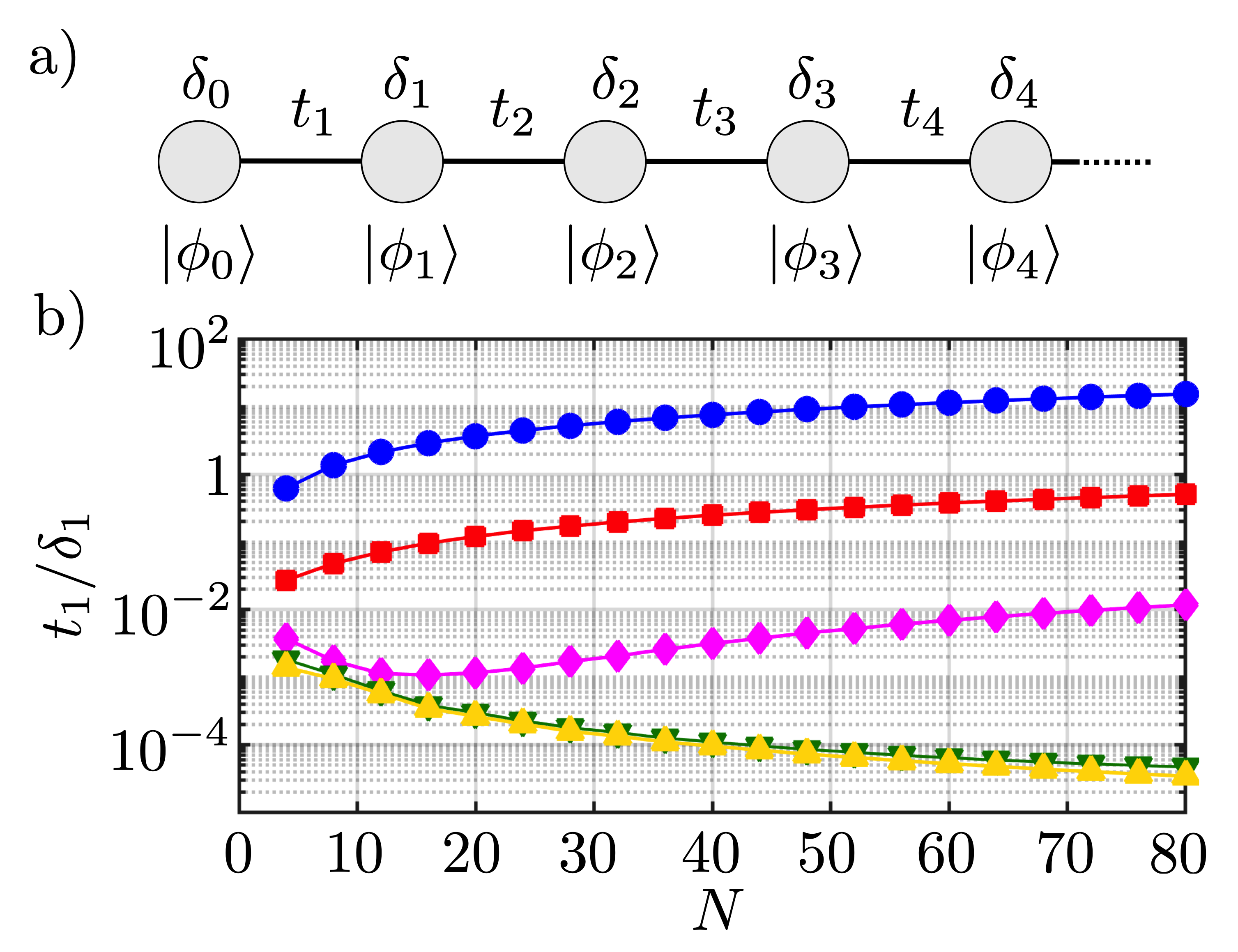}
	\caption{{\bf Analysis of the dynamics of $\ket{11}_\text{L}$.} a) virtual chain for the dynamics of parallel array starting in $\ket{\phi_0}=\ket{11}_\text{L}$ constructed according to the symmetric Lanczos transformation. b) Ratio $t_1/\delta_1$ (we choose a frame where $\delta_0=0$) as function of $N$ for different array's separation $l/a=1$ (blue circles), $l/a=2$ (red squares), $l/a=3$ (pink diamond), $l/a=4$ (green inverted triangles), and $l/a=5$ (yellow triangles).}\label{fig:Lanczos}
\end{figure}

The dynamics of $\ket{11}_\text{L}$ during the $\sqSWAP$-gate operation depends strongly on the array's separation $l$ and length $N$. Generally $\ket{11}_\text{L}$ couples to a large number of two-excitation eigenstates of the system, and the resulting dynamics is quite complicated. To elucidate the evolution of $\ket{11}_\text{L}$ and gain a qualitative understanding of its dynamics, it is convenient to implement a Lanczos transformation~\cite{Haydock1980}. 
According to such transformation, we can imagine the dynamics of $\ket{11}_\text{L}$ as the dynamics of a chain of neighbouring coupled sites (see \figref{fig:Lanczos}.a).
Each site $j$ is represented by a state $\ket{\phi_j}$ reached by the dynamics such that states corresponding to different sites are orthogonal. In particular, $\ket{\phi_0}\equiv \ket{11}_\text{L}$. On the basis $\ket{\phi_j}$ the dynamics of the system starting in $\ket{\phi_0}$ is represented by a tridiagonal Hamiltonian
\be\label{eq:H_tridiag}
	\frac{\Hop_\text{chain}}{\hbar}= \begin{pmatrix}
	\delta_0 & t_1 & 0 & \ldots  & 0 &\\
	t_1 & \delta_1 & \ddots & 0 & \vdots \\
	0 & \ddots & \ddots & \ddots  & 0 \\
	\vdots & 0 & \ddots  &  \ddots  &  t_M \\
	0 & \ldots & 0 & t_M & \delta_M 
	\end{pmatrix},
\ee 
which describes the nearest neighbour hopping of excitations at rate $t_j$ between the sites $j$ and $j+1$, as well as the energy $\delta_j$ of each site.
In the particular case considered here, the state directly coupled to $\ket{11}_\text{L}$ by \eqnref{eq:Ham_2Arrays_Pinned} reads $\ket{\phi_1}= \ket{S_2}\equiv(\ket{2q_a,0}+\ket{0,2q_a})/\sqrt{2}$.
and the coupling rate is simply given by $t_1 =\, _\text{L}\bra{11}\Hop_\text{AB}\ket{\phi_1}$.
In general, the dynamics of $\ket{11}_\text{L}$ cannot be reduced to an effective two-level dynamics on the subspace $\{\ket{11}_\text{L},\ket{S_2}\}$, for the coupling between $\ket{S_2}$ and $\ket{\phi_3}$ is typically strong. Such reduction to an effective two level dynamics is justified only at large separation $l/a$ as shown by the agreement between the black dashed markers and the colored markes in \figref{fig:Fig3}.d,e.
The two level dynamics represented by the first block in \eqnref{eq:H_tridiag} gives however a qualitative explanation to the trend observed in \figref{fig:Fig3}.e. In fact, the decrease in the gate fidelity over the array's size $N$ at small separation $l/a$ can be understood as coming from a reduction in the detuning $\delta_1$ (\figref{fig:Lanczos}.b).

% =================================================
% =================================================
\section{Perpendicular Polarized Atoms}\label{app:Perp_Polarization}
% =================================================
% =================================================

\begin{figure}
	\includegraphics[width=0.8\columnwidth]{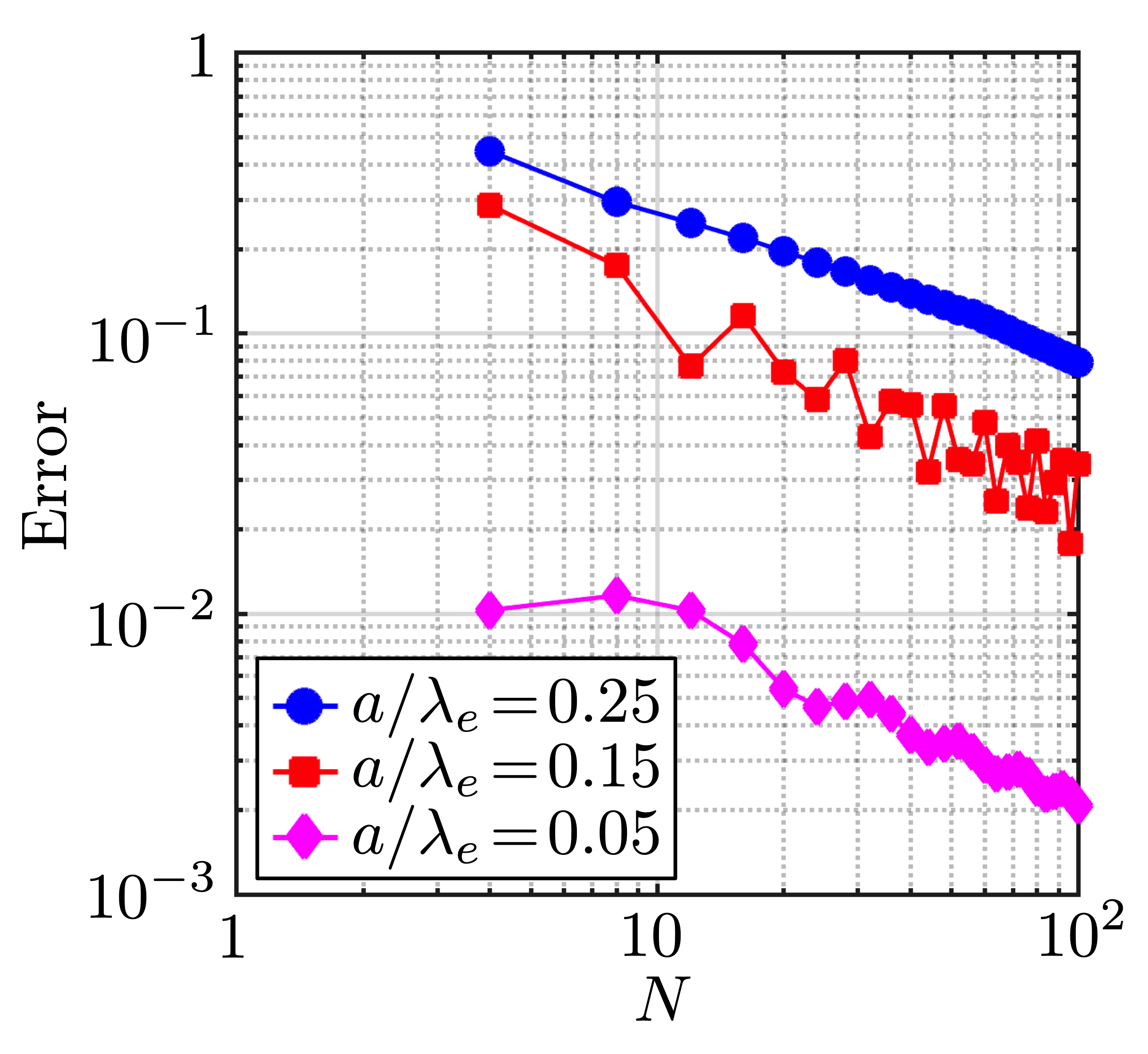}
	\caption{{\bf Array with transverse atomic polarisation}. Optimal Error for preparing the most subradiant state of a transverly polarised single array ($\dd\perp \uez$) as function of $N$ for different lattice spacing (see legend).}\label{Fig:Fig_p-pol}
\end{figure}

In this appendix, we consider the case of an array of pinned atoms which are polarised along a direction orthogonal to the array's direction.  
Mathematically, the only difference between this case and the one of atoms polarised along the array's direction is the value of the coupling in~\eqnref{eq:G0}, which now includes a long range coupling term which scales as $1/r$, $r$ being the distance between two atoms.
Single and two qubit gate can be realised following the same protocol as presented in~\secref{sec:PinnedAtoms}. As an example, we show in~\figref{Fig:Fig_p-pol} the dependence on $N$ of the error for preparing the most-subradiant state of a single array. 
The error scaling is qualitatively the same as for the case of atoms polarized along the array [Cfr.~\figref{Fig:Fig2}.b]. 
We observe an oscillating behaviour of the error as function of $N$ for $a< \lambda_e/4$, which can be attributed to a non-monotonic dependence on $a/\lambda_e$ of the dark mode decay rate $\Gamma_{q_a}$ when $a\lesssim \lambda_e/4$~\cite{Kornovan2019}. Such behaviour is characteristic of the case of transversly polarised arrays. It is believed to originate in the long range behaviour of the coupling and in the existence of modes with same energy but different wave-vector in the array's dispersion relation~\cite{Kornovan2019,Masson2020}. 

% ==================================================
% ==================================================
\section{Details on the Numerical Simulations}\label{app:Numerics}
% ==================================================
% ==================================================

\begin{figure*}
	\includegraphics[width=2\columnwidth]{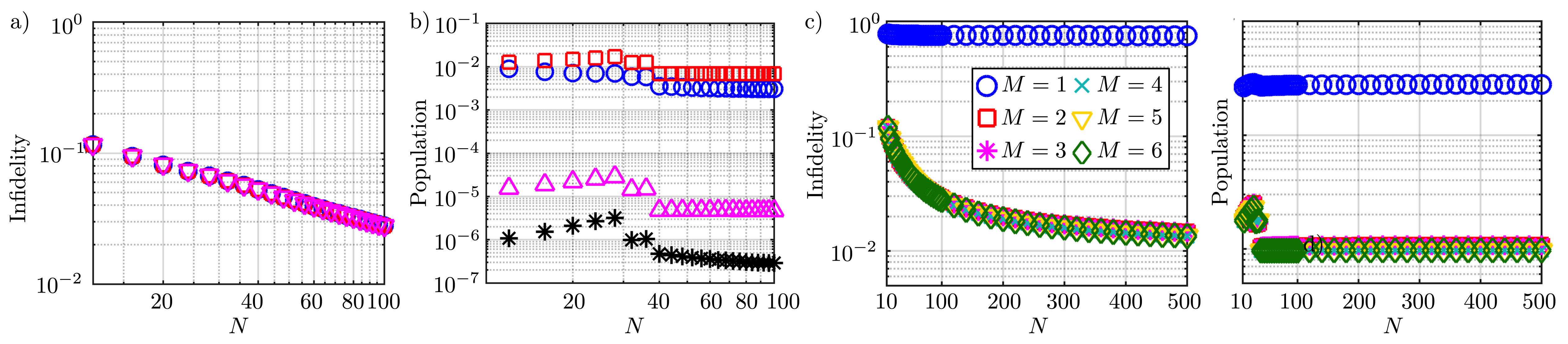}
	\caption{{\bf Numerical details for subradiant state excitation in a driven single array at the optimal Rabi frequency in \figref{Fig:Fig2}.d.} a) Comparison of the Infidelity obtained with full diagonalization of $\tilde{H}$ (blue circles), Krylov-Schur diagonalization truncating after the second (red squares) and third (magenta triangle) exictation manifold. b) Total population in the ground state (red squared), two-excitations manifold (blue circles), three-excitations manifold (magenta triangles), and states other than the target in the one-excitation manifold (black stars). c) Left (right) panel: simulation of the driven array infidelity (total population in other manifold) in function of $N$ for different values of $M$. In all panels we set $a=\lambda_e/4$.}\label{Fig:Numerics}
\end{figure*}

In the following section, we explain how we obtained the numerical results presented in Fig.~(\ref{Fig:Fig2}-\ref{fig:Fig3}) in the main text.
We simulated the dynamics generated by the Sch\"odinger equation ($\hbar=1$)
\be\label{eq:Sch_EQ}
	\pa{t}\ket{\psi} = -\im\Hop\ket{\psi}
\ee
where $\Hop$ is the non hermitian Hamiltonian for the evolution of the driven single, $\Hop=\Hop_\text{1array}+\Vop_\text{1array}$ array or parallel arrays, $\Hop=\Hop_0+\Vop_0$.
The state of the system at time $t>0$ reads
\be\label{eq:psi_t}
	\ket{\psi(t)}= e^{-\im \Hop t}\ket{\psi(0)},
\ee
and can be obtained from a given initial state $\ket{\psi(0)}$ by numerical diagonalisation of $\Hop$.
For array's length larger than just a dozen of atoms it is impossible to diagonalise exactly the full $2^N\times2^N$ Hamiltonian $\Hop$ of the system, and we thus resort on the truncation of the Hilbert space of the system.
Within the blockade regime, we expect higher excitation manifold to be only marginally populated. We checked numerically in \figref{Fig:Numerics}.a-b that for a single array driven at the Rabi frequency which maximises the target state population, $\W_0^\text{opt}$ 
(see \figref{Fig:Fig2}.d), non-negligible population transfer occurs only up to the second excitation manifold.
Hence, the results in Fig.~(\ref{Fig:Fig2}-\ref{fig:Fig3}) are obtained by numerically diagonalising the Hamiltonian $\tilde{H}$ describing the dynamics of the system on the Hilbert subspace containing at most two excitations. The dimension of the truncated Hilbert space scales as $N^2$. According to \figref{Fig:Numerics}.b, this truncation gives an error which decreases with $N$ and is negligible compared to the values of fidelities obtained in the range of parameters we simulated.
We numerically diagonalized $\tilde{H}$ with two different methods. The first method is the exact diagonalisation of the truncated Hamiltonin which yields the full spectrum of $\tilde{H}$. The spectrum can be used to calculate the transition amplitude between the initial state $\ket{\psi(0)}$ and a target state $\ket{\psi_\text{targ}}$ at any time $t$ as
\be\label{eq:Tr_Ampl}
	\braket{\psi_\text{targ}}{\psi(t)} = \sum_{\nu=1}^{1+N+\binom{N}{2}} \braket{\psi_\text{targ}}{\nu}\braket{\bar{\nu}}{\psi(0)}e^{-\im E_\nu t}.
\ee
Here we labelled $\ket{\nu}$ ($\bra{\bar{\nu}}$) the right (left) eigenvector of $\tilde{H}$ and $E_\nu$ their corresponding complex eigenvalue.
The size occupied by $\tilde{H}$ in the memory ultimately limits the array size that can be simulated with this method. Furthermore, as the interaction between the emitters in the array are long range, see Eq.~(\ref{eq:G_ij}-\ref{eq:G0}), the dimension of the Hamiltonian matrix $\tilde{H}$ cannot be reduced efficiently using sparse matrix without truncating the dipole couplings after a fixed number of sites.
To simulate $N>200$ array sizes as shown in \figref{Fig:Fig2}.c-d in the main text,
we diagonalize $\tilde{H}$ on a Krylov subspace of dimension $M\ll N^2$. The transition amplitude in \eqnref{eq:Tr_Ampl} can then be calculated as 
\be\label{eq:Tr_Ampl_Krylov}
	\braket{\psi_\text{targ}}{\psi(t)}\simeq \sum_{\nu=1}^{M} \braket{\psi_\text{targ}}{r_\nu}\braket{\bar{r}_\nu}{\psi(0)}\,e^{-\im R_\nu t},
\ee
where  $\ket{r_\nu}$ ($\ket{\bar{r}_\nu}$) and $R_\nu$ are the Ritz right (left) eigenvectors and eigenvalues obtained from a Krylov-Schur diagonalization procedure using \Matlab~\cite{ChenCh7}.
Thanks to the mode matching of the driving  and the finite decay rate of the eigenstates of $\tilde{H}$, the state evolved from $\ket{\psi(0)}=\ket{0}$ at $t\sim 1/\W_0^\text{opt}$ overlaps with the first handful of states ordered for increasing decay rate (see~\figref{Fig:Numerics}.c).
We stress that, as the Ritz spectrum approximate the exact spectrum up to an estimated tolerance at least $\lesssim 10^{-10}$, the main error in calculating the transition amplitudes is due to the truncation in \eqnref{eq:Tr_Ampl_Krylov}. The result displayed  in Fig.~(\ref{Fig:Fig2}-\ref{fig:Fig3}) have been calculated with $M=30$.

The simulation for the fast motion regime Fig.~(\ref{Fig:Fig4}-\ref{Fig:Fig5}), are obtained using the same truncation as described above but with the Hamiltonian \eqnref{eq:Ham_fam} instead. Due to a reduced array length used for those simulations, there was no need to use the Krylov-Schur method.


\begin{thebibliography}{69}%
\makeatletter
\providecommand \@ifxundefined [1]{%
 \@ifx{#1\undefined}
}%
\providecommand \@ifnum [1]{%
 \ifnum #1\expandafter \@firstoftwo
 \else \expandafter \@secondoftwo
 \fi
}%
\providecommand \@ifx [1]{%
 \ifx #1\expandafter \@firstoftwo
 \else \expandafter \@secondoftwo
 \fi
}%
\providecommand \natexlab [1]{#1}%
\providecommand \enquote  [1]{``#1''}%
\providecommand \bibnamefont  [1]{#1}%
\providecommand \bibfnamefont [1]{#1}%
\providecommand \citenamefont [1]{#1}%
\providecommand \href@noop [0]{\@secondoftwo}%
\providecommand \href [0]{\begingroup \@sanitize@url \@href}%
\providecommand \@href[1]{\@@startlink{#1}\@@href}%
\providecommand \@@href[1]{\endgroup#1\@@endlink}%
\providecommand \@sanitize@url [0]{\catcode `\\12\catcode `\$12\catcode
  `\&12\catcode `\#12\catcode `\^12\catcode `\_12\catcode `\%12\relax}%
\providecommand \@@startlink[1]{}%
\providecommand \@@endlink[0]{}%
\providecommand \url  [0]{\begingroup\@sanitize@url \@url }%
\providecommand \@url [1]{\endgroup\@href {#1}{\urlprefix }}%
\providecommand \urlprefix  [0]{URL }%
\providecommand \Eprint [0]{\href }%
\providecommand \doibase [0]{https://doi.org/}%
\providecommand \selectlanguage [0]{\@gobble}%
\providecommand \bibinfo  [0]{\@secondoftwo}%
\providecommand \bibfield  [0]{\@secondoftwo}%
\providecommand \translation [1]{[#1]}%
\providecommand \BibitemOpen [0]{}%
\providecommand \bibitemStop [0]{}%
\providecommand \bibitemNoStop [0]{.\EOS\space}%
\providecommand \EOS [0]{\spacefactor3000\relax}%
\providecommand \BibitemShut  [1]{\csname bibitem#1\endcsname}%
\let\auto@bib@innerbib\@empty
%</preamble>
\bibitem [{\citenamefont {Chang}\ \emph {et~al.}(2018)\citenamefont {Chang},
  \citenamefont {Douglas}, \citenamefont {Gonz\'alez-Tudela}, \citenamefont
  {Hung},\ and\ \citenamefont {Kimble}}]{Chang2018}%
  \BibitemOpen
  \bibfield  {author} {\bibinfo {author} {\bibfnamefont {D.~E.}\ \bibnamefont
  {Chang}}, \bibinfo {author} {\bibfnamefont {J.~S.}\ \bibnamefont {Douglas}},
  \bibinfo {author} {\bibfnamefont {A.}~\bibnamefont {Gonz\'alez-Tudela}},
  \bibinfo {author} {\bibfnamefont {C.-L.}\ \bibnamefont {Hung}},\ and\
  \bibinfo {author} {\bibfnamefont {H.~J.}\ \bibnamefont {Kimble}},\ }\href
  {https://doi.org/10.1103/RevModPhys.90.031002} {\bibfield  {journal}
  {\bibinfo  {journal} {Rev. Mod. Phys.}\ }\textbf {\bibinfo {volume} {90}},\
  \bibinfo {pages} {031002} (\bibinfo {year} {2018})}\BibitemShut {NoStop}%
\bibitem [{\citenamefont {Dicke}(1954)}]{Dicke1954}%
  \BibitemOpen
  \bibfield  {author} {\bibinfo {author} {\bibfnamefont {R.~H.}\ \bibnamefont
  {Dicke}},\ }\href {https://doi.org/10.1103/PhysRev.93.99} {\bibfield
  {journal} {\bibinfo  {journal} {Phys. Rev.}\ }\textbf {\bibinfo {volume}
  {93}},\ \bibinfo {pages} {99} (\bibinfo {year} {1954})}\BibitemShut {NoStop}%
\bibitem [{\citenamefont {Lehmberg}(1970{\natexlab{a}})}]{Lehmberg1970a}%
  \BibitemOpen
  \bibfield  {author} {\bibinfo {author} {\bibfnamefont {R.~H.}\ \bibnamefont
  {Lehmberg}},\ }\href {https://doi.org/10.1103/PhysRevA.2.883} {\bibfield
  {journal} {\bibinfo  {journal} {Phys. Rev. A}\ }\textbf {\bibinfo {volume}
  {2}},\ \bibinfo {pages} {883} (\bibinfo {year}
  {1970}{\natexlab{a}})}\BibitemShut {NoStop}%
\bibitem [{\citenamefont {Lehmberg}(1970{\natexlab{b}})}]{Lehmberg1970b}%
  \BibitemOpen
  \bibfield  {author} {\bibinfo {author} {\bibfnamefont {R.~H.}\ \bibnamefont
  {Lehmberg}},\ }\href {https://doi.org/10.1103/PhysRevA.2.889} {\bibfield
  {journal} {\bibinfo  {journal} {Phys. Rev. A}\ }\textbf {\bibinfo {volume}
  {2}},\ \bibinfo {pages} {889} (\bibinfo {year}
  {1970}{\natexlab{b}})}\BibitemShut {NoStop}%
\bibitem [{\citenamefont {Antezza}\ and\ \citenamefont
  {Castin}(2009{\natexlab{a}})}]{AntezzaPRL2009}%
  \BibitemOpen
  \bibfield  {author} {\bibinfo {author} {\bibfnamefont {M.}~\bibnamefont
  {Antezza}}\ and\ \bibinfo {author} {\bibfnamefont {Y.}~\bibnamefont
  {Castin}},\ }\href {https://doi.org/10.1103/PhysRevLett.103.123903}
  {\bibfield  {journal} {\bibinfo  {journal} {Phys. Rev. Lett.}\ }\textbf
  {\bibinfo {volume} {103}},\ \bibinfo {pages} {123903} (\bibinfo {year}
  {2009}{\natexlab{a}})}\BibitemShut {NoStop}%
\bibitem [{\citenamefont {Antezza}\ and\ \citenamefont
  {Castin}(2009{\natexlab{b}})}]{AntezzaPRA2009}%
  \BibitemOpen
  \bibfield  {author} {\bibinfo {author} {\bibfnamefont {M.}~\bibnamefont
  {Antezza}}\ and\ \bibinfo {author} {\bibfnamefont {Y.}~\bibnamefont
  {Castin}},\ }\href {https://doi.org/10.1103/PhysRevA.80.013816} {\bibfield
  {journal} {\bibinfo  {journal} {Phys. Rev. A}\ }\textbf {\bibinfo {volume}
  {80}},\ \bibinfo {pages} {013816} (\bibinfo {year}
  {2009}{\natexlab{b}})}\BibitemShut {NoStop}%
\bibitem [{\citenamefont {Bettles}\ \emph
  {et~al.}(2016{\natexlab{a}})\citenamefont {Bettles}, \citenamefont
  {Gardiner},\ and\ \citenamefont {Adams}}]{BettlesPRL2016}%
  \BibitemOpen
  \bibfield  {author} {\bibinfo {author} {\bibfnamefont {R.~J.}\ \bibnamefont
  {Bettles}}, \bibinfo {author} {\bibfnamefont {S.~A.}\ \bibnamefont
  {Gardiner}},\ and\ \bibinfo {author} {\bibfnamefont {C.~S.}\ \bibnamefont
  {Adams}},\ }\href {https://doi.org/10.1103/PhysRevLett.116.103602} {\bibfield
   {journal} {\bibinfo  {journal} {Phys. Rev. Lett.}\ }\textbf {\bibinfo
  {volume} {116}},\ \bibinfo {pages} {103602} (\bibinfo {year}
  {2016}{\natexlab{a}})}\BibitemShut {NoStop}%
\bibitem [{\citenamefont {Bettles}\ \emph
  {et~al.}(2016{\natexlab{b}})\citenamefont {Bettles}, \citenamefont
  {Gardiner},\ and\ \citenamefont {Adams}}]{BettlesPRA2016}%
  \BibitemOpen
  \bibfield  {author} {\bibinfo {author} {\bibfnamefont {R.~J.}\ \bibnamefont
  {Bettles}}, \bibinfo {author} {\bibfnamefont {S.~A.}\ \bibnamefont
  {Gardiner}},\ and\ \bibinfo {author} {\bibfnamefont {C.~S.}\ \bibnamefont
  {Adams}},\ }\href {https://doi.org/10.1103/PhysRevA.94.043844} {\bibfield
  {journal} {\bibinfo  {journal} {Phys. Rev. A}\ }\textbf {\bibinfo {volume}
  {94}},\ \bibinfo {pages} {043844} (\bibinfo {year}
  {2016}{\natexlab{b}})}\BibitemShut {NoStop}%
\bibitem [{\citenamefont {Shahmoon}\ \emph {et~al.}(2017)\citenamefont
  {Shahmoon}, \citenamefont {Wild}, \citenamefont {Lukin},\ and\ \citenamefont
  {Yelin}}]{Shahmoon2017}%
  \BibitemOpen
  \bibfield  {author} {\bibinfo {author} {\bibfnamefont {E.}~\bibnamefont
  {Shahmoon}}, \bibinfo {author} {\bibfnamefont {D.~S.}\ \bibnamefont {Wild}},
  \bibinfo {author} {\bibfnamefont {M.~D.}\ \bibnamefont {Lukin}},\ and\
  \bibinfo {author} {\bibfnamefont {S.~F.}\ \bibnamefont {Yelin}},\ }\href
  {https://doi.org/10.1103/PhysRevLett.118.113601} {\bibfield  {journal}
  {\bibinfo  {journal} {Phys. Rev. Lett.}\ }\textbf {\bibinfo {volume} {118}},\
  \bibinfo {pages} {113601} (\bibinfo {year} {2017})}\BibitemShut {NoStop}%
\bibitem [{\citenamefont {Rui}\ \emph {et~al.}(2020)\citenamefont {Rui},
  \citenamefont {Wei}, \citenamefont {Rubio-Abadal}, \citenamefont {Hollerith},
  \citenamefont {Zeiher}, \citenamefont {Stamper-Kurn}, \citenamefont {Gross},\
  and\ \citenamefont {Bloch}}]{Rui2020}%
  \BibitemOpen
  \bibfield  {author} {\bibinfo {author} {\bibfnamefont {J.}~\bibnamefont
  {Rui}}, \bibinfo {author} {\bibfnamefont {D.}~\bibnamefont {Wei}}, \bibinfo
  {author} {\bibfnamefont {A.}~\bibnamefont {Rubio-Abadal}}, \bibinfo {author}
  {\bibfnamefont {S.}~\bibnamefont {Hollerith}}, \bibinfo {author}
  {\bibfnamefont {J.}~\bibnamefont {Zeiher}}, \bibinfo {author} {\bibfnamefont
  {D.~M.}\ \bibnamefont {Stamper-Kurn}}, \bibinfo {author} {\bibfnamefont
  {C.}~\bibnamefont {Gross}},\ and\ \bibinfo {author} {\bibfnamefont
  {I.}~\bibnamefont {Bloch}},\ }\href
  {https://doi.org/10.1038/s41586-020-2463-x} {\bibfield  {journal} {\bibinfo
  {journal} {Nature}\ }\textbf {\bibinfo {volume} {583}},\ \bibinfo {pages}
  {369} (\bibinfo {year} {2020})}\BibitemShut {NoStop}%
\bibitem [{\citenamefont {Bettles}\ \emph {et~al.}(2017)\citenamefont
  {Bettles}, \citenamefont {Min\'a\ifmmode~\check{r}\else \v{r}\fi{}},
  \citenamefont {Adams}, \citenamefont {Lesanovsky},\ and\ \citenamefont
  {Olmos}}]{Bettles2017}%
  \BibitemOpen
  \bibfield  {author} {\bibinfo {author} {\bibfnamefont {R.~J.}\ \bibnamefont
  {Bettles}}, \bibinfo {author} {\bibfnamefont {J.~c.~v.}\ \bibnamefont
  {Min\'a\ifmmode~\check{r}\else \v{r}\fi{}}}, \bibinfo {author} {\bibfnamefont
  {C.~S.}\ \bibnamefont {Adams}}, \bibinfo {author} {\bibfnamefont
  {I.}~\bibnamefont {Lesanovsky}},\ and\ \bibinfo {author} {\bibfnamefont
  {B.}~\bibnamefont {Olmos}},\ }\href
  {https://doi.org/10.1103/PhysRevA.96.041603} {\bibfield  {journal} {\bibinfo
  {journal} {Phys. Rev. A}\ }\textbf {\bibinfo {volume} {96}},\ \bibinfo
  {pages} {041603} (\bibinfo {year} {2017})}\BibitemShut {NoStop}%
\bibitem [{\citenamefont {Perczel}\ \emph {et~al.}(2017)\citenamefont
  {Perczel}, \citenamefont {Borregaard}, \citenamefont {Chang}, \citenamefont
  {Pichler}, \citenamefont {Yelin}, \citenamefont {Zoller},\ and\ \citenamefont
  {Lukin}}]{Perczel2017}%
  \BibitemOpen
  \bibfield  {author} {\bibinfo {author} {\bibfnamefont {J.}~\bibnamefont
  {Perczel}}, \bibinfo {author} {\bibfnamefont {J.}~\bibnamefont {Borregaard}},
  \bibinfo {author} {\bibfnamefont {D.~E.}\ \bibnamefont {Chang}}, \bibinfo
  {author} {\bibfnamefont {H.}~\bibnamefont {Pichler}}, \bibinfo {author}
  {\bibfnamefont {S.~F.}\ \bibnamefont {Yelin}}, \bibinfo {author}
  {\bibfnamefont {P.}~\bibnamefont {Zoller}},\ and\ \bibinfo {author}
  {\bibfnamefont {M.~D.}\ \bibnamefont {Lukin}},\ }\href
  {https://doi.org/10.1103/PhysRevLett.119.023603} {\bibfield  {journal}
  {\bibinfo  {journal} {Phys. Rev. Lett.}\ }\textbf {\bibinfo {volume} {119}},\
  \bibinfo {pages} {023603} (\bibinfo {year} {2017})}\BibitemShut {NoStop}%
\bibitem [{\citenamefont {Asenjo-Garcia}\ \emph {et~al.}(2017)\citenamefont
  {Asenjo-Garcia}, \citenamefont {Moreno-Cardoner}, \citenamefont {Albrecht},
  \citenamefont {Kimble},\ and\ \citenamefont {Chang}}]{AsenjoGarcia2017PRX}%
  \BibitemOpen
  \bibfield  {author} {\bibinfo {author} {\bibfnamefont {A.}~\bibnamefont
  {Asenjo-Garcia}}, \bibinfo {author} {\bibfnamefont {M.}~\bibnamefont
  {Moreno-Cardoner}}, \bibinfo {author} {\bibfnamefont {A.}~\bibnamefont
  {Albrecht}}, \bibinfo {author} {\bibfnamefont {H.~J.}\ \bibnamefont
  {Kimble}},\ and\ \bibinfo {author} {\bibfnamefont {D.~E.}\ \bibnamefont
  {Chang}},\ }\href {https://doi.org/10.1103/PhysRevX.7.031024} {\bibfield
  {journal} {\bibinfo  {journal} {Phys. Rev. X}\ }\textbf {\bibinfo {volume}
  {7}},\ \bibinfo {pages} {031024} (\bibinfo {year} {2017})}\BibitemShut
  {NoStop}%
\bibitem [{\citenamefont {Porras}\ and\ \citenamefont
  {Cirac}(2008)}]{Porras2008}%
  \BibitemOpen
  \bibfield  {author} {\bibinfo {author} {\bibfnamefont {D.}~\bibnamefont
  {Porras}}\ and\ \bibinfo {author} {\bibfnamefont {J.~I.}\ \bibnamefont
  {Cirac}},\ }\href {https://doi.org/10.1103/PhysRevA.78.053816} {\bibfield
  {journal} {\bibinfo  {journal} {Phys. Rev. A}\ }\textbf {\bibinfo {volume}
  {78}},\ \bibinfo {pages} {053816} (\bibinfo {year} {2008})}\BibitemShut
  {NoStop}%
\bibitem [{\citenamefont {Lukin}\ \emph {et~al.}(2001)\citenamefont {Lukin},
  \citenamefont {Fleischhauer}, \citenamefont {Cote}, \citenamefont {Duan},
  \citenamefont {Jaksch}, \citenamefont {Cirac},\ and\ \citenamefont
  {Zoller}}]{Lukin2001}%
  \BibitemOpen
  \bibfield  {author} {\bibinfo {author} {\bibfnamefont {M.~D.}\ \bibnamefont
  {Lukin}}, \bibinfo {author} {\bibfnamefont {M.}~\bibnamefont {Fleischhauer}},
  \bibinfo {author} {\bibfnamefont {R.}~\bibnamefont {Cote}}, \bibinfo {author}
  {\bibfnamefont {L.~M.}\ \bibnamefont {Duan}}, \bibinfo {author}
  {\bibfnamefont {D.}~\bibnamefont {Jaksch}}, \bibinfo {author} {\bibfnamefont
  {J.~I.}\ \bibnamefont {Cirac}},\ and\ \bibinfo {author} {\bibfnamefont
  {P.}~\bibnamefont {Zoller}},\ }\href
  {https://doi.org/10.1103/PhysRevLett.87.037901} {\bibfield  {journal}
  {\bibinfo  {journal} {Phys. Rev. Lett.}\ }\textbf {\bibinfo {volume} {87}},\
  \bibinfo {pages} {037901} (\bibinfo {year} {2001})}\BibitemShut {NoStop}%
\bibitem [{\citenamefont {Bekenstein}\ \emph {et~al.}(2020)\citenamefont
  {Bekenstein}, \citenamefont {Pikovski}, \citenamefont {Pichler},
  \citenamefont {Shahmoon}, \citenamefont {Yelin},\ and\ \citenamefont
  {Lukin}}]{Bekenstein2020}%
  \BibitemOpen
  \bibfield  {author} {\bibinfo {author} {\bibfnamefont {R.}~\bibnamefont
  {Bekenstein}}, \bibinfo {author} {\bibfnamefont {I.}~\bibnamefont
  {Pikovski}}, \bibinfo {author} {\bibfnamefont {H.}~\bibnamefont {Pichler}},
  \bibinfo {author} {\bibfnamefont {E.}~\bibnamefont {Shahmoon}}, \bibinfo
  {author} {\bibfnamefont {S.~F.}\ \bibnamefont {Yelin}},\ and\ \bibinfo
  {author} {\bibfnamefont {M.~D.}\ \bibnamefont {Lukin}},\ }\href
  {https://doi.org/10.1038/s41567-020-0845-5} {\bibfield  {journal} {\bibinfo
  {journal} {Nature Physics}\ }\textbf {\bibinfo {volume} {16}},\ \bibinfo
  {pages} {676} (\bibinfo {year} {2020})}\BibitemShut {NoStop}%
\bibitem [{\citenamefont {Moreno-Cardoner}\ \emph {et~al.}(2021)\citenamefont
  {Moreno-Cardoner}, \citenamefont {Goncalves},\ and\ \citenamefont
  {Chang}}]{MorenoCardoner2021}%
  \BibitemOpen
  \bibfield  {author} {\bibinfo {author} {\bibfnamefont {M.}~\bibnamefont
  {Moreno-Cardoner}}, \bibinfo {author} {\bibfnamefont {D.}~\bibnamefont
  {Goncalves}},\ and\ \bibinfo {author} {\bibfnamefont {D.~E.}\ \bibnamefont
  {Chang}},\ }\href@noop {} {\bibinfo {title} {Quantum nonlinear optics based
  on two-dimensional rydberg atom arrays}} (\bibinfo {year} {2021}),\ \Eprint
  {https://arxiv.org/abs/2101.01936} {arXiv:2101.01936 [quant-ph]} \BibitemShut
  {NoStop}%
\bibitem [{\citenamefont {Masson}\ and\ \citenamefont
  {Asenjo-Garcia}(2020)}]{Masson2020}%
  \BibitemOpen
  \bibfield  {author} {\bibinfo {author} {\bibfnamefont {S.~J.}\ \bibnamefont
  {Masson}}\ and\ \bibinfo {author} {\bibfnamefont {A.}~\bibnamefont
  {Asenjo-Garcia}},\ }\href {https://doi.org/10.1103/PhysRevResearch.2.043213}
  {\bibfield  {journal} {\bibinfo  {journal} {Phys. Rev. Research}\ }\textbf
  {\bibinfo {volume} {2}},\ \bibinfo {pages} {043213} (\bibinfo {year}
  {2020})}\BibitemShut {NoStop}%
\bibitem [{\citenamefont {Williamson}\ \emph {et~al.}(2020)\citenamefont
  {Williamson}, \citenamefont {Borgh},\ and\ \citenamefont
  {Ruostekoski}}]{Williamson2020}%
  \BibitemOpen
  \bibfield  {author} {\bibinfo {author} {\bibfnamefont {L.~A.}\ \bibnamefont
  {Williamson}}, \bibinfo {author} {\bibfnamefont {M.~O.}\ \bibnamefont
  {Borgh}},\ and\ \bibinfo {author} {\bibfnamefont {J.}~\bibnamefont
  {Ruostekoski}},\ }\href {https://doi.org/10.1103/PhysRevLett.125.073602}
  {\bibfield  {journal} {\bibinfo  {journal} {Phys. Rev. Lett.}\ }\textbf
  {\bibinfo {volume} {125}},\ \bibinfo {pages} {073602} (\bibinfo {year}
  {2020})}\BibitemShut {NoStop}%
\bibitem [{\citenamefont {Cidrim}\ \emph {et~al.}(2020)\citenamefont {Cidrim},
  \citenamefont {do~Espirito~Santo}, \citenamefont {Schachenmayer},
  \citenamefont {Kaiser},\ and\ \citenamefont {Bachelard}}]{Cidrim2020}%
  \BibitemOpen
  \bibfield  {author} {\bibinfo {author} {\bibfnamefont {A.}~\bibnamefont
  {Cidrim}}, \bibinfo {author} {\bibfnamefont {T.~S.}\ \bibnamefont
  {do~Espirito~Santo}}, \bibinfo {author} {\bibfnamefont {J.}~\bibnamefont
  {Schachenmayer}}, \bibinfo {author} {\bibfnamefont {R.}~\bibnamefont
  {Kaiser}},\ and\ \bibinfo {author} {\bibfnamefont {R.}~\bibnamefont
  {Bachelard}},\ }\href {https://doi.org/10.1103/PhysRevLett.125.073601}
  {\bibfield  {journal} {\bibinfo  {journal} {Phys. Rev. Lett.}\ }\textbf
  {\bibinfo {volume} {125}},\ \bibinfo {pages} {073601} (\bibinfo {year}
  {2020})}\BibitemShut {NoStop}%
\bibitem [{\citenamefont {Bettles}\ \emph {et~al.}(2020)\citenamefont
  {Bettles}, \citenamefont {Lee}, \citenamefont {Gardiner},\ and\ \citenamefont
  {Ruostekoski}}]{Bettles2020}%
  \BibitemOpen
  \bibfield  {author} {\bibinfo {author} {\bibfnamefont {R.~J.}\ \bibnamefont
  {Bettles}}, \bibinfo {author} {\bibfnamefont {M.~D.}\ \bibnamefont {Lee}},
  \bibinfo {author} {\bibfnamefont {S.~A.}\ \bibnamefont {Gardiner}},\ and\
  \bibinfo {author} {\bibfnamefont {J.}~\bibnamefont {Ruostekoski}},\ }\href
  {https://doi.org/10.1038/s42005-020-00404-3} {\bibfield  {journal} {\bibinfo
  {journal} {Communications Physics}\ }\textbf {\bibinfo {volume} {3}},\
  \bibinfo {pages} {141} (\bibinfo {year} {2020})}\BibitemShut {NoStop}%
\bibitem [{\citenamefont {Facchi}\ and\ \citenamefont
  {Pascazio}(2002)}]{Facchi2002}%
  \BibitemOpen
  \bibfield  {author} {\bibinfo {author} {\bibfnamefont {P.}~\bibnamefont
  {Facchi}}\ and\ \bibinfo {author} {\bibfnamefont {S.}~\bibnamefont
  {Pascazio}},\ }\href {https://doi.org/10.1103/PhysRevLett.89.080401}
  {\bibfield  {journal} {\bibinfo  {journal} {Phys. Rev. Lett.}\ }\textbf
  {\bibinfo {volume} {89}},\ \bibinfo {pages} {080401} (\bibinfo {year}
  {2002})}\BibitemShut {NoStop}%
\bibitem [{\citenamefont {Guimond}\ \emph {et~al.}(2019)\citenamefont
  {Guimond}, \citenamefont {Grankin}, \citenamefont {Vasilyev}, \citenamefont
  {Vermersch},\ and\ \citenamefont {Zoller}}]{Guimond2019}%
  \BibitemOpen
  \bibfield  {author} {\bibinfo {author} {\bibfnamefont {P.-O.}\ \bibnamefont
  {Guimond}}, \bibinfo {author} {\bibfnamefont {A.}~\bibnamefont {Grankin}},
  \bibinfo {author} {\bibfnamefont {D.~V.}\ \bibnamefont {Vasilyev}}, \bibinfo
  {author} {\bibfnamefont {B.}~\bibnamefont {Vermersch}},\ and\ \bibinfo
  {author} {\bibfnamefont {P.}~\bibnamefont {Zoller}},\ }\href
  {https://doi.org/10.1103/PhysRevLett.122.093601} {\bibfield  {journal}
  {\bibinfo  {journal} {Phys. Rev. Lett.}\ }\textbf {\bibinfo {volume} {122}},\
  \bibinfo {pages} {093601} (\bibinfo {year} {2019})}\BibitemShut {NoStop}%
\bibitem [{Note1()}]{Note1}%
  \BibitemOpen
  \bibinfo {note} {This is however not necessary for the case of pinned
  atoms.}\BibitemShut {Stop}%
\bibitem [{\citenamefont {Kornovan}\ \emph {et~al.}(2016)\citenamefont
  {Kornovan}, \citenamefont {Sheremet},\ and\ \citenamefont
  {Petrov}}]{Kornovan2016}%
  \BibitemOpen
  \bibfield  {author} {\bibinfo {author} {\bibfnamefont {D.~F.}\ \bibnamefont
  {Kornovan}}, \bibinfo {author} {\bibfnamefont {A.~S.}\ \bibnamefont
  {Sheremet}},\ and\ \bibinfo {author} {\bibfnamefont {M.~I.}\ \bibnamefont
  {Petrov}},\ }\href {https://doi.org/10.1103/PhysRevB.94.245416} {\bibfield
  {journal} {\bibinfo  {journal} {Phys. Rev. B}\ }\textbf {\bibinfo {volume}
  {94}},\ \bibinfo {pages} {245416} (\bibinfo {year} {2016})}\BibitemShut
  {NoStop}%
\bibitem [{\citenamefont {Mewton}\ and\ \citenamefont
  {Ficek}(2007)}]{Mewton2007}%
  \BibitemOpen
  \bibfield  {author} {\bibinfo {author} {\bibfnamefont {C.~J.}\ \bibnamefont
  {Mewton}}\ and\ \bibinfo {author} {\bibfnamefont {Z.}~\bibnamefont {Ficek}},\
  }\href {https://doi.org/10.1088/0953-4075/40/9/s11} {\bibfield  {journal}
  {\bibinfo  {journal} {Journal of Physics B: Atomic, Molecular and Optical
  Physics}\ }\textbf {\bibinfo {volume} {40}},\ \bibinfo {pages} {S181}
  (\bibinfo {year} {2007})}\BibitemShut {NoStop}%
\bibitem [{\citenamefont {Zhang}\ and\ \citenamefont
  {M\o{}lmer}(2019)}]{Zhang2019}%
  \BibitemOpen
  \bibfield  {author} {\bibinfo {author} {\bibfnamefont {Y.-X.}\ \bibnamefont
  {Zhang}}\ and\ \bibinfo {author} {\bibfnamefont {K.}~\bibnamefont
  {M\o{}lmer}},\ }\href {https://doi.org/10.1103/PhysRevLett.122.203605}
  {\bibfield  {journal} {\bibinfo  {journal} {Phys. Rev. Lett.}\ }\textbf
  {\bibinfo {volume} {122}},\ \bibinfo {pages} {203605} (\bibinfo {year}
  {2019})}\BibitemShut {NoStop}%
\bibitem [{Note2()}]{Note2}%
  \BibitemOpen
  \bibinfo {note} {Note that this cannot be simply achieved with a single
  external laser as subwavelength excitation lie outside of the light-cone of
  free space electromagnetic modes.}\BibitemShut {Stop}%
\bibitem [{Note3()}]{Note3}%
  \BibitemOpen
  \bibinfo {note} {This gives an upper bound to the Fidelity as given by the
  full master equation}\BibitemShut {NoStop}%
\bibitem [{\citenamefont {Nielsen}(2002)}]{Nielsen2002}%
  \BibitemOpen
  \bibfield  {author} {\bibinfo {author} {\bibfnamefont {M.~A.}\ \bibnamefont
  {Nielsen}},\ }\href
  {https://doi.org/https://doi.org/10.1016/S0375-9601(02)01272-0} {\bibfield
  {journal} {\bibinfo  {journal} {Physics Letters A}\ }\textbf {\bibinfo
  {volume} {303}},\ \bibinfo {pages} {249} (\bibinfo {year}
  {2002})}\BibitemShut {NoStop}%
\bibitem [{\citenamefont {Pedersen}\ \emph {et~al.}(2007)\citenamefont
  {Pedersen}, \citenamefont {Møller},\ and\ \citenamefont
  {Mølmer}}]{Pedersen2007}%
  \BibitemOpen
  \bibfield  {author} {\bibinfo {author} {\bibfnamefont {L.~H.}\ \bibnamefont
  {Pedersen}}, \bibinfo {author} {\bibfnamefont {N.~M.}\ \bibnamefont
  {Møller}},\ and\ \bibinfo {author} {\bibfnamefont {K.}~\bibnamefont
  {Mølmer}},\ }\href
  {https://doi.org/https://doi.org/10.1016/j.physleta.2007.02.069} {\bibfield
  {journal} {\bibinfo  {journal} {Physics Letters A}\ }\textbf {\bibinfo
  {volume} {367}},\ \bibinfo {pages} {47 } (\bibinfo {year}
  {2007})}\BibitemShut {NoStop}%
\bibitem [{Note4()}]{Note4}%
  \BibitemOpen
  \bibinfo {note} {We found the coupling strength between $| {11} \rangle
  _\protect \text {L}$ and $| {S_2} \rangle $ to be $g_2\simeq 2g_{q_a}$ for
  all the values $a/\lambda _e$ in the range shown. This is particularly
  dramatic as at the $T_\protect \text {g}$ the population in $| {S_2} \rangle
  $ is maximised.}\BibitemShut {Stop}%
\bibitem [{Note5()}]{Note5}%
  \BibitemOpen
  \bibinfo {note} {We consider only integer values of $l/a$. Better results can
  be obtained if this assumption is relaxed.}\BibitemShut {Stop}%
\bibitem [{\citenamefont {Cooper}\ and\ \citenamefont
  {Stacey}(1974)}]{Cooper1974}%
  \BibitemOpen
  \bibfield  {author} {\bibinfo {author} {\bibfnamefont {J.}~\bibnamefont
  {Cooper}}\ and\ \bibinfo {author} {\bibfnamefont {D.~N.}\ \bibnamefont
  {Stacey}},\ }\href {https://doi.org/10.1088/0022-3700/7/16/013} {\bibfield
  {journal} {\bibinfo  {journal} {Journal of Physics B: Atomic and Molecular
  Physics}\ }\textbf {\bibinfo {volume} {7}},\ \bibinfo {pages} {2143}
  (\bibinfo {year} {1974})}\BibitemShut {NoStop}%
\bibitem [{\citenamefont {Power}(1974)}]{Power1974}%
  \BibitemOpen
  \bibfield  {author} {\bibinfo {author} {\bibfnamefont {E.~A.}\ \bibnamefont
  {Power}},\ }\href {https://doi.org/10.1088/0022-3700/7/16/014} {\bibfield
  {journal} {\bibinfo  {journal} {Journal of Physics B: Atomic and Molecular
  Physics}\ }\textbf {\bibinfo {volume} {7}},\ \bibinfo {pages} {2149}
  (\bibinfo {year} {1974})}\BibitemShut {NoStop}%
\bibitem [{\citenamefont {Berman}(1997)}]{Berman1997}%
  \BibitemOpen
  \bibfield  {author} {\bibinfo {author} {\bibfnamefont {P.~R.}\ \bibnamefont
  {Berman}},\ }\href {https://doi.org/10.1103/PhysRevA.55.4466} {\bibfield
  {journal} {\bibinfo  {journal} {Phys. Rev. A}\ }\textbf {\bibinfo {volume}
  {55}},\ \bibinfo {pages} {4466} (\bibinfo {year} {1997})}\BibitemShut
  {NoStop}%
\bibitem [{\citenamefont {Zhu}\ \emph {et~al.}(2016)\citenamefont {Zhu},
  \citenamefont {Cooper}, \citenamefont {Ye},\ and\ \citenamefont
  {Rey}}]{Zhu2016}%
  \BibitemOpen
  \bibfield  {author} {\bibinfo {author} {\bibfnamefont {B.}~\bibnamefont
  {Zhu}}, \bibinfo {author} {\bibfnamefont {J.}~\bibnamefont {Cooper}},
  \bibinfo {author} {\bibfnamefont {J.}~\bibnamefont {Ye}},\ and\ \bibinfo
  {author} {\bibfnamefont {A.~M.}\ \bibnamefont {Rey}},\ }\href
  {https://doi.org/10.1103/PhysRevA.94.023612} {\bibfield  {journal} {\bibinfo
  {journal} {Phys. Rev. A}\ }\textbf {\bibinfo {volume} {94}},\ \bibinfo
  {pages} {023612} (\bibinfo {year} {2016})}\BibitemShut {NoStop}%
\bibitem [{Note6()}]{Note6}%
  \BibitemOpen
  \bibinfo {note} {The perturbative approached followed here was presented
  in~\cite {Guimond2019}. In the present case the form of the dipole coupling
  rate is difference due to the Raman transition.}\BibitemShut {Stop}%
\bibitem [{\citenamefont {Ludlow}\ \emph {et~al.}(2015)\citenamefont {Ludlow},
  \citenamefont {Boyd}, \citenamefont {Ye}, \citenamefont {Peik},\ and\
  \citenamefont {Schmidt}}]{Ludlow2015}%
  \BibitemOpen
  \bibfield  {author} {\bibinfo {author} {\bibfnamefont {A.~D.}\ \bibnamefont
  {Ludlow}}, \bibinfo {author} {\bibfnamefont {M.~M.}\ \bibnamefont {Boyd}},
  \bibinfo {author} {\bibfnamefont {J.}~\bibnamefont {Ye}}, \bibinfo {author}
  {\bibfnamefont {E.}~\bibnamefont {Peik}},\ and\ \bibinfo {author}
  {\bibfnamefont {P.~O.}\ \bibnamefont {Schmidt}},\ }\href
  {https://doi.org/10.1103/RevModPhys.87.637} {\bibfield  {journal} {\bibinfo
  {journal} {Rev. Mod. Phys.}\ }\textbf {\bibinfo {volume} {87}},\ \bibinfo
  {pages} {637} (\bibinfo {year} {2015})}\BibitemShut {NoStop}%
\bibitem [{\citenamefont {Olmos}\ \emph {et~al.}(2013)\citenamefont {Olmos},
  \citenamefont {Yu}, \citenamefont {Singh}, \citenamefont {Schreck},
  \citenamefont {Bongs},\ and\ \citenamefont {Lesanovsky}}]{Olmos2013}%
  \BibitemOpen
  \bibfield  {author} {\bibinfo {author} {\bibfnamefont {B.}~\bibnamefont
  {Olmos}}, \bibinfo {author} {\bibfnamefont {D.}~\bibnamefont {Yu}}, \bibinfo
  {author} {\bibfnamefont {Y.}~\bibnamefont {Singh}}, \bibinfo {author}
  {\bibfnamefont {F.}~\bibnamefont {Schreck}}, \bibinfo {author} {\bibfnamefont
  {K.}~\bibnamefont {Bongs}},\ and\ \bibinfo {author} {\bibfnamefont
  {I.}~\bibnamefont {Lesanovsky}},\ }\href
  {https://doi.org/10.1103/PhysRevLett.110.143602} {\bibfield  {journal}
  {\bibinfo  {journal} {Phys. Rev. Lett.}\ }\textbf {\bibinfo {volume} {110}},\
  \bibinfo {pages} {143602} (\bibinfo {year} {2013})}\BibitemShut {NoStop}%
\bibitem [{\citenamefont {Zhou}\ \emph {et~al.}(2010)\citenamefont {Zhou},
  \citenamefont {Xu}, \citenamefont {Chen},\ and\ \citenamefont
  {Chen}}]{Zhou2010}%
  \BibitemOpen
  \bibfield  {author} {\bibinfo {author} {\bibfnamefont {X.}~\bibnamefont
  {Zhou}}, \bibinfo {author} {\bibfnamefont {X.}~\bibnamefont {Xu}}, \bibinfo
  {author} {\bibfnamefont {X.}~\bibnamefont {Chen}},\ and\ \bibinfo {author}
  {\bibfnamefont {J.}~\bibnamefont {Chen}},\ }\href
  {https://doi.org/10.1103/PhysRevA.81.012115} {\bibfield  {journal} {\bibinfo
  {journal} {Phys. Rev. A}\ }\textbf {\bibinfo {volume} {81}},\ \bibinfo
  {pages} {012115} (\bibinfo {year} {2010})}\BibitemShut {NoStop}%
\bibitem [{\citenamefont {Betzig}\ \emph {et~al.}(2006)\citenamefont {Betzig},
  \citenamefont {Patterson}, \citenamefont {Sougrat}, \citenamefont
  {Lindwasser}, \citenamefont {Olenych}, \citenamefont {Bonifacino},
  \citenamefont {Davidson}, \citenamefont {Lippincott-Schwartz},\ and\
  \citenamefont {Hess}}]{Betzig2006}%
  \BibitemOpen
  \bibfield  {author} {\bibinfo {author} {\bibfnamefont {E.}~\bibnamefont
  {Betzig}}, \bibinfo {author} {\bibfnamefont {G.~H.}\ \bibnamefont
  {Patterson}}, \bibinfo {author} {\bibfnamefont {R.}~\bibnamefont {Sougrat}},
  \bibinfo {author} {\bibfnamefont {O.~W.}\ \bibnamefont {Lindwasser}},
  \bibinfo {author} {\bibfnamefont {S.}~\bibnamefont {Olenych}}, \bibinfo
  {author} {\bibfnamefont {J.~S.}\ \bibnamefont {Bonifacino}}, \bibinfo
  {author} {\bibfnamefont {M.~W.}\ \bibnamefont {Davidson}}, \bibinfo {author}
  {\bibfnamefont {J.}~\bibnamefont {Lippincott-Schwartz}},\ and\ \bibinfo
  {author} {\bibfnamefont {H.~F.}\ \bibnamefont {Hess}},\ }\href
  {https://doi.org/10.1126/science.1127344} {\bibfield  {journal} {\bibinfo
  {journal} {Science}\ }\textbf {\bibinfo {volume} {313}},\ \bibinfo {pages}
  {1642} (\bibinfo {year} {2006})}\BibitemShut {NoStop}%
\bibitem [{\citenamefont {Hell}(2007)}]{Hell2007}%
  \BibitemOpen
  \bibfield  {author} {\bibinfo {author} {\bibfnamefont {S.~W.}\ \bibnamefont
  {Hell}},\ }\href {https://doi.org/10.1126/science.1137395} {\bibfield
  {journal} {\bibinfo  {journal} {Science}\ }\textbf {\bibinfo {volume}
  {316}},\ \bibinfo {pages} {1153} (\bibinfo {year} {2007})}\BibitemShut
  {NoStop}%
\bibitem [{\citenamefont {Plankensteiner}\ \emph {et~al.}(2015)\citenamefont
  {Plankensteiner}, \citenamefont {Ostermann}, \citenamefont {Ritsch},\ and\
  \citenamefont {Genes}}]{Plankensteiner2015}%
  \BibitemOpen
  \bibfield  {author} {\bibinfo {author} {\bibfnamefont {D.}~\bibnamefont
  {Plankensteiner}}, \bibinfo {author} {\bibfnamefont {L.}~\bibnamefont
  {Ostermann}}, \bibinfo {author} {\bibfnamefont {H.}~\bibnamefont {Ritsch}},\
  and\ \bibinfo {author} {\bibfnamefont {C.}~\bibnamefont {Genes}},\ }\href
  {https://doi.org/10.1038/srep16231} {\bibfield  {journal} {\bibinfo
  {journal} {Scientific Reports}\ }\textbf {\bibinfo {volume} {5}},\ \bibinfo
  {pages} {16231} (\bibinfo {year} {2015})}\BibitemShut {NoStop}%
\bibitem [{\citenamefont {He}\ \emph {et~al.}(2020{\natexlab{a}})\citenamefont
  {He}, \citenamefont {Ji}, \citenamefont {Wang}, \citenamefont {Qiu},
  \citenamefont {Zhao}, \citenamefont {Ma}, \citenamefont {Huang},
  \citenamefont {Wu},\ and\ \citenamefont {Chang}}]{He2020a}%
  \BibitemOpen
  \bibfield  {author} {\bibinfo {author} {\bibfnamefont {Y.}~\bibnamefont
  {He}}, \bibinfo {author} {\bibfnamefont {L.}~\bibnamefont {Ji}}, \bibinfo
  {author} {\bibfnamefont {Y.}~\bibnamefont {Wang}}, \bibinfo {author}
  {\bibfnamefont {L.}~\bibnamefont {Qiu}}, \bibinfo {author} {\bibfnamefont
  {J.}~\bibnamefont {Zhao}}, \bibinfo {author} {\bibfnamefont {Y.}~\bibnamefont
  {Ma}}, \bibinfo {author} {\bibfnamefont {X.}~\bibnamefont {Huang}}, \bibinfo
  {author} {\bibfnamefont {S.}~\bibnamefont {Wu}},\ and\ \bibinfo {author}
  {\bibfnamefont {D.~E.}\ \bibnamefont {Chang}},\ }\href
  {https://doi.org/10.1103/PhysRevLett.125.213602} {\bibfield  {journal}
  {\bibinfo  {journal} {Phys. Rev. Lett.}\ }\textbf {\bibinfo {volume} {125}},\
  \bibinfo {pages} {213602} (\bibinfo {year} {2020}{\natexlab{a}})}\BibitemShut
  {NoStop}%
\bibitem [{\citenamefont {He}\ \emph {et~al.}(2020{\natexlab{b}})\citenamefont
  {He}, \citenamefont {Ji}, \citenamefont {Wang}, \citenamefont {Qiu},
  \citenamefont {Zhao}, \citenamefont {Ma}, \citenamefont {Huang},
  \citenamefont {Wu},\ and\ \citenamefont {Chang}}]{He2020b}%
  \BibitemOpen
  \bibfield  {author} {\bibinfo {author} {\bibfnamefont {Y.}~\bibnamefont
  {He}}, \bibinfo {author} {\bibfnamefont {L.}~\bibnamefont {Ji}}, \bibinfo
  {author} {\bibfnamefont {Y.}~\bibnamefont {Wang}}, \bibinfo {author}
  {\bibfnamefont {L.}~\bibnamefont {Qiu}}, \bibinfo {author} {\bibfnamefont
  {J.}~\bibnamefont {Zhao}}, \bibinfo {author} {\bibfnamefont {Y.}~\bibnamefont
  {Ma}}, \bibinfo {author} {\bibfnamefont {X.}~\bibnamefont {Huang}}, \bibinfo
  {author} {\bibfnamefont {S.}~\bibnamefont {Wu}},\ and\ \bibinfo {author}
  {\bibfnamefont {D.~E.}\ \bibnamefont {Chang}},\ }\href
  {https://doi.org/10.1103/PhysRevResearch.2.043418} {\bibfield  {journal}
  {\bibinfo  {journal} {Phys. Rev. Research}\ }\textbf {\bibinfo {volume}
  {2}},\ \bibinfo {pages} {043418} (\bibinfo {year}
  {2020}{\natexlab{b}})}\BibitemShut {NoStop}%
\bibitem [{\citenamefont {S\o{}rensen}\ and\ \citenamefont
  {M\o{}lmer}(1999)}]{Sorensen1999}%
  \BibitemOpen
  \bibfield  {author} {\bibinfo {author} {\bibfnamefont {A.}~\bibnamefont
  {S\o{}rensen}}\ and\ \bibinfo {author} {\bibfnamefont {K.}~\bibnamefont
  {M\o{}lmer}},\ }\href {https://doi.org/10.1103/PhysRevLett.82.1971}
  {\bibfield  {journal} {\bibinfo  {journal} {Phys. Rev. Lett.}\ }\textbf
  {\bibinfo {volume} {82}},\ \bibinfo {pages} {1971} (\bibinfo {year}
  {1999})}\BibitemShut {NoStop}%
\bibitem [{\citenamefont {S\o{}rensen}\ and\ \citenamefont
  {M\o{}lmer}(2000)}]{Sorensen2000}%
  \BibitemOpen
  \bibfield  {author} {\bibinfo {author} {\bibfnamefont {A.}~\bibnamefont
  {S\o{}rensen}}\ and\ \bibinfo {author} {\bibfnamefont {K.}~\bibnamefont
  {M\o{}lmer}},\ }\href {https://doi.org/10.1103/PhysRevA.62.022311} {\bibfield
   {journal} {\bibinfo  {journal} {Phys. Rev. A}\ }\textbf {\bibinfo {volume}
  {62}},\ \bibinfo {pages} {022311} (\bibinfo {year} {2000})}\BibitemShut
  {NoStop}%
\bibitem [{\citenamefont {Shahmoon}\ \emph {et~al.}(2019)\citenamefont
  {Shahmoon}, \citenamefont {Lukin},\ and\ \citenamefont
  {Yelin}}]{Shahmoon2019}%
  \BibitemOpen
  \bibfield  {author} {\bibinfo {author} {\bibfnamefont {E.}~\bibnamefont
  {Shahmoon}}, \bibinfo {author} {\bibfnamefont {M.~D.}\ \bibnamefont
  {Lukin}},\ and\ \bibinfo {author} {\bibfnamefont {S.~F.}\ \bibnamefont
  {Yelin}},\ }in\ \href@noop {} {\emph {\bibinfo {booktitle} {Advances In
  Atomic, Molecular, and Optical Physics}}},\ Vol.~\bibinfo {volume} {68}\
  (\bibinfo  {publisher} {Elsevier},\ \bibinfo {year} {2019})\ pp.\ \bibinfo
  {pages} {1--38}\BibitemShut {NoStop}%
\bibitem [{\citenamefont {Shahmoon}\ \emph {et~al.}(2020)\citenamefont
  {Shahmoon}, \citenamefont {Lukin},\ and\ \citenamefont
  {Yelin}}]{Shahmoon2020}%
  \BibitemOpen
  \bibfield  {author} {\bibinfo {author} {\bibfnamefont {E.}~\bibnamefont
  {Shahmoon}}, \bibinfo {author} {\bibfnamefont {M.~D.}\ \bibnamefont
  {Lukin}},\ and\ \bibinfo {author} {\bibfnamefont {S.~F.}\ \bibnamefont
  {Yelin}},\ }\href {https://doi.org/10.1103/PhysRevA.101.063833} {\bibfield
  {journal} {\bibinfo  {journal} {Phys. Rev. A}\ }\textbf {\bibinfo {volume}
  {101}},\ \bibinfo {pages} {063833} (\bibinfo {year} {2020})}\BibitemShut
  {NoStop}%
\bibitem [{\citenamefont {Marzoli}\ \emph {et~al.}(1994)\citenamefont
  {Marzoli}, \citenamefont {Cirac}, \citenamefont {Blatt},\ and\ \citenamefont
  {Zoller}}]{Marzoli1994}%
  \BibitemOpen
  \bibfield  {author} {\bibinfo {author} {\bibfnamefont {I.}~\bibnamefont
  {Marzoli}}, \bibinfo {author} {\bibfnamefont {J.~I.}\ \bibnamefont {Cirac}},
  \bibinfo {author} {\bibfnamefont {R.}~\bibnamefont {Blatt}},\ and\ \bibinfo
  {author} {\bibfnamefont {P.}~\bibnamefont {Zoller}},\ }\href
  {https://doi.org/10.1103/PhysRevA.49.2771} {\bibfield  {journal} {\bibinfo
  {journal} {Phys. Rev. A}\ }\textbf {\bibinfo {volume} {49}},\ \bibinfo
  {pages} {2771} (\bibinfo {year} {1994})}\BibitemShut {NoStop}%
\bibitem [{\citenamefont {Gonz{\'a}lez-Tudela}\ \emph
  {et~al.}(2015)\citenamefont {Gonz{\'a}lez-Tudela}, \citenamefont {Hung},
  \citenamefont {Chang}, \citenamefont {Cirac},\ and\ \citenamefont
  {Kimble}}]{GonzalezTudela2015}%
  \BibitemOpen
  \bibfield  {author} {\bibinfo {author} {\bibfnamefont {A.}~\bibnamefont
  {Gonz{\'a}lez-Tudela}}, \bibinfo {author} {\bibfnamefont {C.~L.}\
  \bibnamefont {Hung}}, \bibinfo {author} {\bibfnamefont {D.~E.}\ \bibnamefont
  {Chang}}, \bibinfo {author} {\bibfnamefont {J.~I.}\ \bibnamefont {Cirac}},\
  and\ \bibinfo {author} {\bibfnamefont {H.~J.}\ \bibnamefont {Kimble}},\
  }\href {https://doi.org/10.1038/nphoton.2015.54} {\bibfield  {journal}
  {\bibinfo  {journal} {Nature Photonics}\ }\textbf {\bibinfo {volume} {9}},\
  \bibinfo {pages} {320} (\bibinfo {year} {2015})}\BibitemShut {NoStop}%
\bibitem [{\citenamefont {Hung}\ \emph {et~al.}(2016)\citenamefont {Hung},
  \citenamefont {Gonz{\'a}lez-Tudela}, \citenamefont {Cirac},\ and\
  \citenamefont {Kimble}}]{Hung2016}%
  \BibitemOpen
  \bibfield  {author} {\bibinfo {author} {\bibfnamefont {C.-L.}\ \bibnamefont
  {Hung}}, \bibinfo {author} {\bibfnamefont {A.}~\bibnamefont
  {Gonz{\'a}lez-Tudela}}, \bibinfo {author} {\bibfnamefont {J.~I.}\
  \bibnamefont {Cirac}},\ and\ \bibinfo {author} {\bibfnamefont {H.~J.}\
  \bibnamefont {Kimble}},\ }\href {https://doi.org/10.1073/pnas.1603777113}
  {\bibfield  {journal} {\bibinfo  {journal} {Proceedings of the National
  Academy of Sciences}\ }\textbf {\bibinfo {volume} {113}},\ \bibinfo {pages}
  {E4946} (\bibinfo {year} {2016})}\BibitemShut {NoStop}%
\bibitem [{\citenamefont {Stenholm}(1986)}]{Stenholm1986}%
  \BibitemOpen
  \bibfield  {author} {\bibinfo {author} {\bibfnamefont {S.}~\bibnamefont
  {Stenholm}},\ }\href {https://doi.org/10.1103/RevModPhys.58.699} {\bibfield
  {journal} {\bibinfo  {journal} {Rev. Mod. Phys.}\ }\textbf {\bibinfo {volume}
  {58}},\ \bibinfo {pages} {699} (\bibinfo {year} {1986})}\BibitemShut
  {NoStop}%
\bibitem [{\citenamefont {Javanainen}(1986)}]{Javanainen1986}%
  \BibitemOpen
  \bibfield  {author} {\bibinfo {author} {\bibfnamefont {J.}~\bibnamefont
  {Javanainen}},\ }\href {https://doi.org/10.1103/PhysRevLett.56.1798}
  {\bibfield  {journal} {\bibinfo  {journal} {Phys. Rev. Lett.}\ }\textbf
  {\bibinfo {volume} {56}},\ \bibinfo {pages} {1798} (\bibinfo {year}
  {1986})}\BibitemShut {NoStop}%
\bibitem [{\citenamefont {Javanainen}(1988)}]{Javanainen1988}%
  \BibitemOpen
  \bibfield  {author} {\bibinfo {author} {\bibfnamefont {J.}~\bibnamefont
  {Javanainen}},\ }\href {https://doi.org/10.1364/JOSAB.5.000073} {\bibfield
  {journal} {\bibinfo  {journal} {J. Opt. Soc. Am. B}\ }\textbf {\bibinfo
  {volume} {5}},\ \bibinfo {pages} {73} (\bibinfo {year} {1988})}\BibitemShut
  {NoStop}%
\bibitem [{\citenamefont {Palmer}\ and\ \citenamefont
  {Beige}(2010)}]{Palmer2010}%
  \BibitemOpen
  \bibfield  {author} {\bibinfo {author} {\bibfnamefont {R.~N.}\ \bibnamefont
  {Palmer}}\ and\ \bibinfo {author} {\bibfnamefont {A.}~\bibnamefont {Beige}},\
  }\href {https://doi.org/10.1103/PhysRevA.81.053411} {\bibfield  {journal}
  {\bibinfo  {journal} {Phys. Rev. A}\ }\textbf {\bibinfo {volume} {81}},\
  \bibinfo {pages} {053411} (\bibinfo {year} {2010})}\BibitemShut {NoStop}%
\bibitem [{\citenamefont {Bethe}(1947)}]{Bethe1947}%
  \BibitemOpen
  \bibfield  {author} {\bibinfo {author} {\bibfnamefont {H.~A.}\ \bibnamefont
  {Bethe}},\ }\href {https://doi.org/10.1103/PhysRev.72.339} {\bibfield
  {journal} {\bibinfo  {journal} {Phys. Rev.}\ }\textbf {\bibinfo {volume}
  {72}},\ \bibinfo {pages} {339} (\bibinfo {year} {1947})}\BibitemShut
  {NoStop}%
\bibitem [{\citenamefont {Agarwal}(1973)}]{Agarwal1973}%
  \BibitemOpen
  \bibfield  {author} {\bibinfo {author} {\bibfnamefont {G.~S.}\ \bibnamefont
  {Agarwal}},\ }\href {https://doi.org/10.1103/PhysRevA.7.1195} {\bibfield
  {journal} {\bibinfo  {journal} {Phys. Rev. A}\ }\textbf {\bibinfo {volume}
  {7}},\ \bibinfo {pages} {1195} (\bibinfo {year} {1973})}\BibitemShut
  {NoStop}%
\bibitem [{\citenamefont {Fleming}\ \emph {et~al.}(2010)\citenamefont
  {Fleming}, \citenamefont {Cummings}, \citenamefont {Anastopoulos},\ and\
  \citenamefont {Hu}}]{Fleming2010}%
  \BibitemOpen
  \bibfield  {author} {\bibinfo {author} {\bibfnamefont {C.}~\bibnamefont
  {Fleming}}, \bibinfo {author} {\bibfnamefont {N.~I.}\ \bibnamefont
  {Cummings}}, \bibinfo {author} {\bibfnamefont {C.}~\bibnamefont
  {Anastopoulos}},\ and\ \bibinfo {author} {\bibfnamefont {B.~L.}\ \bibnamefont
  {Hu}},\ }\href {https://doi.org/10.1088/1751-8113/43/40/405304} {\bibfield
  {journal} {\bibinfo  {journal} {Journal of Physics A: Mathematical and
  Theoretical}\ }\textbf {\bibinfo {volume} {43}},\ \bibinfo {pages} {405304}
  (\bibinfo {year} {2010})}\BibitemShut {NoStop}%
\bibitem [{Note7()}]{Note7}%
  \BibitemOpen
  \bibinfo {note} {The same kind of approximation was originally adopted
  in~\cite {Fermi1932} to preserve causality in the field-mediated interaction
  between two atoms. Such \protect \emph {ad hoc} correction is also discussed
  more recently in~\cite {Ujihara2002,Haizhen2004,Dolce2006}}\BibitemShut
  {NoStop}%
\bibitem [{\citenamefont {Sternheim}\ and\ \citenamefont
  {Walker}(1972)}]{Sternheim1972}%
  \BibitemOpen
  \bibfield  {author} {\bibinfo {author} {\bibfnamefont {M.~M.}\ \bibnamefont
  {Sternheim}}\ and\ \bibinfo {author} {\bibfnamefont {J.~F.}\ \bibnamefont
  {Walker}},\ }\href {https://doi.org/10.1103/PhysRevC.6.114} {\bibfield
  {journal} {\bibinfo  {journal} {Phys. Rev. C}\ }\textbf {\bibinfo {volume}
  {6}},\ \bibinfo {pages} {114} (\bibinfo {year} {1972})}\BibitemShut {NoStop}%
\bibitem [{\citenamefont {Haydock}(1980)}]{Haydock1980}%
  \BibitemOpen
  \bibfield  {author} {\bibinfo {author} {\bibfnamefont {R.}~\bibnamefont
  {Haydock}},\ }\href
  {https://doi.org/https://doi.org/10.1016/0010-4655(80)90101-0} {\bibfield
  {journal} {\bibinfo  {journal} {Computer Physics Communications}\ }\textbf
  {\bibinfo {volume} {20}},\ \bibinfo {pages} {11} (\bibinfo {year}
  {1980})}\BibitemShut {NoStop}%
\bibitem [{\citenamefont {Kornovan}\ \emph {et~al.}(2019)\citenamefont
  {Kornovan}, \citenamefont {Corzo}, \citenamefont {Laurat},\ and\
  \citenamefont {Sheremet}}]{Kornovan2019}%
  \BibitemOpen
  \bibfield  {author} {\bibinfo {author} {\bibfnamefont {D.~F.}\ \bibnamefont
  {Kornovan}}, \bibinfo {author} {\bibfnamefont {N.~V.}\ \bibnamefont {Corzo}},
  \bibinfo {author} {\bibfnamefont {J.}~\bibnamefont {Laurat}},\ and\ \bibinfo
  {author} {\bibfnamefont {A.~S.}\ \bibnamefont {Sheremet}},\ }\href
  {https://doi.org/10.1103/PhysRevA.100.063832} {\bibfield  {journal} {\bibinfo
   {journal} {Phys. Rev. A}\ }\textbf {\bibinfo {volume} {100}},\ \bibinfo
  {pages} {063832} (\bibinfo {year} {2019})}\BibitemShut {NoStop}%
\bibitem [{\citenamefont {Chen}\ \emph {et~al.}()\citenamefont {Chen},
  \citenamefont {Demmel}, \citenamefont {Gu}, \citenamefont {Saad},
  \citenamefont {Lehoucq}, \citenamefont {Sorensen}, \citenamefont {Maschhoff},
  \citenamefont {Bai}, \citenamefont {Day}, \citenamefont {Freund},
  \citenamefont {Sleijpen}, \citenamefont {van~der Vorst},\ and\ \citenamefont
  {Li}}]{ChenCh7}%
  \BibitemOpen
  \bibfield  {author} {\bibinfo {author} {\bibfnamefont {T.}~\bibnamefont
  {Chen}}, \bibinfo {author} {\bibfnamefont {J.}~\bibnamefont {Demmel}},
  \bibinfo {author} {\bibfnamefont {M.}~\bibnamefont {Gu}}, \bibinfo {author}
  {\bibfnamefont {Y.}~\bibnamefont {Saad}}, \bibinfo {author} {\bibfnamefont
  {R.}~\bibnamefont {Lehoucq}}, \bibinfo {author} {\bibfnamefont
  {D.}~\bibnamefont {Sorensen}}, \bibinfo {author} {\bibfnamefont
  {K.}~\bibnamefont {Maschhoff}}, \bibinfo {author} {\bibfnamefont
  {Z.}~\bibnamefont {Bai}}, \bibinfo {author} {\bibfnamefont {D.}~\bibnamefont
  {Day}}, \bibinfo {author} {\bibfnamefont {R.}~\bibnamefont {Freund}},
  \bibinfo {author} {\bibfnamefont {G.}~\bibnamefont {Sleijpen}}, \bibinfo
  {author} {\bibfnamefont {H.}~\bibnamefont {van~der Vorst}},\ and\ \bibinfo
  {author} {\bibfnamefont {R.}~\bibnamefont {Li}},\ }\bibinfo {title} {7.
  non-hermitian eigenvalue problems},\ in\ \href
  {https://doi.org/10.1137/1.9780898719581.ch7} {\emph {\bibinfo {booktitle}
  {Templates for the Solution of Algebraic Eigenvalue Problems}}},\ pp.\
  \bibinfo {pages} {149--231}\BibitemShut {NoStop}%
\bibitem [{\citenamefont {Fermi}(1932)}]{Fermi1932}%
  \BibitemOpen
  \bibfield  {author} {\bibinfo {author} {\bibfnamefont {E.}~\bibnamefont
  {Fermi}},\ }\href {https://doi.org/10.1103/RevModPhys.4.87} {\bibfield
  {journal} {\bibinfo  {journal} {Rev. Mod. Phys.}\ }\textbf {\bibinfo {volume}
  {4}},\ \bibinfo {pages} {87} (\bibinfo {year} {1932})}\BibitemShut {NoStop}%
\bibitem [{\citenamefont {Ujihara}\ and\ \citenamefont
  {Dung}(2002)}]{Ujihara2002}%
  \BibitemOpen
  \bibfield  {author} {\bibinfo {author} {\bibfnamefont {K.}~\bibnamefont
  {Ujihara}}\ and\ \bibinfo {author} {\bibfnamefont {H.~T.}\ \bibnamefont
  {Dung}},\ }\href {https://doi.org/10.1103/PhysRevA.66.053807} {\bibfield
  {journal} {\bibinfo  {journal} {Phys. Rev. A}\ }\textbf {\bibinfo {volume}
  {66}},\ \bibinfo {pages} {053807} (\bibinfo {year} {2002})}\BibitemShut
  {NoStop}%
\bibitem [{\citenamefont {Haizhen}\ and\ \citenamefont
  {Ujihara}(2004)}]{Haizhen2004}%
  \BibitemOpen
  \bibfield  {author} {\bibinfo {author} {\bibfnamefont {Z.}~\bibnamefont
  {Haizhen}}\ and\ \bibinfo {author} {\bibfnamefont {K.}~\bibnamefont
  {Ujihara}},\ }\href
  {https://doi.org/https://doi.org/10.1016/j.optcom.2004.06.016} {\bibfield
  {journal} {\bibinfo  {journal} {Optics Communications}\ }\textbf {\bibinfo
  {volume} {240}},\ \bibinfo {pages} {153} (\bibinfo {year}
  {2004})}\BibitemShut {NoStop}%
\bibitem [{\citenamefont {Dolce}\ \emph {et~al.}(2006)\citenamefont {Dolce},
  \citenamefont {Passante},\ and\ \citenamefont {Persico}}]{Dolce2006}%
  \BibitemOpen
  \bibfield  {author} {\bibinfo {author} {\bibfnamefont {I.}~\bibnamefont
  {Dolce}}, \bibinfo {author} {\bibfnamefont {R.}~\bibnamefont {Passante}},\
  and\ \bibinfo {author} {\bibfnamefont {F.}~\bibnamefont {Persico}},\ }\href
  {https://doi.org/https://doi.org/10.1016/j.physleta.2006.02.018} {\bibfield
  {journal} {\bibinfo  {journal} {Physics Letters A}\ }\textbf {\bibinfo
  {volume} {355}},\ \bibinfo {pages} {152} (\bibinfo {year}
  {2006})}\BibitemShut {NoStop}%
\end{thebibliography}%
\end{document}